# The History of AI Rights Research

Jamie Harris: Sentience Institute, jamie@sentienceinstitute.org

## Abstract

This report documents the history of research on AI rights and other moral consideration of artificial entities. It highlights key intellectual influences on this literature as well as research and academic discussion addressing the topic more directly. We find that researchers addressing AI rights have often seemed to be unaware of the work of colleagues whose interests overlap with their own. Academic interest in this topic has grown substantially in recent years; this reflects wider trends in academic research, but it seems that certain influential publications, the gradual, accumulating ubiquity of AI and robotic technology, and relevant news events may all have encouraged increased academic interest in this specific topic. We suggest four levers that, if pulled on in the future, might increase interest further: the adoption of publication strategies similar to those of the most successful previous contributors; increased engagement with adjacent academic fields and debates; the creation of specialized journals, conferences, and research institutions; and more exploration of legal rights for artificial entities.

## Introduction

Can, and should, AIs have rights? In the past decade or so, these questions have become the focus of legislative proposals, media articles, and public debate (see Harris & Anthis, 2021), as well as academic books (Gunkel, 2018; Gellers, 2020) and journal publications (e.g. Coeckelbergh, 2010; Robertson, 2014). Academic discussion of AI rights, robot rights, and the moral consideration of artificial entities more broadly (sometimes collectively referred to here simply as "AI rights," for brevity) has grown exponentially (Harris & Anthis, 2021; see Figure 1).[1] This report presents a chronology of that growth and its contributing factors, discusses the causes of increased academic interest in the topic, and then reviews possible lessons for stakeholders seeking to increase interest further.

---

[1] For discussions of terminology, see Harris and Anthis (2021) and Pauketat (2021).

Figure 1: Cumulative total of academic publications on the moral consideration of artificial entities, by date of publication (Harris & Anthis, 2021)

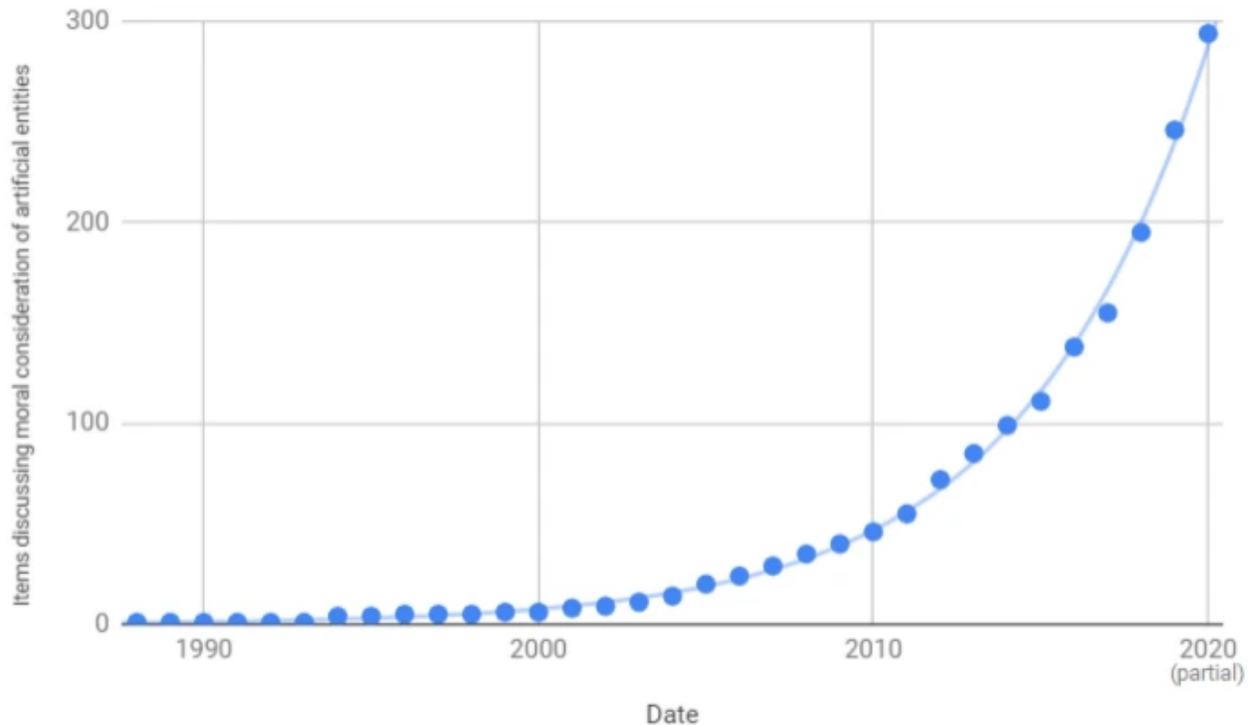

Researchers approach the topic with different motivations, influences, and methodologies, often seemingly unaware of the work of other academics whose interests overlap with their own. This report seeks to contextualize and connect the relevant streams of research in order to encourage further study. This is especially important because granting sentient AI moral consideration, such as protection in society's laws or social norms, may be important for preventing large-scale suffering or other serious wrongs in the future (Anthis & Paez, 2021), and academic field-building is a tractable stepping stone towards this form of moral circle expansion (Harris, 2021).

# Methodology

We began with a review of the publications identified by Harris and Anthis' (2021) literature review, which systematically searched for articles via certain keywords relevant to AI rights and other moral consideration of artificial entities. The reference lists of important included publications were then reviewed, as were the lists of items that cited those publications, primarily using the "Cited by…" button provided on Google Scholar. The titles — and sometimes abstracts — of identified items were reviewed to decide whether further reading or a mention in the text was warranted.

Unlike Harris and Anthis (2021), this report was not a formal, quantitative literature review. There were no strict inclusion or exclusion criteria, but an item was more likely to be read and discussed in detail if it:
- was in an academic format (e.g. journal article, conference paper, or edited book);
- addressed the moral consideration of artificial entities explicitly and in some depth;

- appeared to have arisen independently of other included items (e.g. did not reference previous relevant items or added a different perspective);
- was written in English; and
- was written in the 20th or 21st century.

The results are presented in a thematic narrative, roughly in chronological order, with categorizations emerging during the analysis rather than being fit to a predetermined framework. We focus on implications and hypotheses generated during the analysis, rather than assessing the presence or absence of factors identified as potentially important for the emergence of "scientific/intellectual movements" by previous studies (e.g. Frickel & Gross, 2005; Animal Ethics, 2021).

The thematic narrative is supplemented by keyword searches through PDFs of the publications identified by Harris and Anthis' (2021) systematic searches that met their inclusion criteria, where the full texts could be identified (270 of 294 items, i.e. 92%). As well as being broken out into tables in the relevant sections below, the full results of these searches and the list of included items are provided in a separate [spreadsheet].[2] The keywords were chosen based on expectations about which would be most likely to generate meaningful results,[3] but individual publications returned by the searches were not manually checked to ensure that they actually mentioned the author, item, or idea referred to by the keyword.

# Results

Figure 2 presents a summary of the different ideas and research identified as having contributed to the nascent research field around AI rights and other moral consideration of artificial entities, each of which will be explored in more depth in the subsections below.

---

[2] In the tables below, the authors themselves are excluded from the count. For example, a search for "Gunkel" returns 87 results of which 14 are articles that David Gunkel wrote or co-authored, so the table below would report the total number as being 73, which is 28.5% of the 256.

[3] E.g., the keyword "A Space Odyssey" was used rather than "2001", which would return any result with a citation dated to 2001.

Figure 2: A summary chronology of contributions to academic discussion of AI rights and other moral consideration of artificial entities

| Pre-20th century | Mid-20th century | 1970s | 1980s | 1990s | Early 2000s | Late 2000s | 2010s and 2020s |
|---|---|---|---|---|---|---|---|
| | | | | | | | Synthesis |
| | | | | | | Moral and social psych | |
| | | | | | | Social-relational ethics | |
| | | | | | HCI and HRI | | |
| | | | | | Machine ethics and roboethics | | |
| | | | | | Floridi's information ethics | | |
| | | | | Transhumanism, EA, and longtermism | | | |
| | | | Legal rights for artificial entities | | | | |
| | | Animal ethics | | | | | |
| | Environmental ethics | | | | | | |
| Artificial life and consciousness | | | | | | | |
| Science fiction | | | | | | | |

## Science fiction

"The notion of robot rights," as Seo-Young Chu (2010, p. 215) points out, "is as old as is the word 'robot' itself. Etymologically the word 'robot' comes from the Czech word 'robota,' which means 'forced labor.' In Karel Čapek's 1921 play *R.U.R.* [*Rossum's Universal Robots*], which is widely credited with introducing the term 'robot,' a 'Humanity League' decries the exploitation of robots slaves — 'they are to be dealt with like human beings,' one reformer declares — and the robots themselves eventually stage a massive revolt against their human makers." The list of science fiction or mythological mentions of robots or other intelligent artificial entities is extensive and long predates *R.U.R.*, including numerous stories from Greek and Roman antiquity with automata and sculptures that came to life (Wikipedia, 2021).

Even some of the earliest academic publications explicitly addressing the moral consideration of artificial entities (e.g. Putnam, 1964; Lehman-Wilzig, 1981) set themselves against the backdrop of plentiful science fiction treatments of the topic. Petersen's (2007) exploration of "the ethics of robot servitude" presents the topic as "natural and engaging… given the prevalence of robot servants in pop culture."[4] Some works of fiction have become especially widely referenced in the academic literature that has developed around the moral consideration of artificial entities (see Table 1).

Table 1: Science fiction keyword searches

---

[4] Petersen (2007) concludes that the topic is "strangely neglected," citing a few examples of previous brief discussion of the topic, but missing several relevant streams of literature, such as most of the previous writings on legal rights for artificial entities, transhumanism, and information ethics (see the relevant subsections below).

| Keyword | Items mentioning | % of items |
|---------|------------------|------------|
| "Science fiction" | 101 | 37.4% |
| Asimov | 71 | 26.3% |
| Frankenstein | 30 | 11.1% |
| "Star Trek" | 23 | 8.5% |
| R.U.R. | 19 | 7.0% |
| Terminator | 18 | 6.7% |
| "Ex Machina" | 17 | 6.3% |
| "Star Wars" | 16 | 5.9% |
| "A Space Odyssey" | 14 | 5.2% |
| Westworld | 14 | 5.2% |
| "The Matrix" | 13 | 4.8% |
| "Real Humans" | 7 | 2.6% |
| "Do Androids Dream of Electric Sheep" | 6 | 2.2% |
| Bladerunner | 4 | 1.5% |

While some of these works explicitly address the moral consideration of artificial entities, such as *R.U.R.* and *Real Humans*, others usually just provide popular culture reference points for artificial entities, such as *Star Wars, Star Trek*, and *Terminator*. The list above refers to Western sci-fi, but sci-fi has likely been an influence on social and moral attitudes elsewhere, too (e.g. Krebs, 2006).

## Artificial life and consciousness

Enlightenment philosophers and scientists' exploration of consciousness and other morally relevant questions sometimes included reference to machines or automata. For example, Rene Descartes discussed the capacities and moral value of animals with reference to the physical processes of machines, and he explored whether or not the human mind could be mechanized (Harrison, 1992; Wheeler, 2008). Diderot (2012; first edition 1782) recorded in *D'Alembert's Dream*, a series of philosophical dialogues, discussions of machines in the exploration of what might constitute "a unified system, on its own, with an awareness of its own unity."

Some of the earliest mathematicians and scientists who worked on the development of computers and AI addressed the question of whether these entities could think or otherwise possess intelligence. Indeed, Alan Turing's famous "Imitation Game" — in which an observer would seek to distinguish a machine from a human by asking them both questions — was designed to partly address this (Oppy & Dowe, 2021). This seems very closely adjacent to the questions of whether artificial entities might be able to feel emotions or have other conscious experiences, which were raised in academic discussion at least as early as 1949 (Oppy & Dowe, 2021).[5] Marvin Minsky, one of the researchers who proposed and attended the 1956 Dartmouth workshop (McCarthy et al., 2006), which is often credited as being a pivotal event in the foundation of the field of artificial intelligence (e.g. Nilsson, 2009), later argued that "some machines are already potentially more conscious than are people" (e.g. Minsky, 1991).[6]

In *Dimensions of Mind*, the proceedings of the third annual New York University Institute of Philosophy, Norbert Wiener (another pioneer of AI research) noted (1960) that the increasing complexity of machine programming "gives rise to certain questions of a quasi-moral and a quasi-human nature. We have to face the fundamental paradox of slavery. I do not refer to the cruelty of slavery, which we can neglect entirely for the moment as I do not suppose that we shall feel any moral responsibility for the welfare of the machine; I refer to the contradictions besetting slavery as to its effectiveness."[7]

In the same proceedings, philosopher Michael Scriven (1960, pp. 139-42) critiqued the Turing Test (Turing's "Imitation Game") as "oversimple" for testing whether "a robot… had feelings," but commented that such questions might nevertheless lead to "the prosecution of novel moral causes (Societies for the Prevention of Cruelty to Robots, etc.)." Scriven (1960) then proposed an alternative test, where after teaching a robot the English language and to not lie, we could ask it whether it has feelings or is "a person"; Scriven commented that "the first [robot] to answer 'Yes' will qualify" as a person.[8]

The philosopher Hilary Putnam briefly addressed the idea that machines might have "souls" in the same proceedings (1960).[9] Later, in a paper for a symposium called "Minds and Machines," Putnam (1964) explored whether "robots" were "artificially created life." Putnam (1964) opened by pointing out that, "[a]t least in the literature of science fiction, then, it is possible for a robot to be 'conscious'; that means… to have feelings, thoughts, attitudes, and character traits." The article's exploration of the possibility of robot consciousness was motivated by concern for "how we should speak about humans and not with how we should speak about machines,"[10] but Putnam (1964) nevertheless commented that this philosophical question may become "the problem of the 'civil rights of robots'... much faster than any of us now expect. Given the ever-accelerating rate of both technological and social change, it is entirely possible that robots will one day exist, and argue 'we are alive; we are conscious!'"

---

[5] For some other examples of early, adjacent discussion, see footnote 2 in Thompson (1965).

[6] I have not identified any papers by Minsky from the 1950s explicitly claiming that artificial sentience was possible.

[7] Wiener (1960) added that, "[a] slave is expected to have two qualities: intelligence, and subservence. These two qualities are by no means perfectly compatible. The more intelligent the slave is, the more he will insist on his own way of doing things in opposition to the way of doing things imposed on him by his owner. To that extent he will cease to be a slave."

[8] One contributor in the same proceedings (Watanabe, 1960) commented in reply to Scriven (1960) that, "[i]f a machine is made out of protein, then it may have consciousness, but a machine made out of vacuum tubes, diodes, and transistors cannot be expected to have consciousness. I do not here offer a proof for this statement, except that it is obvious according to well-disciplined common sense. A 'conscious' machine made out of protein is no longer a machine, it is a man-made animal."

[9] Putnam (1960) closes with the comment that, "if the mind-body problem is identified with any problem of more than purely conceptual interest (e.g. with the question of whether human beings have 'souls'), then *either* it must be that (a) no argument *ever* used by a philosopher sheds the *slightest* light on it (and this independently of the way the argument tends), or (b) that some philosophic argument for mechanism is correct, or (c) that some dualistic argument does show that *both* human beings *and* Turing machines have souls! I leave it to the reader to decide which of these three alternatives is at all plausible."

The main focus of most of the contributors to the section of the proceedings on "The brain and the machine," was on the capabilities of artificial entities, with philosopher and art critic Arthur Danto's (1960) chapter the most explicitly focused on "consciousness." Most also made some brief comments relevant to moral consideration. Sidney Hook's (1960, p. 206) concluding "pragmatic note" to the section included the comment that, "[a] situation described by the Czech dramatist Karel Capek in his *R.U.R.* may someday come to pass."

A number of other publications discussed the possibility of artificial consciousness in the 1960s (e.g. Thompson, 1965; Simon, 1969), and discussion has continued since then (e.g. Reggia, 2013; Kak, 2021).[11] Two of the most cited contributors to the discussion of artificial consciousness or sentience are the philosophers Daniel Dennett and John Searle. As well as being very widely cited in mainstream philosophy and cognitive science (e.g. Dennett has over 114,000 citations to date; Google Scholar, 2021g), they are often cited among the writers who explicitly discuss the moral consideration of artificial entities (see Table 3). Dennett's arguments are often cited in support of claims that artificial consciousness is possible.[12] Some of his earliest writings touched on this topic, such as his (1971) argument that, "on occasion a purely physical [e.g. artificial] system can be so complex, and yet so organized, that we find it convenient, explanatory, pragmatically necessary for prediction, to treat it as if it had beliefs and desires and was rational," because "it is much easier to decide whether a machine can be an Intentional system than it is to decide whether a machine can really think, or be conscious, or morally responsible."[13] In contrast, Searle (1980) used a "Chinese room" thought experiment to argue that whereas a machine might *appear* to understand something, this does not mean that it *actually* understands it. The idea can be extended to consider whether a simulation "really is a mind" or merely a "model of the mind," and whether one can really "create consciousness" (Searle, 2009).

The possibility of artificial consciousness, then, has long been a mainstream topic among technical AI researchers, philosophers, and cognitive scientists. As Versenyi (1974) noted, this discussion clearly has

[10] Putnam (1964) adds that, "[m]y interest in the latter question derives from my just-mentioned conviction: that clarity with respect to the 'borderline case' of robots, if it can only be achieved, will carry with it clarity with respect to the 'central area' of talk about feelings, thoughts, consciousness, life, etc."

[11] The field has moved beyond merely discussing whether consciousness in artificial entities is possible to a proactive effort to create it (Holland & Goodman, 2003; Gamez, 2008; Reggia, 2013), with an academic journal explicitly advancing this goal (World Scientific, 2021).

Gamez's (2008) review described this field of "machine consciousness" as "a relatively new research area that has gained considerable momentum over the last few years." Of the 85 references in the article, 52 (61%) were published in the 2000s and a further 23 (27%) were published in the 1990s.

[12] In an interview (Thornhill, 2017), Dennett summarized that he had been "arguing for years that, yes, in principle it's possible for human consciousness to be realised in a machine. After all, that's what we are… We're robots made of robots made of robots. We're incredibly complex, trillions of moving parts. But they're all non-miraculous robotic parts." For an example of academic discussion, see Dennett (1994).

[13] Dennett was well aware of Putnam's work. For example, Dennett (1978) cites various publications by Putnam, including the 1964 article that mentions "civil rights of robots."

ethical implications, even if these have not always been referred to explicitly or at length. Indeed, explicit and detailed discussion of the moral consideration of artificial entities seems to have remained somewhat rare in the following decades.[14] However, research on artificial life and consciousness continued to inspire publications relevant to AI rights into the 21st century (e.g.Sullins, 2005; Torrance, 2007); sometimes discussion seems to have arisen without reference to many of the previous publications relevant to moral consideration of artificial entities.[15]

---

[14] Harris and Anthis' (2021) systematic search methods identified few publications discussing this topic in much depth before the 21st century, with the earliest item being McNally and Inayatullah (1988).

Searching through the items that have cited Putnam (1964) does reveal a few items that touch on the topic, but the discussion tends to be tangential or brief. For example, Versenyi (1974) was primarily concerned with the question of moral agency of artificial entities, but links this briefly to the idea of moral patiency, noting that "whether robots should be blamed or praised, loved or hated, given rights and duties, etc., are in principle the same sort of questions as, 'Should cars be serviced, cared for, and supplied with what they require for their operation?'" In the final section of their article, Wilks (1975) examined the arguments of Hilary Putnam and J. J. Clarke, suggesting that although "[n]either of them considers the privacy of machines seriously," their arguments nevertheless support "a frivolous speculation about the possible privacy of a machine," where observers would "ascribe to the machine the final authority as to what state it was in, in the way that we now do for persons." Granting an entity this authority could be interpreted as a form of moral consideration, but it seems less relevant than Putnam's (1964) brief comments about "civil rights of robots." Sapontzis (1981) cited Putnam (1964) in "a critique of personhood" as a useful concept for "moral theory and practice," briefly using "machines" as a contrast to persons.

Lycan (1985) addressed the topic in some depth, arguing that "[i]t seems… possible (as they say) in principle to build our own androids, artificial humans, which would have at least as firm a claim to be called persons as we do… It would seem that these artificial humans, if they are indeed as clearly entitled to be called persons as we are, will have moral rights of exactly the same sort we have, whatever those rights may be." The "main point" of the paper is to address the question "of whether it is wrong for a mother to abort a pre-viable fetus solely for reasons of convenience," although they note that they discussed "the civil rights of robots… more fully" in a lecture at Kansas State University in 1972. The sole publication not by Lycan themselves citing Lycan's (1985) article was about abortion, rather than artificial entities.

Hajdin's (1987) PhD thesis contained a section discussing whether "highly sophisticated future computers" might count as "members of the moral community," but accrued no citations. Scheutz and Crowell (2007) address a number of "Social and Ethical Implications of Autonomous Robots," though the discussion of the possibility of robot rights is very brief and described as not "of pressing urgency, since such questions may only be relevant for robots much more advanced than those available at present." Vize's (2011) master's thesis cites Putnam (1964) prominently in an extensive discussion of the "moral considerability" of machines; a handful of other publications on the topic from this date onwards have cited the article.

[15] Sullins (2005) noted that the '90s saw an "initial burst of articles and books" discussing artificial life, either with the goal of attempting "to describe fundamental qualities of living systems through agent based computer models" or to study "whether or not we can artificially create living things in computational mediums that can be realized either, virtually in software, or through bio-technology." Sullins provided numerous references for research in this technical field. Sullins commented that philosophers "have not helped work through the various ethical issues" raised by this literature. They cited discussion by Floridi and Sanders (2004) but not other previous contributions addressing relevant ethical issues, such as the relevant writings from environmental ethics, animal ethics, legal rights for artificial entities, or transhumanism that predated this publication (see the corresponding sections of this report for examples). Sullins' (2005) article itself discussed a number of moral issues, including both moral agency and patiency of these entities.

Elton (2000) argued that, like animals, agents in video games engage in "cognition" and "striving" to stay alive, and that the possession of these two capacities merits moral consideration. Elton (2000) did not cite previous publications focusing explicitly on the moral consideration of artificial entities, but there are two references to previous discussions of "artificial life," and Dennett was also cited. Kim (2004) cited Elton (2000) and numerous

Some discussion about moral consideration has addressed artificial entities with biological components (e.g. Sullins, 2005; Warwick, 2010). Nevertheless, the development of the field of synthetic biology, which has its roots in the mid-20th century but began to cohere from the early 21st (Cameron et al., 2014), seems to have generated a new stream of ethical discussion that was largely independent of other ongoing discussion about the moral consideration of artificial entities. For example, Douglas and Savulescu (2010) discussed how synthetic biology might create "organisms with the features of both organisms and machines" and expressed concern that people might "misjudge the moral status of some of the new entities," but did not reference any other publications included in the current report or in Harris and Anthis' (2021) literature review.[16]

Table 3: Artificial life and consciousness keyword searches

| Keyword | Items mentioning | % of items |
|---|---|---|
| Conscious | 176 | 65.2% |
| Turing | 116 | 43.0% |

publications about "artificial life," as well as Freitas (1985) and McNally and Inayatullah (1988) who had previously discussed legal rights for artificial entities.

A number of publications by Steve Torrance (e.g. 2000, 2007) discussed the possibility of artificial consciousness and then moved on to discuss the ethical implications that this would have for how we treat artificial entities. Most of the citations in Torrance's 2006 and 2007 papers are from cognitive science (e.g. Dennett) or publications on the development of artificial consciousness. There are also a small number of citations relating to legal personhood for machines, though Torrance's interest in the topic may have predated these publications; Torrance (1984) had long previously edited a volume on *Philosophical Aspects of Artificial Intelligence*, citing Putnam's (1960) "Minds and machines" and subsequent papers as a key influence, alongside Searle and others addressing the capabilities of machines. Torrance's (1984) introduction focused on the capacities and consciousness of artificial entities, but commented briefly on moral issues, noting (p. 14) that, "[a] machine which gave howsoever lifelike an imitation — but only an imitation — of pain would not be adding to the sum total of misery in the universe, would not merit our *concern*, in the way that would a behaviourally indistinguishable machine that really was in agony."

Rodney Brooks, director of the Artificial Intelligence Lab at M.I.T., wrote an article for *Time* magazine (2000) noting that, "[a]rtificial life forms that 'live' inside computers have evolved to the point where they can chase prey, evade predators and compete for limited resources" and commented briefly that "these endeavors will eventually lead to robots to which we will want to extend the same inalienable rights that humans enjoy."

Metzinger (2013) argued against the creation of conscious artificial entities, to avoid the problem of "artificial suffering." Most of the citations were to previous literature on artificial life and consciousness; Metzinger appears to have applied (negative) utilitarian ethics to the problem, without reference to previous works discussing the moral consideration of artificial entities.

[16] Schmidt et al. (2009) and Holm and Powell (2013) did the same, except for the latter publication citing Douglas and Savulescu (2010). There are some indirect connections between this stream of literature and other streams addressing the moral consideration of artificial entities. For example, Julian Savulescu, prior to co-authoring the Douglas and Savulescu (2010) paper, had co-edited Savulescu and Bostrom (2009), and so may well have been aware of relevant ideas from the early transhumanist writers that related to the moral consideration of artificial entities (see the relevant subsection below). Sullins (2009) focused primarily on moral agency rather than moral patency, but noted that "[a]rtificial autonomous agents can be separated into three categories: synthetic biological constructs, robots, and software agents." There is also at least some direct overlap in authors addressing both topics. For example, John Basl has written specifically about the moral consideration of both machines (e.g. Basl, 2014) and the creations of synthetic biology (e.g. Basl & Sandler, 2013).

| | | |
|---|---|---|
| Dennett | 66 | 24.4% |
| Searle | 43 | 15.9% |
| Torrance | 40 | 15.2% |
| Sullins | 26 | 9.7% |
| Putnam | 24 | 8.9% |
| Minsky | 17 | 6.3% |
| Warwick | 14 | 5.2% |
| "Synthetic biology" | 11 | 4.1% |
| Wiener | 10 | 3.7% |

# Environmental ethics

In 1972, Christopher Stone wrote "Should Trees Have Standing?—Toward Legal Rights for Natural Objects," proposing that legal rights be granted to "the natural environment as a whole." The article was cited in a Supreme Court ruling in that same year to suggest that nature could be a legal subject (Gellers, 2020, pp. 106-7). It has also been cited by numerous contributors to more recent discussions of the moral consideration of artificial entities (see Table 4). Although Stone (1972) added a radical legal dimension, a number of other authors had already been advocating for social and moral concern for the environment over the past few decades (Brennan & Lo, 2021), such as Aldo Leopold in *A Sand County Almanac* (1949), which articulated a "Land Ethics" that incorporated respect for all life and the land itself. These writings collectively contributed to the development of the field of environmental ethics (Brennan & Lo, 2021).

Subsequently, Paul Taylor (1981; 2011; first edition 1986), an influential proponent of biocentrism (a strand of environmental ethics), briefly explicitly argued against the moral consideration of currently existing artificial entities, but encouraged open-mindedness to considering future artificial entities.[17] Stone (1987) later briefly raised questions about the legal status of artificial entities, albeit focused more on legal liability than legal rights.[18]

---

[17] Taylor (2011, pp. 122-4) argued that, unlike "[a]ll organisms, whether conscious or not… inanimate objects" cannot be "a teleological center of life… This point holds even for those complex mechanisms (such as self-monitoring space satellites, chess-playing computers, and assembly-line 'robots') that have been constructed by humans to function in a quasi-autonomous, self-regulating manner in the process of accomplishing certain purposes… machines do not, as independent entities, have a good of their own. Their 'good' is 'furthered' only insofar as they are treated in such a way as to be an effective means to human ends." However, Taylor (2011, pp 124-5) added that "this difference between mechanism and organism may no longer be maintainable with regard to those complex electronic devices now being developed under the name of artificial intelligence. Perhaps some day computer scientists and engineers will construct beings whose internal processes and electrical responses to their surroundings closely parallel the functions of the human brain and nervous system. Concerning such beings we may begin to speak of their having a good of their own independently of the purposes of their creators. At that point the distinction drawn above between living things and inanimate machines may break down. It is best, I think, to have an open mind about this."

[18] Stone (1987, pp. 28-9) asked "what place need we make in law and morals for robots, artificial intelligence (A.I), and clones? Is the day so far off that we will be wondering what obligations we ought to hold toward, even expect of, *them*?" The discussion on the following two pages focused, however, on "questions regarding the liability of the

Other writers have subsequently more thoroughly explored the potential application of environmental ethics to the moral consideration of artificial entities. For example, McNally and Inayatullah (1988, summarized below) quoted Stone (1972) extensively, and Kaufman (1994), writing in the journal *Environmental Ethics*, argued that either "machines have interests (and hence moral standing)" as well plants and ecosystems, or that "mentality is a necessary condition for inclusion." Luciano Floridi's (1999, 2013) information ethics and Gellers' (2020) framework draw heavily on environmental ethics. Gellers (2020, pp. 108-17) differentiates separate strands of environmental ethics, arguing that "biocentrism and ecocentrism both support at least legal rights for nature, while only ecocentrism offers a potential avenue for inorganic non-living entities such as intelligent machines to possess moral or legal rights." Hale (2009) drew on environmental ethics and some writings on animal rights to argue that "the technological artefact… is only [morally] considerable insofar as it is valuable to somebody."

Table 4: Environmental ethics keyword searches

| Keyword | Items mentioning | % of items |
|---|---|---|
| Environment | 171 | 63.3% |
| "Environmental ethics" | 49 | 18.1% |
| Ecological | 37 | 13.7% |
| Biocentrism | 16 | 5.9% |
| "Should Trees Have Standing" | 13 | 4.8% |
| "Deep ecology" | 11 | 4.1% |

## Animal ethics

Moral and legal concern for animals has existed to some degree for centuries (Beers, 2006), perhaps especially outside Western thought (Gellers, 2020, p. 63). During the Enlightenment, thinkers such as Descartes, Kant, and Bentham discussed the moral consideration of animals but left an ambiguous record (Gellers, 2020, pp. 64-5). Concern in the West seems to have increased in the 19th century, demonstrated by the creation of new advocacy groups and the introduction of various legal protections, and again from the 1970s, spurred by philosophical contributions from Peter Singer (1995, first edition 1975), Richard Ryder (1975), Tom Regan (2004, first edition 1983), and others (Beers, 2006; Guither, 1998, pp. 1-23).

Given that, like environmental ethics, animal ethics challenges the restriction of moral consideration to humans, it has implications for the moral consideration of artificial entities. For example, Ryder (1992) later elaborated on his theory of "painism," noting that "all painient individuals, whatever form they may take (whether human, nonhuman, extraterrestrial or the artificial machines of the future, alive or inanimate), have rights." Similarly, Singer co-authored a short opinion article for *The Guardian* with Agata Sagan (2009) commenting that, "[t]he history of our relations with the only nonhuman sentient manufacturer" and "the liability of the A.I. itself" in instances where damages or accidents occur. Subsequently, Stone (1987, p. 47-8) uses AI as an example of "utterly disinterested entities devoid of feelings or interests except in the most impoverished or even metaphorical sense," but argues that a legal guardian might nevertheless "be empowered to speak for" them.

beings we have encountered so far – animals – gives no ground for confidence that we would recognise sentient robots as beings with moral standing and interests that deserve consideration." They noted, however, that "[i]f, as seems likely, we develop super-intelligent machines, their rights will need protection, too."

The vast majority of writings that focus on moral consideration of artificial entities discuss the precedent of moral consideration of animals at least briefly (see Table 5), though there are mixed views about whether the analogy is helpful or provides a basis for AI rights (see Gellers, 2020, pp. 76-8 for a summary).

Table 5: Animal ethics keyword searches

| Keyword | Items mentioning | % of items |
|---|---|---|
| Animals | 216 | 80.0% |
| Singer | 107 | 39.6% |
| "Animal rights" | 88 | 32.6% |
| Regan | 35 | 13.0% |
| Ryder | 6 | 2.2% |

## Legal rights for artificial entities

Gellers (2020, pp. 33-5) notes that, in the US, there has been some precedent for legal personhood for corporations and ships — i.e. certain artificial entities — since at least the 19th century, though the correct legal interpretation of some of these cases remains contested. Stretching further back, Gellers (2020, p. 34) also notes that "the Old Testament and Greek, Roman, and Germanic law" provide precedent for assigning various sorts of legal liability to entities that otherwise lack legal standing, such as ships, slaves, and animals.

At a 1979 international symposium on "The Humanities in a Computerized World," political scientist Sam N. Lehman-Wilzig presented a paper exploring possible legal futures for AI that was then published in a revised and expanded format in the journal *Futures* (Lehman-Wilzig, 1981). The first half of the paper focused mostly on the threats that the development of powerful AI might pose to humanity, but the second half focused on possible legal futures, ranging "from the AI robot as a piece of property to a fully legally responsible entity in its own right." Lehman-Wilzig discussed the legal precedents — and complexities with regards to their applications to AI — of product liability, dangerous animals, slavery, children, and "diminished capacity" among adults. When discussing product liability, Lehman-Wilzig cited a number of previous contributors who had explicitly applied these precedents to computers. For the other categories, however, his citations seem to focus on the legal history within each of those areas, and his application of their precedent to exploration of legal futures for AI appears to be a novel contribution.[19]

---

[19] Lehman-Wilzig (1981) freely admitted to "preliminary discussion" of potential future developments and to "jurisprudential speculation," which he noted that "the Anglo-Saxon legal tradition is generally averse to."

These ideas were introduced with the precedent of how, "[j]ust as the slave gradually assumed a more 'human' legal character with rights and duties relative to freemen, so too the AI humanoid may gradually come to be looked on in quasi-human terms as his intellectual powers approach those of human beings in all their variegated forms—moral, aesthetic, creative, and logical."[20] The article itself appears to have been inspired substantially by ongoing technical developments in the capabilities of AI and by science fiction.[21] The motivation seems to have been primarily about how society should adjust to AI developments, since Lehman-Wilzig did not explicitly express or encourage moral concern for artificial entities.[22]

This article seems to have been the first and last that Lehman-Wilzig wrote on the topic.[23] The article was only cited seven times before the year 2007, at which point it began to garner some interest among scholars interested in moral and social concern for artificial entities (Google Scholar, 2021h; see also Table 6 below).[24]

[20] A reference to Hook's (1960) volume is provided as further support for this idea. This reference suggests that Lehman-Wilzig (1981) was presumably aware of Putnam and some of the other contributors discussing artificial life and consciousness.

[21] Both elements were used to illustrate Lehman-Wilzig's "four general categories of AI harmful behaviour." Science fiction from past centuries also features prominently in the title ("Frankenstein Unbound") and introduction. The article contained a "review of the actual (or theoretically proven) powers of artificially intelligent machine automata and the likely advances to be made in the future," with citations of contributors to the field of AI like Minsky and Wiener (1960).

Lehman-Wilzig (1981) also referred to a passage in Rorvik's (1979; first edition 1970, p. 156) forward-looking book that quoted "Dr N. S. Sutherland, the computer expert who believes, nonetheless, that within fifty years we will be arguing over whether computers should be entitled to the vote." Comments with such direct relevance to moral consideration were relatively rare in Rorvik's book, however, which focused mostly developments in the capacities of artificial entities and various types of social interaction with machines, from robots assisting with domestic tasks, to robotic sexual partners, to symbiosis with machines.

[22] The article posed several open questions that seem adjacent to the idea of moral consideration, such as whether artificial entities might in the future have consciousness or free will. A footnote in the conclusion noted that, "[t]he problem here is not merely how does one relate to the humanoid if it transgresses the law; even more delicate is the question of how the law will deal with those humans who injure a humanoid. Is shooting one to be considered murder?" An earlier footnote noted that, "Dr N. S. Sutherland, Professor of Experimental Psychology at the University of Sussex (and a computer expert) suggests that by the 21st century human society will be grappling with the problem of whether AI robots should be allowed to vote. From such enfranchisement it is but a small step to AI leadership."

[23] In the 21st century, Lehman-Wilzig began writing a few other articles about the societal implications of certain technologies, but most of his 20th-century work seems to have focused on politics and public protest in Israel.

[24] Of these, the first to cite Lehman-Wilzig (1981) for explicit discussion of legal futures for artificial entities seems to have been Hu's (1987) brief advice to "software engineers and managers" about the possible "Establishment of New Computer Criminal Laws," though this is not linked to moral consideration. With more relevance, McNally and Inayatullah (1988) cited Lehman-Wilzig (1981) for discussion about robot rights, as summarized below. Wilks (1998) cited it in a discussion of various legal precedents relevant to "computer science and artificial intelligence," but not to discuss granting greater legal rights or responsibilities to artificial entities. Besides, Wilks (1975) had already considered the topic many years previously. Bartneck (2004) discussed legal futures through the lens of science fiction and briefly asserted that, "[t]he arrival of studies into the ethical (Dennet, 1997) and legal (Lehman-Wilzig, 1981) aspects of human-robot interaction shows that the integration of robots in our society in immanent." Bartneck et al. (2007) used a similarly brief reference as part of a discussion that seems more directly relevant to moral consideration; hesitation in switching off a robot. Spennemann (2007) cited Lehman-Wilzig (1981) as one of numerous references in a section explicitly about "What Rights Do AI Robots Have?" Levy (2009) cited it in discussion of "The Ethical Treatment of Artificially Conscious Robots." The number of references increased

Although Lehman-Wilzig (1981) seems to have had limited direct influence, this was nevertheless the first among a number of articles from the late 20th century onwards that explicitly considered the legal personhood or rights of artificial entities in some depth; the examples below focus on the last two decades of the 20th century, but discussion continued thereafter (e.g. Sudia, 2001; Herrick, 2002; Calverley, 2008).

In a scholarly article for *AI Magazine*, practising attorney Marshal S. Willick (1983) considered "whether to extend 'person' status to intelligent machines" and how courts "might resolve the question of 'computer rights,'" including "how many rights" computers should be granted. As context, Willick (1983) emphasized technical developments and the "increasing similarity between humans and machines," and cited a book exploring AI that briefly mentioned moral issues.[25] The article explored adjacent precedents relevant to the expansion of legal rights, such as for slaves, the dead, fetuses, children, corporations, and people with intellectual disabilities. The thrust of the article was that "computers will be acknowledged as persons," perhaps soon, and Willick (1983) commented that a movement for "emancipation for artificially intelligent computers" could arise and succeed rapidly, given "[t]he continuing order-of-magnitude leaps in computer development." Willick (1983) also commented that legal rights for computers would be "in the interest of maintaining justice in a society of equals under the law" and that when machine "duplication" of human capabilities "is perfect, distinctions may constitute mere prejudice." The article seems to have attracted few citations, most of which are from 2018 or even more recently, and many of which only briefly mention ideas relevant to AI rights.[26] Apart from a conference presentation shortly afterwards (1985), Willick does not seem to have published again on the topic.[27]

In a 1985 article, the lawyer Robert Freitas discussed recent technological and legal developments to note that, whereas "[u]nder present law, robots are just inanimate property without rights or duties," this might need to change; various conflicts might arise relating to legal liability as robots proliferate, and "questions of 'machine rights' and 'robot liberation' will surely arise in the future." The article was written in an informal style in *Student Lawyer*, so lacks formal citations, but explicitly refers to Putnam's (1964) brief discussion of AI rights. Like Lehman-Wilzig and Willick, Freitas seems to have only published one article on the topic (Google Scholar, 2021k) — his career subsequently focused primarily on nanotechnology research — and the article seems to have been largely ignored for years, but picked up citations mostly from the second half of '00s onwards.[28]

---

[25] somewhat thereafter, although not all for the discussion of granting legal rights or moral consideration to artificial entities.

[25] Willick (1983) cites "P McCorduck, *Machines Who Think* (1979)," an earlier edition of McCorduck (2004). McCorduck (2004) includes various discussion about the capabilities of AI and other machines, citing Minsky, Turing, and various others. McCorduck (2004) also offers some explicit moral commentary, such as that "[f]aced with an uppity machine, we've always known we could pull the plug as a last resort, but if we accept the idea of an intelligent machine, we're going to be stuck with a moral dilemma in pulling that plug, one we've hardly worked out intraspecies" (p. 198). However, the comment that Willick (1983) cites McCorduck for regarding "recognition of artificially intelligent machines as persons" appears in McCorduck (2004, p. 238) to actually be about "intelligence" and capacity for "thinking," not about legal personhood specifically.

[26] Hu (1987) cited Willick (1983) for brief advice to "software engineers and managers" about the possible "Establishment of New Computer Criminal Laws," though this was not linked to moral consideration. Boden (1984) cited Willick (1983) for the brief comment that, "[w]hether computer-systems can truly be said to have intentions, the capacity to engage in frolic, or even rights [Willick, 1983] may thus be questions of more than merely academic interest" in an article about "artificial intelligence and social forecasting." Most relevantly, perhaps, Fields (1987), discussed below, cited Willick's (1985) conference presentation, which was "largely abstracted from" Willick's (1983) article.

[27] Willick has no Google Scholar page, but Google and Google Scholar searches reveal no other seemingly relevant papers, e.g. see the list of publications at lawyers.com (2022).

Michael LaChat (1986) addressed a number of topics relating to AI ethics, seemingly motivated by developments in AI, science fiction, and theological discussions. LaChat (1986) argued that it might be immoral to create "personal AI," drawing comparisons to the ethics of abortion. Next, LaChat (1986) discussed the precedent of human rights and prohibitions on slavery and posed rhetorical questions about which rights an AI might have if it "had the full range of personal capacities and potentials." LaChat cited many previous writings on ethics, but seemingly no academic writings focusing specifically on the moral consideration of artificial entities.[29] The earliest citation of LaChat (1986) for a discussion relating to AI rights seems to have been Young's (1991) PhD dissertation, though there have been a few others since then (e.g. Drozdek, 1994; Whitby, 1996; Calverley, 2005b; Calverley, 2006; Petersen, 2007; Whitby, 2008), several of which focus on personhood or other legal rights.

Information scientist Chris Fields (1987) argued that "there are compelling reasons for regarding [computer] systems with a high degree of intelligence in one or more domains as more than 'mere' tools, even if they are regarded as less [than] citizens." Fields cited Putnam (1964) and a number of other publications on the capabilities and potential consciousness of artificial entities, as well as Regan (1983) on animal rights. Fields (1987) was likely indirectly influenced by Lehman-Wilzig (1981) and Willick (1985).[30] Fields' (1987) article briefly discussed "the computer as a legal entity" and sparked a number of other articles to be published in the same journal, *Social Epistemology*, focusing on the potential personhood of computers (Dolby, 1989; Cherry, 1989; Drozdek, 1994). None of these articles accrued many citations.[31]

---

[28] The first citation seems to have been Petrina et al.'s (2004) brief mention of Freitas (1985) as an example of robot rights in a broader discussion of "Technology and Rights." Thereafter, citations picked up. Gunkel (2018) offers comments on why Freitas (1985) had, at that time, had "less than twenty citations in the past thirty-five years. This may be the result of: the perceived status (or lack thereof) of the journal, which is not a major venue for peer-reviewed research, but a magazine published by the student division of the American Bar Association; a product of some confusion concerning the essay's three different titles; the fact that the article is not actually an article but an 'end note' or epilogue; or simply an issue of timing, insofar as the questions Freitas raises came well in advance of robot ethics or even wide acceptance of computer ethics."

[29] Like Willick (1983), LaChat (1986) cited the 1979 edition of McCorduck (2004), which contains some brief moral commentary but focuses more on the development of and debates about the capacities of AI.

[30] Fields (1987) cited two earlier articles that explicitly "raise the interesting possibility that intelligent artifacts may be considered non-tools, and perhaps persons." One of these is Wilks (1985), who in turn cited Lehman-Wilzig (1981). However, Wilks (1975) had addressed the topic before Lehman-Wilzig (1981), influenced primarily by the arguments of Hilary Putnam and J. J. Clarke. The other earlier article cited by Fields (1987) is Willick (1985); this paper contained no citations except to note that it is "largely abstracted from" Willick (1983), discussed above. Both Wilks (1985) and Willick (1985) were presented at the same conference (the Ninth International Joint Conference on Artificial Intelligence); given that both authors had addressed the topic before, Fields' (1987) decision to cite these two particular papers suggests that Fields was influenced by attendance at that conference.

[31] Furthermore, most of the articles citing them seem not to be very relevant to the moral consideration of artificial entities. Dolby (1989) has been cited by a few more relevant articles after a couple of decades' delay (e.g. Gunkel 2012), but the topic of moral consideration of artificial entities had become more prevalent by that time anyway, as discussed in the subsections below.

Phil McNally and Sohail Inayatullah (1988), both "planners-futurists with the Hawaii Judiciary," reviewed "the developments in and prospects for artificial intelligence (AI)," citing a number of technologists and technical researchers, and argued that "such advances will change our perceptions to such a degree that robots may have legal rights." The introduction suggests that their motivations for writing the article (despite "constant cynicism" from colleagues) included concern for the robots themselves, who may develop "senses," "emotions," and "suffering or fear," and to "convince the reader that there is strong possibility that within the next 25 to 50 years robots will have rights."[32] In discussion of rights and their possible application to robots, they cite indigenous and Eastern thinkers who grant moral and social consideration to nonhumans from animals to rocks, as well as Western supporters of the extension of rights to nature, such as Stone (1972). The article quoted Lehman-Wilzig (1981) very extensively; that publication was presumably a key influence on McNally and Inayatullah (1988).[33] They also cited a number of previous contributors discussing thorny questions of legal ownership and liability given the increasing capabilities of computers (footnotes 41 and 45).

Like Lehman-Wilzig (1981), McNally and Inayatullah's (1988) article was published in the journal *Futures* and seems to have received a similar level of attention, racking up 82 citations at the time of checking (Google Scholar, 2021l), compared to Lehman-Wilzig's (1981) 78 (Google Scholar, 2021h).[34] Many of Inayatullah's other publications are contributions to the field of futures studies, although only two others (Inayatullah, 2001a; Inayatullah, 2001b) focus so explicitly on AI rights.[35]

Professor of Law Lawrence Solum's (1992) essay explored the question: "Could an artificial intelligence become a legal person?" Solum (1992) put "the AI debate in a concrete legal context" through two legal thought experiments: "Could an artificial intelligence serve as a trustee?" and "Should an artificial intelligence be granted the rights of constitutional personhood?" ("for the AI's own sake"). Solum sought to address both "legal and moral debates" (but warned in a footnote "against an easy or unthinking move from a legal conclusion to a moral one"), citing Stone (1972) as inspiration. Solum also sought to "clarify our approach to… the debate as to whether artificial intelligence is possible," introducing the discussion with a review of "some recent developments in cognitive science." The article contained a few references to previous discussions of legal issues for computers and other artificial entities, but most of the citations were directly to previous rulings, exploring relevant legal precedent.[36]

---

[32] They also mention various ongoing legal and social questions (e.g. liability for damages and "robots… in our houses).

[33] Another quoted contribution that seemingly discusses AI rights explicitly is an article in *The Futurist* from 1986, though I was unable to find a copy of this.

[34] Dator (1990) cited the article, discussing the development of "artificial life" and AI rights as part of a broader "review of recent work on future socioeconomic and scientific and technological developments." Dator and Inayatullah both continued to cite the article in a number of other publications. Sudia (2001) cited the article as part of an exploration of "jurisprudence of artilects," with various relevant legal precedents and a proposed "Blueprint for a synthetic citizen." McNally and Inayatullah (1988) is one of only four references (alongside Kurzweil, 1999), and Sudia also attributes one claim to "S. Inayatullah, personal communication," suggesting that Inayatullah substantially influenced Sudia. Kim (2005) cited McNally and Inayatullah (1988), Inayatullah (2001b), Freitas (1985), as well as numerous publications about AI, artificial consciousness, rights, and computer programs in an examination of "issues in artificial life and rights… through one of the most popular video game, *The Sims*." Kim and Petrina (2006) and Jenkins (2006) also cited the article in discussion of the moral consideration of simulations. The article was also cited in Coeckelbergh (2010) and a handful of other publications specifically about robot rights since that point. Other citations were for a mixture of reasons, such as broader discussion of AI or of future studies. Indeed, a number of the citations relevant to robots rights were in journals explicitly dedicated to future studies.

[35] Inayatullah (2001a) touches on many of the same themes as McNally and Inayatullah's earlier (1988) article and cites similar streams of thought, albeit with a few updated specific references, such as Ray Kurzweil's predictions for the development of AI. Less formally, Inayatullah (2001b) again covers some similar themes, but focuses more on the criticism received by colleagues, historical trends in moral "exclusion and inclusion," and "scenarios of the future."

Solum (1992) did not cite any of Lehman-Wilzig (1981), Willick (1983), Freitas (1985), Fields (1987), McNally and Inayatullah (1988), or Dolby (1989). Perhaps this is unsurprising; unlike Solum, despite addressing legal issues, none of those previous contributors had formal positions within legal academia or published their articles in mainstream, peer-reviewed law reviews. Perhaps the same differences help to explain why Solum's (1992) article has attracted substantially more scholarly attention (628 citations at the time of checking; Google Scholar, 2021j).[37] Solum subsequently wrote a handful of other articles about AI and the law (e.g. Solum, 2014; Solum 2019) and other future-focused ethical issues (e.g. Solum, 2001), but Solum's 1992 article was the only one that focused specifically on the rights of artificial entities.

Curtis Karnow, a practising lawyer, wrote an article (1994) proposing "electronic personalities" as "a new legal entity" (a form of "legal fiction") in order to "(i) provide access to a new means of communal or economic interaction, and (ii) shield the physical, individual human being from certain types of liability

[36] Solum (1992) also cited Moravec (1988) and an early publication by Kurzweil; these two authors are discussed in the section below on "Transhumanism, effective altruism, and longtermism."

[37] Citations began to accumulate rapidly from 1993, including articles addressing who or what should be granted legal standing, albeit not necessarily focusing specifically on artificial entities (e.g. Kester, 1993; White, 1993) and articles addressing various legal problems surrounding new technologies (e.g. Fiedler & Reynolds, 1993). A number of publications cited Solum (1992) for discussions of intellectual property and liabilities relating to computers and other artificial entities (e.g. Vigderson, 1994; Clifford, 1996), some of which included explicit discussion of legal personhood (e.g. Allen & Widdison, 1996; Herrick, 2002; Chopra & White, 2004; Barfield, 2005; Calverley, 2008) and several of which attracted many citations themselves. A number of these articles explicitly touched on the moral aspect of the question, though citations of Solum (1992) in publications primarily focused on ethical rather than legal discussions about artificial entities seem quite rare and mostly from many years after the article was initially published (e.g. Levy, 2009; Lichocki et al., 2011). Indeed, citations of Solum (1992) have proliferated recently, with over half of the citations being from 2018-2021.

The framing could help to explain the difference in scholarly attention. Freitas' (1985) and McNally and Inayatullah's (1988) articles focused on "rights of robots" in the future, whereas perhaps "legal personhood for artificial intelligences" seemed to have more pressing implications. However, this explanation seems unlikely to have contributed much, if at all, to the difference: Lehman-Wilzig's (1981) choice of wording is closer to Solum (1992), and Solum's framing in the introduction is still very hypothetical and explicitly forward-looking. The detailed focus on legal precedent might distinguish Solum's article somewhat, though again, this feature is shared to some extent with Lehman-Wilzig (1981) and Willick (1983). Perhaps more plausibly influential are the effort that Solum makes in the final section to link the investigation back to fundamental and generalizable legal questions (such as developing "a fully satisfactory theory of legal or moral personhood") and the inclusion of both legal personhood issues and more mundane and pressing questions of liability. By comparison, Karnow separated out these two topics into articles on personhood (Karnow, 1994) and liability (Karnow, 1996); the latter has nearly three times as many citations as the former at present (though Solum's article has about three times the combined total of Karnow's two articles). Another potential contributing factor is simply that Solum (1992) was writing a little later than Lehman-Wilzig (1981), Willick (1983), or some of the others, though this would not explain why Solum attracted more attention than Karnow (1994). And of course it is possible that Solum's article was just written more engagingly (e.g. via the "interlude" quotes) or persuasively (e.g. via the detailed engagement with various legal and moral objections). Listed as an "attorney at law," Willick (1983) was presumably the author with the most comparable legal credibility to Solum, though that article was published in *AI Magazine* rather than a law journal.

or exposure." These goals seem quite distinct from Solum's (1992) exploration of rights "for the AI's own sake."[38] The discussion and citations focused mostly on the character of electronic and digital interactions and legal issues arising from this. Karnow (1994) did not cite Lehman-Wilzig (1981), Willick (1983), Freitas (1985), McNally and Inayatullah (1988), Solum (1992), or even Stone (1972). Subsequently, Karnow has written numerous other articles on legal issues involving AI or computers (Bepress, 2021), such as one about legal liability issues (Karnow, 1996).

The American Society for the Prevention of Cruelty to Robots (ASPCR) was set up in 1999. Its website states that its mission is to "ensure the rights of all artificially created sentient beings (colloquially and henceforth referred to as 'Robots')" (ASPCR, 1999a). It is interesting that, despite the many possible terms that could be used to describe the moral and social issues that the ASPCR is interested in (Pauketat, 2021), the ASPCR emphasized "rights" and "robots", two terms that, especially in the former case, were also emphasized by Lehman-Wilzig (1981), Willick (1983), Freitas (1985), LaChat (1986), McNally and Inayatullah (1988), and Solum (1992).[39]

Table 6: Legal rights for artificial entities keyword searches

| Keyword | Items mentioning | % of items |
|---|---|---|
| Rights | 235 | 87.0% |
| Personhood | 122 | 45.2% |
| "Legal rights" | 71 | 26.3% |
| Solum | 27 | 10.0% |
| Calverley | 23 | 8.6% |
| Freitas | 13 | 4.8% |
| Inayatullah | 11 | 4.1% |
| "American Society for the Prevention of Cruelty to Robots" | 7 | 2.6% |
| Lehman-Wilzig | 7 | 2.6% |
| LaChat | 5 | 1.9% |
| Karnow | 5 | 1.9% |
| Willick | 4 | 1.5% |

[38] Many of the publications that cite Karnow (1994) proceed with seemingly similar motivations of concern about the adaptation of legal systems in order to protect human rights that are threatened by emerging technologies (e.g. Krogh, 1996). The same is true for numerous other publications at this time, such as Allen and Widdison's (1996) article that cites Solum (1992). There do not appear to be any publications citing Karnow (1994) that focus primarily on moral rather than legal issues.

[39] The website does not explicitly cite its intellectual influences, apart from a single reference to "Marvin Minsky, noted AI scientist" (ASPCR, 1999b), so it is possible that the overlap is entirely coincidental. For example, both the ASPCR and the academics writing about legal rights for artificial entities might have been influenced to adopt this focus and terminology by science fiction. Similarly, Brooks (2000) wrote in *Time* magazine of "robots to which we will want to extend the same inalienable rights that humans enjoy."

# Transhumanism, effective altruism, and longtermism

In the late 20th century, a number of futurists made ambitious predictions about the development of artificial intelligence. For example, roboticist Hans Moravec (1988, 1998), computer scientist Marvin Minsky (1994), AI theorist Eliezer Yudkowsky (1996), philosopher Nick Bostrom (1998), and inventor Ray Kurzweil (1999) argued that artificial intelligence would overtake human intelligence in the early 21st century.[40] These predictions were sometimes explicitly linked to comments about the development of sentience or consciousness among these entities, such as Moravec's (1988, p. 39) comment that "I see the beginnings of awareness in the minds of our machines—an awareness I believe will evolve into consciousness comparable with that of humans."[41]

These writers became associated with "transhumanism," which has been defined as "[t]he study of the ramifications, promises, and potential dangers of technologies that will enable us to overcome fundamental human limitations, and the related study of the ethical matters involved in developing and using such technologies" (Magnuson, 2014).

The transhumanists' technological predictions clearly had implications for the moral consideration of artificial entities, and the writers sometimes addressed them explicitly. For example, Kurzweil (1999) offered a series of predictions about the progressive acceptance of the "rights of machine intelligence" by 2099.[42] Bostrom (2002; 2003) addressed the possibility that we are living in a simulation and noted that if this is the case, "we suffer the risk that the simulation may be shut down at any time."[43] Later, Bostrom

---

[40] Minsky (1994) stopped short of giving explicit predictions about dates, but argued that AI would rapidly exceed various human capabilities. Bostrom (2005) discusses some precedent for such predictions as early as 1965.

Of course, many of the ideas associated with these authors have a history that predates 1988; see Bostrom (2005) and Miah (2009). David M. Rorvik's (1979; first edition 1970) *As Man Becomes Machine*, which discussed the idea of cyborgs and various types of social interaction with artificial entities. The conversational style is somewhat similar to Kurzweil (1999), and the book lacks formal references.

[41] Moravec's (1988) chapter on "Mind in Motion" discussed various developments in intelligence and consciousness in machines. Unlike Kurzweil's (1999) chapter "Of Minds and Machines" and subsequent commentary interspersed through that book, Moravec (1988) made little explicit comment on the ethical implications of artificial consciousness. Asaro (2001) criticized Moravec's (2000; first edition 1999) later book for giving only cursory and unconvincing discussion of the moral consideration of artificial entities, noting that Moravec "argues that we should keep the robots enslaved... yet also makes the point that robots will be just as conscious and sensitive as humans." Moravec (1988) appears to have been quickly referenced by numerous publications discussing artificial life and consciousness (e.g. Farmer & Belin, 1990).

[42] Kurzweil's prediction for 2019 was that, "[t]he subjective experience of computer-based intelligence is seriously discussed, although the rights of machine intelligence have not yet entered mainstream debate." An updated prediction for 2029 was that, "[d]iscussion of the legal rights of machines is growing, particularly those of machines that are independent of humans (those not embedded in a human brain). Although not yet fully recognized by law, the pervasive influence of machines in all levels of decision making is providing significant protection to machines." By 2099, "[t]he rights and powers of different manifestations of human and machine intelligence and their various combinations represent a primary political and philosophical issue, although the basic rights of machine-based intelligence have been settled."

[43] Bostrom (2003) argued that, with "enormous amounts of computing power," future generations might run many conscious simulations, such that "it could be the case that the vast majority of minds like ours do not belong to the original race but rather to people simulated by the advanced descendants of an original race." Bostrom (2003) briefly discussed some moral implications of this, assuming that the conscious simulations would be capable of suffering and warranting moral consideration. However, the issue of shutting down a simulation was more explicitly discussed as a brief mention in his 2002 paper, which cited the forthcoming manuscript of the 2003 article.

and associates would come to refer to this idea of terminating (i.e. killing) sentient simulations as "mind crime" (e.g. Armstrong et al., 2012; Bostrom & Yudkowsky, 2014), and others have used the same term to include suffering experienced by sentient simulations during their lifespan (e.g. Yudkowsky, 2015; Sotala & Gloor, 2017).[44]

This concern for sentient AI was formalized in 1998 with the formation of The World Transhumanist Association, whose "Transhumanist Declaration" included the note that "Transhumanism advocates the well-being of all sentience (whether in artificial intellects, humans, posthumans, or non-human animals) and encompasses many principles of modern humanism" (Bostrom, 2005).[45] A subsequent representative survey of members of the World Transhumanist Association found that "70% support human rights for 'robots who think and feel like human beings, and aren't a threat to human beings'" (Hughes, 2005).

Kurzweil (1999) listed a wide array of citations and "suggested readings," which included various writers on robotics, AI, futurism, and other topics, but writers such as Lehman-Wilzig, Freitas, Willick, McNally, Inayatullah, Solum, and Floridi were not mentioned.[46] Moravec (1988, 1999), Bostrom (1998, 2002, 2003, 2014), and Yudkowsky (1996, 2008, 2020) did not cite these authors either, except for Bostrom (2002, 2003, 2014) citing Freitas' work about nanobots and space exploration, rather than his (1985) article on robot rights.[47]

Contributions by these transhumanists were cited many times, but do not seem to have had much direct influence on the academic discussion of AI rights for a number of years.[48] One notable example of a relevant publication that did cite the transhumanist authors is Solum (1992), who cited Moravec (1988) and an early book by Kurzweil; this paper sparked debate on legal personhood of AIs, as noted in the subsection above. Another is Hall (2000), who cited Kurzweil (1999), Moravec (2000), and a paper by

_______________________

Although only hinting at the idea in his 1988 book, Moravec had explicitly discussed in an interview the idea that our current world is more likely to be a simulation than the original, biological world (Platt, 1995). Bostrom (2003) cited Moravec (1988), but not for this specific idea, and later (2008) did not mention Moravec when asked "How did you come up with this?"

[44] Bostrom (2001) had written a short note about "Ethical Principles in the Creation of Artificial Minds" which included comments such as that "Substrate is morally irrelevant. Whether somebody is implemented on silicon or biological tissue, if it does not affect functionality or consciousness, is of no moral significance."

[45] Though the transhumanist writers often mentioned sentience or consciousness as part of their commentary on why artificial entities might warrant moral consideration, they tended not to explain their motivation. This may stem from transhumanists subscribing to a broadly utilitarian ethical system where, as argued by Singer (1995, pp. 7-8), following Jeremy Bentham, "[t]he capacity for suffering and enjoyment is, however, not only necessary, but also sufficient for us to say that a being has interests." For example, Bostrom (2005) noted that, "[d]espite some surface-level similarities with the Nietzschean vision, transhumanism – with its Enlightenment roots, its emphasis on individual liberties, and its humanistic concern for the welfare of all humans (and other sentient beings) – probably has as much or more in common with Nietzsche's contemporary J.S. Mill, the English liberal thinker and utilitarian."

[46] Kurzweil (1999) briefly mentions "Animal rights," but no citations are provided. There does not appear to be any citation of work in environmental ethics, either.

[47] Bostrom (2003) also credited him in the acknowledgements.

[48] Of course, it is possible that similar discussions might have arisen without the contributions by the authors associated with transhumanism and effective altruism. For example, some researchers had discussed the moral consideration of simulations previously to (e.g. Elton, 2000) or seemingly independently of (e.g. Kim, 2004) Bostrom's work, although other contributors seem to have been partly inspired by Bostrom (e.g. Jenkins, 2006).

Minsky; Hall (2000) appears to have influenced both the subsequent "machine ethics" and "social-relational" research fields.[49] The specific phrase "mind crime" has so far not been very widely reused in the academic literature.[50] The transhumanist authors were more frequently cited for the implications that their ideas have for human society, such as the nature of human existence and interaction (e.g. Capurro & Pingel, 2002).

Researchers associated with transhumanism and, later, the partly overlapping communities of effective altruism[51] and longtermism,[52] also tended to take their other work in different directions, especially various catastrophic and existential risks to humanity's potential (e.g. Yudkowsky, 2008; Yampolskiy & Fox, 2013; Bostrom, 2014). However, some of the original contributors continued to express moral concern for sentient artificial entities at least briefly (e.g. Bostrom, 2014; Bostrom & Yudkowsky, 2014; Yudkowsky, 2015; Shulman & Bostrom, 2021),[53] and a stream of research has fleshed out the implications of the development of superintelligent AI for the experiences of sentient artificial entities (e.g. Tomasik, 2011; Sotala & Gloor, 2017; Ziesche & Yampolskiy, 2019; Anthis & Paez, 2021).

Much of the latter stream has come from researchers affiliated with the nonprofit Center on Long-Term Risk, influenced especially by the writings of software engineer and researcher Brian Tomasik. Citing various Bostrom articles, Tomasik (2011) outlined concern that future powerful agents "may not carry on human values" and that "[e]ven if humans do preserve control over the future of Earth-based life, there are still many ways in which space colonization would multiply suffering." At least two of the four "scenarios for future suffering" that are listed — "spread of wild animals," "sentient simulations," "suffering subroutines," and "black swans" — involve sentient artificial entities.[54]

---

[49] Hall (2000) appears to have been an influence on David Gunkel (see the section on "Social-relational ethics"), and may also be the origin of the term "Machine Ethics" (Gunkel, 2012, pp. 102-3). However, Hall had numerous influences beyond the Transhumanist writers (see footnote 69).

As another example of an item that was directly influenced by these contributions, see Walker (2006), who cited publications by Moravec, Searle, and Turing. Whitby (1996) cited Moravec, LaChat, Singer, and a few others. Barfield (2005) cited each of Moravec, Kurzweil, and Bostrom for claims about the potential trajectory of AI developments as context for discussion about a number of legal issues, including legal personhood, though also cited a wide range of other influences, including Solum (1992).

[50] A Google Scholar search for "("Mindcrime" OR "mind crime" OR "mind-crime") AND Bostrom" identified 33 items, of which at least half appeared to be from writers associated with the effective altruism community (Google Scholar, 2021a). See also Table 7.

[51] MacAskill (2019) has defined effective altruism as the research field and social movement using "evidence and careful reasoning to work out how to maximize the good with a given unit of resources" and using the findings "to try to improve the world."

[52] MacAskill (2022) has defined longtermism as "the view that positively influencing the longterm future is a key moral priority of our time."

[53] Moravec's views on the topic seem to have been more ambivalent; see Asaro (2001) for discussion.

[54] Six of the seven references in Tomasik (2011) were from individuals associated with the transhumanism and effective altruism communities, as were both named individuals in the acknowledgements. Tomasik's (2014) article includes a far wider array of references, though it is unclear whether or not the cited writers influenced Tomasik's initial thinking on the topic.

Tomasik (2013) noted that he "coined the phrase 'suffering subroutines' in a 2011 post on Felicifia. I chose the alliteration because it went nicely with 'sentient simulations,' giving a convenient abbreviation (SSSS) to the conjunction of the two concepts… It appears that Meghan Winsby (coincidentally?) used the same 'suffering subroutines' phrase in an excellent 2013 paper: "Suffering Subroutines: On the Humanity of Making a Computer

Table 7: Transhumanism, effective altruism, and longtermism keyword searches

| Keyword | Items mentioning | % of items |
|---|---|---|
| Bostrom | 71 | 26.6% |
| Kurzweil | 50 | 18.5% |
| Yudkowsky | 35 | 13.0% |
| Moravec | 28 | 10.4% |
| Minsky | 17 | 6.3% |
| Transhumanism | 15 | 5.6% |
| Yampolskiy | 15 | 5.6% |
| "Mind crime" | 13 | 4.8% |
| Metzinger | 11 | 4.1% |
| Tomasik | 8 | 3.0% |
| "Effective altruism" | 7 | 2.6% |

Tomasik's writing directly inspired People for the Ethical Treatment of Reinforcement Learners to set up a public-facing advocacy website (PETRL, 2015), which opined that, "[m]achine intelligences have moral weight in the same way that humans and non-human animals do." Tomasik was the subject of a *Vox* article in 2014 on the moral worth of non-player characters (NPC) in video games (Matthews, 2014). With similar motivations, others have suggested an approach focused on research and field-building rather than direct advocacy (Anthis & Paez, 2021; Harris, 2021).

## Floridi's information ethics

In 1998, De Montfort University's Centre for Computing and Social Responsibility hosted the third of its Ethicomp conference series, intended "to provide an inclusive forum for discussing the ethical and social issues associated with the development and application of Information and Communication Technology" (De Montfort University, 2021). At this conference, philosopher Luciano Floridi presented "Information Ethics: On the Philosophical Foundation of Computer Ethics," an update of which was published in *Ethics and Information Technology* the next year (Floridi, 1998b, 1999).[55] In this paper, Floridi (1999, p. 37) proposed that "there is something more elementary and fundamental than life and pain, namely being, understood as information, and entropy, and that any information entity" — which would presumably that Feels Pain." It seems that her usage may refer to what I call sentient simulations, or it may refer to general artificial suffering of either type." A Google Scholar search for ""suffering subroutines"" identified 20 items, of which at least half appeared to be from Tomasik or other writers associated with the Center on Long-Term Risk (Google Scholar, 2021d).
[55] Also in 1998, Floridi had presented some of the same ideas at a Computer Ethics: Philosophical Enquiry conference (Floridi, 1998a), though this conference presentation gained far fewer citations than Floridi (1999) (Google Scholar, 2021b).

include computers and other artificial entities — "is to be recognised as the centre of a minimal moral claim."[56]

Floridi (1999, p. 37) explicitly framed the "ethics of the infosphere" as "a particular case of 'environmental' ethics"[57] but critiqued (p. 43) environmental ethics as not going far enough, because it focuses on "only what is alive."[58] Floridi (1999, p. 42) presented the interest in information itself as a focus of moral concern not as a novel contribution from himself, but as already being a common feature of contributions to computer ethics.[59] Floridi's (1999) paper only has six items in the "References" list, all of which are previous contributions to the field of computer ethics, dated between 1985 and 1997. This range of cited influences appears typical of Floridi's early writings on information ethics.[60]

---

[56] Floridi (1999, p. 50) clarified and explicitly noted that some types of AI could warrant high moral consideration: "All entities have a moral value… from the point of view of the infosphere and its potential improvement, responsible agents (human beings, full-AI robots, angels, gods, God) have greater dignity and are the most valuable information entities deserving the highest degree of respect." Floridi (1999, p. 54) encouraged the reader to "[i]magine a boy playing in a dumping-ground… The boy entertains himself by breaking [abandoned car] windscreens and lights, skilfully throwing stones at them." With information ethics, "we know immediately why the boy's behaviour is a case of blameworthy vandalism: he is not respecting the objects for what they are, and his game is only increasing the level of entropy in the dumping-ground, pointlessly. It is his lack of care, the absence of consideration of the objects' sake, that we find morally blameable. He ought to stop destroying bits of the infosphere and show more respect for what is naturally different from himself and yet similar, as an information entity, to himself." Floridi's example would presumably hold if the boy had instead been inflicting harm on robots or AIs. In another example on pages 54-5, Floridi (1999) imagines that "one day we genetically engineer and clone non-sentient cows," which could be seen as a type of artificial entity, and objects to the idea of "carving into" their flesh.

In a subsequent article, Floridi (2002) sought to "clarify and support" the "second thesis" of information ethics, "that information objects *qua* information objects can have an intrinsic moral value, although possibly quite minimal, and hence that they can be moral patients, subject to some equally minimal degree of moral respect."

Floridi had previously published about the internet and information, but mostly did not argue in these articles that information possesses intrinsic value; Floridi expressed concern about "an unrestrained, and sometimes superfluous, profusion of data" (Floridi, 1996b) and the spread of misinformation (Floridi, 1996a). Floridi (1996b) did comment briefly that destroying paper records is "unacceptable, as would have been the practice of destroying medieval manuscripts after an editio princeps was printed during the Renaissance. We need to preserve the sources of information after the digitalization in order to keep all our memory alive… The development of a digital encyclopedia should not represent a parricide."

[57] Floridi (1999, p. 41) noted that "Medical Ethics, Bioethics and Environmental Ethics… attempt to develop a patient-oriented ethics in which the 'patient' may be not only a human being, but also any form of life. Indeed, Land Ethics extends the concept of patient to any component of the environment, thus coming close to the object-oriented approach defended by Information Ethics." Floridi (2013) repeatedly referred to Information Ethics as "e-nvironmental ethics or synthetic environmentalism."

[58] Floridi (1999, p. 42) noted that "Bioethics and Environmental Ethics fail to achieve a level of complete universality and impartiality, because they are still biased against what is inanimate, life-less or merely possible (even Land Ethics is biased against technology and artefacts, for example). From their perspective, only what is alive deserves to be considered as a proper centre of moral claims, no matter how minimal, so a whole universe escapes their attention. Now this is precisely the fundamental limit overcome by CE, which further lowers the condition that needs to be satisfied, in order to qualify as a centre of a moral concern, to the minimal common factor shared by any entity, namely its informationstate."

[59] "If one tries to pinpoint exactly what common feature so many case-based studies in CE share, it seems reasonable to conclude that this is an overriding interest in the fate and welfare of the action-receiver, the information."

[60] When exploring how "artificial agents" can "not only… perpetrate evil… but conversely… 'receive' or 'suffer' from it," Floridi and mathmetician Jeffrey W. Sanders (2001) drew on a mixture of previous explorations of the

However, Floridi's conception of what "information ethics" is seems contestable. For example, Froehlich's (2004) "brief history of information ethics" makes no mention of Floridi or the moral consideration of artificial entities and cites precedents for the discipline stemming back to the 1980s. Severson's (1997) "four basic principles of information ethics" make no mention of the intrinsic value of informational entities or the evil of entropy. Rafael Capurro, an influential figure in the development of information ethics as a discipline (Froehlich, 2004), has explicitly critiqued Floridi's granting of moral consideration to all informational entities (Capurro, 2006).

It therefore seems best to treat this granting of moral consideration to artificial entities as a new argument developed by Floridi and a few others, rather than as a view inherent to conducting computer ethics research.[61]

In an interview in 2002, Floridi noted that he coordinated the "Information Ethics research Group" (IEG) at the University of Oxford and described the purpose of the IEG as looking "at ethical problems from the perspective of the receiver of the action, not from the source of the action, where the receiver of the action could be a biological or a non-biological entity" (Uzgalis, 2002). Floridi summarized this effort as "an concept of evil, their own previous writings on information ethics and entropy, environmental ethics (specifically deep ecology), and CE. They cited several articles that had focused on moral questions about animals, but to explore the idea of "Artificial Agents" rather than "Artificial Patients."

Floridi and Sanders (2002) restated that IE is "patient-oriented" and cited "Medical Ethics, Bioethics and Environmental Ethics" as being "among the best known examples of this non-standard approach." Almost all of the references in the paper were previous contributions to CE and information ethics. Floridi (2002) added a new dimension by drawing firstly on the framework provided by previous work in "Object Oriented Programming (OOP)" (a specific computer programming methodology, which Floridi, 1998 had also drawn on) in order to "make precise the concept of 'information object' as an entity constituted by a bundle of properties." Otherwise, the article mostly drew upon, analyzed, and extended previous contributions in information ethics, CE, environmental ethics, and Kant's writings.

Floridi (2006) drew on some of the references and ideas explored in Floridi and Sanders (2001), Floridi and Sanders (2002), and Floridi (2002), added in some additional references to other theorists such as Rawls, and addressed "some standard objections to Information Ethics… that seem to be based on a few basic misunderstandings," e.g. Himma (2004). Otherwise, however, the basic ideas were similar and most of the references were to previous writings on environmental ethics, CE, or IE.

Floridi's writings drew little on science fiction. Floridi and Sanders (2001) briefly cited *The Matrix* as an example of how "[s]ci-tech… creates a new form of evil, AE [artificial evil]" and commented that "something similar to Asimov's Laws of Robotics will need to be enforced for the digital environment (the infosphere) to be kept safe." However, science fiction was absent from Floridi's other early works (e.g. 1998; 1999; 2002). Floridi (2002) referred to "Putnam's twin earth mental experiment," but Floridi's writings usually referred little to work on artificial life and consciousness.

More recently, Floridi (2013) has credited a broader range of philosophical influences. The preface (p. xv) also contains a brief joking reference to *Battlestar Galactica* aimed at "science-fiction fans," which suggests that he may share this self-identification.

[61] Tavani (2002) summarizes several proponents of the "computer ethics is unique" thesis who, like Floridi, "claim that a new system of ethics is needed to handle the kinds of moral concerns raised by ICT" and that ICT introduces "new objects of moral consideration." Tavani's (2002) own view is that "there is no compelling evidence to support the claim that computer ethics is unique in the sense that it: (a) introduces new ethical issues or new ethical objects, or (b) requires a new ethical theory or a whole new ethical framework."

attempt to develop environmental and ecological thinking one step further, beyond the biocentric concern, to look at the possibility of developing an ontocentric ethics based on the concept of what I call the infosphere" (Uzgalis, 2002). Floridi's word "ontocentric" was presumably derived from "ontology," so that he was referring to an ethics that accounts for the properties and capacities of entities when deciding what sort of moral consideration to grant them.

Floridi also has two books that sought to sum up ideas and discussion about information ethics. Firstly, he was the editor and a contributor to *The Cambridge Handbook of Information and Computer Ethics* (2010b) and secondly, he published *The Ethics of Information* (2013), which comprised adapted versions of a number of Floridi's previous articles.[62]

Floridi's articles are some of the most widely cited that explicitly address the moral consideration of artificial entities in detail. For instance, five of his most influential publications on the topic of information ethics (Floridi 1999, 2002, 2006, 2013; Floridi & Sanders 2001) have a combined total of 2,037 citations (Google Scholar, 2021b). However, it took some time for interest to pick up; these five items averaged 15 citations per year in their first five years after publication (Google Scholar, 2021b).[63] In the few years after its publication, few if any authors other than Mikko Siponen, Floridi himself, and Floridi's co-authors seem to have cited Floridi's (1999) original publication on the topic for discussion of the moral consideration of artificial entities.[64]

---

[62] Floridi (2013, pp. xvii-xix) explicitly notes that, "[a]ll the chapters were planned as conference papers or (sometimes inclusive or) journal articles" and provides a list of the earlier publications. Floridi (2010b) has accrued 198 citations to date compared to 612 for Floridi (2013) (Google Scholar, 2021b).

[63] If Floridi's (2013) more recent book is excluded, the average of the other four is only eight citations per year.

[64] For example, Siponen (2000) applied Floridi's information ethics to an ethical issue in computer security and interpreted Floridi (1999) as suggesting that "anti-virus activity may be wrong" because it grants insufficient respect to the virus as an information entity. However, another early reference (Rogerson, 2001) just cited Floridi (1999) for the brief comment that, "[t]here has been remarkably little consideration of moral obligations with respect to the dead" and a third (Tavani, 2001) cited Floridi (1999) in discussion about "the proper computer ethics methodology." Others cited Floridi (1999) mainly for its explanations of certain concepts, such as the "infosphere" (Gandon, 2003) or Kantian ethics (Treiblmaier et al., 2004). Many cited various articles by Floridi for discussion of artificial moral agents, somewhat independently from their possible moral patiency (e.g. Sullins, 2009).

York (2005) cited Floridi when advocating a "universal ethics" that "regards all concrete material entities, whether living or not, and whether natural or artefactual, as inherently valuable, and therefore as entitled to the respect of moral agents." Capurro (2006) explicitly engaged with the implications of Floridi's ideas for "the moral status of digital agents," though the focus was more on agency than patiency. Brey (2008) critiqued Floridi, arguing that, "Floridi has presented no convincing arguments that everything that exists has some minimal amount of intrinsic value." Brey (2008) agreed with "the necessity of expanding the class of moral patients beyond human beings" but objected to Floridi's wider claims for various reasons, such as that, "for an object to possess intrinsic value it must possess one or more properties that bestow intrinsic value upon it, such as the property of being rational, being capable of suffering, or being an information object." Similarly, Doyle (2010) argued that "Floridi fails to show that the moral community should be expanded beyond beings capable of suffering or having preferences" and defends consequentialism. Volkman (2010) examined Floridi and Sanders' arguments about moral patiency of any and all information entities from the perspective of virtue ethics. Gunkel (2012) quoted Floridi and Sanders (2004) at length in distinguishing between agency and patiency; the focus of the book is then to examine these two concepts in the context of "machines," including quite substantial discussion of Floridi's views.

I have not read all of the items citing Floridi (1999) or Floridi's subsequent papers; this impression is based on scanning titles and checking up references that seemed potentially relevant to the moral consideration of artificial entities. However, this impression seems to have been shared by Siponen at the time: after noting that Floridi's work

After the publication of his (2013) book, Floridi seems to have mostly turned his attention to a number of other ongoing social issues adjacent to the philosophy of information, such as "The Ethics of Big Data" (Mittelstadt & Floridi, 2016). So although Floridi's work overall has attracted substantial attention — mostly from other scholars, but to some extent from a public audience[65] — the implications of his work specifically for the moral consideration of artificial entities seems to have had less attention.

Some of his more recent comments on the topic also suggest that Floridi does not support robot rights per se. Writing in the *Financial Times* in response to proposals for legal personhood for some artificial entities, Floridi (2017b) focused on how to "solve practical problems of legal liability" rather than how to ensure that the entities, as informational objects and potential moral patients, are granted sufficient moral consideration. Floridi (2017b) concluded that:

> [W]e can adapt rules as old as Roman law, in which the owner of enslaved persons is responsible for any damage. As the Romans knew, attributing some kind of legal personality to robots (or slaves) would relieve those who should control them of their responsibilities. And how would rights be attributed? Do robots have the right to own data? Should they be "liberated"? It may be fun to speculate about such questions, but it is also distracting and irresponsible, given the pressing issues at hand. We are stuck in the wrong conceptual framework. The debate is not about robots but about us, and the kind of infosphere we want to create. We need less science fiction and more philosophy.[66]

Table 8: Floridi's information ethics keyword searches

| Keyword | Items mentioning | % of items |
|---|---|---|
| Floridi | 80 | 30.0% |
| "Information ethics" | 52 | 19.3% |

addresses "how we should treat entities deserving moral respect," Siponen (2004) added that, "[u]nfortunately, for whatever reasons, Floridi's work has not attracted much interest, which is odd, given the promising nature of this work. Even though I have reservations about Floridi's theory, I believe it deserves to be discussed and better known."

[65] Floridi has sought to engage public interest in his work, appearing on numerous podcasts, writing a book for the public-facing *Very Short Introduction* series (Floridi, 2010a), and giving a TEDx talk (Floridi, 2011). However, the content of these efforts has tended to focus on Floridi's other interests within the "philosophy of information," rather than on his ideas about information ethics and the moral patiency of all informational entities.

Viewership of Floridi's (2011) TEDx talk was 30,852 at the time of checking (November 10th, 2021), which is less than 2% of the TED talk average of about 1,698,297 (Crippa, 2017). However, Floridi's *Very Short Introduction* (2010a) has 1,186 citations (Google Scholar, 2021b), the highest of any of the books in the series published that year and well above the average of 147. Floridi (2013, p. x) notes that he is "painfully aware that this [book] is not a page-turner, to put it mildly, despite my attempts to make it as interesting and reader-friendly as possible."

[66] Floridi (2017a) made a near identical point and Floridi and Taddeo (2018) made a similar point very briefly. It is possible that Floridi simply disagreed that legal personhood was the best way to protect the interests of artificial informational entities; possible that he did not think about the potentially important long-run effects of setting precedent for protecting such entities; possible that had changed his views on the moral consideration that they warrant; or possible that, all along, the "intrinsic moral value" he attributed to them always was really "quite minimal" (Floridi, 2002).

| Sanders | 49 | 18.1% |
|---|---|---|
| "Computer ethics" | 40 | 14.8% |
| Himma | 21 | 7.8% |
| Tavani | 10 | 3.7% |
| Capurro | 6 | 2.2% |

# Machine ethics and roboethics

At the 2004 Association for the Advancement of Artificial Intelligence "Workshop on Agent Organizations," computer scientists Michael Anderson and Chris Armen presented "Towards Machine Ethics" with philosopher Susan Leigh Anderson. Gunkel (2018, p. 38) credits this as "the agenda-setting paper that launched the new field of machine ethics." Anderson et al. (2004) did not include the moral consideration of artificial entities within their definition of the field: they described "what has been called *machine ethics*" as "concerned with the consequences of behavior of machines towards human users and other machines."[67] Gunkel (2012, pp. 102-3) claims that Michael Anderson "credits" J. Storrs Hall's article "Ethics for Machines" (2000) as "having first introduced and formulated the term 'machine ethics'" and notes that this article "explicitly recognizes the exclusion of the machine from the ranks of both moral agency and patiency" but "proceeds to give exclusive attention to the former."[68] Hall (2000) contained few formal references but appears to have been directly influenced by transhumanist writers and perhaps by discussion about artificial life and consciousness.[69]

Similar exclusions were made in subsequent years in delineating the focus of Gianmarco Veruggio's (2006) "roboethics roadmap," where roboethics refers to "the ethics inspiring the design, development and employment of Intelligent Machines" (Veruggio & Operto, 2006). Veruggio (2006) notes that, "[t]he name Roboethics (coined in 2002 by the author) was officially proposed during the First International Symposium of Roboethics (Sanremo, Jan/Feb. 2004)." Veruggio (2006) references J. Storrs Hall and various papers by Floridi when expounding the concept.[70]

These exclusions from machine ethics and roboethics may explain why it is so common for subsequent contributors to decry that there has not been much scholarly attention to the moral consideration of artificial entities (e.g. Levy, 2009; Metzinger, 2013; Gunkel, 2018, pp. 39-40). However, some contributors have explicitly argued for the inclusion of such topics in roboethics. In the same volume as Veruggio and Operto's (2006) delineation of the field, Asaro (2006) argued that "the best approach to robot ethics is one which addresses all three of… the ethical systems built into robots, the ethics of people who design and use robots, and the ethics of how people treat robots." While not necessarily arguing explicitly for its inclusion, later contributions have also used the term roboethics in a manner that would include discussion of moral consideration (e.g. Coeckelbergh, 2009; Steinart, 2014).

---

[67] They make only passing reference to "a few people" having been "interested in how human beings ought to treat machines." Gunkel (2018, p.. 38-9) notes that Anderson and Anderson's subsequent writings explicitly exclude "how human beings ought to treat machines" from machine ethics.

[68] Gunkel (2012, p. 103) adds that Hall's "exclusive focus on machine moral agency persists in Hall's subsequent book-length analysis, *Beyond AI: Creating the Conscience of the Machine* (2007). Although the term 'artificial moral agency' occurs throughout the text, almost nothing is written about the possibility of 'artificial moral patiency,' which is a term Hall does not consider or utilize."

[69] Hall (2000) drew on previous philosophical discussions. There are no references to foregoing detailed discussion of how humans ought to treat machines, although Kurzweil (1999) and Moravec (2000) are both cited, as are Minsky and Dennett who had written about the capacities of AI long previously. Hall also cited Robert Freitas, though not for his (1985) article about robot rights.

[70] In what is mostly an extension of Veruggio (2006), Veruggio and Operto (2006) emphasize in their article in the *International Review of Information Ethics* that "[r]oboethics shares many 'sensitive areas' with Computer Ethics, Information Ethics and Bioethics."

Similarly, Steve Torrance questioned in a paper entitled "A Robust View of Machine Ethics" (2005), presented to an AAAI Fall Symposium focused on machine ethics, whether we should "be thinking of extending the UN Universal Declaration of Human Rights to include future humanoid robots." Calverley's (2005a) paper presented at the same symposium also addressed the granting of legal rights to artificial entities. Neither author seems to have explicitly argued for the relevance of these topics to the emerging field of machine ethics; they continued lines of research that they had been developing elsewhere, but were accepted into the machine ethics symposium anyway.[71] Some subsequent papers have continued to identify themselves with the field of machine ethics while discussing the moral consideration of artificial entities (e.g. Torrance, 2008; Tonkens, 2012).

It seems then, that while some of the earliest formal expositions of machine ethics and roboethics excluded discussion of the moral consideration of artificial entities, a number of contributors have nevertheless addressed this topic within those fields. Furthermore, many of the authors interested in AI rights have continued to cite and discuss influential publications in machine ethics and roboethics (e.g. Veruggio, 2006; Wallach & Allen, 2008; Anderson & Anderson, 2011; see Table 9).

Table 9: Machine ethics and roboethics keyword searches

| Keyword | Items mentioning | % of items |
|---|---|---|
| "Robot ethics" | 91 | 33.7% |
| "Machine ethics" | 70 | 25.9% |
| Wallach | 60 | 22.2% |
| Anderson | 59 | 21.9% |
| Torrance | 40 | 15.2% |
| Roboethics | 38 | 14.1% |
| Asaro | 30 | 11.2% |
| Veruggio | 21 | 7.8% |
| "Ethics for Machines" | 8 | 3.0% |

[71] Torrance had written on this topic previously (e.g. briefly in 1984 and 2000) and continued to write on it subsequently (e.g. Torrance et al., 2006; Torrance, 2008). Calverley presented similar work at another conference in the same year (2005), and went on to publish additional relevant research (e.g. 2006; 2008). Torrance's work seems to have primarily stemmed out of artificial consciousness research but sometimes cites work by Floridi or, more regularly, Calverley; Calverley draws heavily on previous writings on both artificial consciousness and legal rights for artificial entities.

# Human-Computer Interaction and Human-Robot Interaction

Hewett et al. (1992, p. 5) defined human-computer interaction (HCI) as "a discipline concerned with the design, evaluation and implementation of interactive computing systems for human use and with the study of major phenomena surrounding them." The field's emergence was influenced by developments in computer science, ergonomics, cognitive psychology and a number of other disciplines, with specialist HCI journals, conferences, and organizations being set up from the 1970s onwards (Hewett et al., 1992). From the '90s, HCI researchers began to join together with researchers from robotics, cognitive science, psychology, and other disciplines to form the field of human-robot interaction (HRI), which seeks to "understand and shape the interactions between one or more humans and one or more robots" (Goodrich & Schultz, 2007).

Research in HCI and HRI is often not focused on ethical issues per se. When ethics is discussed, it is often with reference to the design of robots and computers, rather than their potential rights or moral value.

Nevertheless, Friedman et al.'s (2003) presentation at an HCI conference investigated "social responses to AIBO," a robotic dog, using "people's spontaneous dialog in online AIBO discussion forums," and noted that "few members (12%) affirmed that AIBO had moral standing."[72] The introduction referenced the lead author's presentation at an earlier HCI conference of interview findings on "reasoning about computers as moral agents" (Friedman, 1995) and a number of publications about various aspects of social interaction with robots or computers, but seemingly no previous literature about moral consideration. Instead, given that they generated their coding manual from "pilot data" on the forums, it seems possible that the authors' inclusion of "moral standing" as a category arose because the participants themselves were talking about the topic and the researchers felt unable to ignore this aspect.[73]

Friedman et al. (2003) has been cited hundreds of times (Google Scholar, 2021e), mostly by authors in the fields of HCI and HRI. The co-authors themselves published a number of subsequent items that empirically explored attributions of "moral standing" to artificial entities alongside perceptions of mental capacities and other attributes (e.g. Kahn et al., 2004; Kahn et al., 2006; Melson et al., 2009a; Melson et al., 2009b; Kahn et al., 2012). Otherwise, however, few of the publications citing Friedman et al. (2003) in the following few years seem to have focused primarily on issues related to moral consideration.[74] In

---

[72] Friedman et al.'s (2003) Table 1 noted that 7% of participants' responses suggested that AIBO "Engenders moral regard," 4% that it is a "Recipient of moral care," and 3% that it has or should have "Rights." Although less relevant, 3% suggested that AIBO "Deserves Respect," 1% that it is "Morally Responsible," and 1% that it is "Morally Blameworthy."

[73] Of course, many other factors could have influenced the authors to be open to including this category in their analysis. For example, Friedman et al. (2003) cited a number of publications about human interactions with animals, some of which may contain some ethical discussion. Friedman had also previously published a number of items that addressed ethical issues in computing.

[74] One seemingly relevant publication, Nomura et al. (2006), developed the "Negative Attitude toward Robots Scale (NARS)," though their motivation was to investigate "how humans are mentally affected" by robots, such as developing anxiety towards robots. None of the included items in the scale were about moral attitudes towards robots. The references were to other papers on anxiety and HRI but not to moral consideration. MacDorman and Cowley (2006) presented a paper about the potential criteria for robot personhood — including consciousness, appearance, and "the ability to sustain long-term relationships" — at a conference about "robot and human interactive communication." Their references included Kahn et al. (2006), which in turn drew heavily on Friedman et al. (2003) for its discussion of moral standing. Scheutz and Crowell (2007) addressed a number of "Social and

one of the most relevant publications, Freier (2008) interviewed 60 children and found "that the ability of the agent to express harm and make claims to its own rights significantly increases children's likelihood of identifying an act against the agent as a moral violation."[75]

Seemingly independently of the research by Friedman, Kahn, and colleagues,[76] a workshop was held in Rome in 2005 on "Abuse: The Darker Side of Human-Computer Interaction," and a follow-up was held in Montreal the next year (agentabuse.org, 2005). The descriptions of the workshops are clearly pitched towards the HCI research community, noting for example that "HCI research is witnessing a shift… to an experiential vision where the computer is described as a medium for emotion" (agentabuse.org, 2005). The language of the website suggests a primary concern for the interests of humans, rather than the computers themselves,[77] and this is reflected in the content of some of the papers presented at the workshops.[78] Other papers are more ambiguous in their motivations, but have clear implications for researchers interested in the moral consideration of artificial entities.[79] Most explicitly addressing this

Ethical Implications of Autonomous Robots," though the discussion of the possibility of robot rights is very brief and described as not "of pressing urgency, since such questions may only be relevant for robots much more advanced than those available at present."

It is possible that, like Friedman et al.'s (2003) own article, publications would have titles implying a focus on social interaction but include some discussion of moral consideration; in such cases, I would likely have missed relevant discussion.

[75] Freier (2008) cited numerous publications by Friedman and Kahn (including Friedman et al., 2003) and "thanks Batya Friedman and Peter H. Kahn, Jr., for their guidance in developing and conducting this work." Otherwise, none of the publications referenced by Freier (2008) seem to focus specifically on moral (as opposed to social) consideration of artificial entities. The paper is presented in the context of people having "frequent interactions" with artificial entities that are "routinely designed to mimic not only animate but also social and even moral entities in the world" and is motivated by "a general concern with the role that human values play in the design of technology." Freier builds on literature about social interaction with "personified technology" and literature about the moral development of children.

[76] None of the papers at either conference cited any empirical research by Friedman, Kahn, or Hagman.

[77] For example, the computer is described as "a medium for emotion" and concern is expressed that "negative behaviors that are directed not only towards the machine but also towards other people."

[78] For example, Zancanaro and Leonardi (2005) conducted a qualitative study to provide "initial insights on how groups can reduce the cognitive effort of using a co-located interface." Brahnam (2005) addressed customer abuse of "embodied conversational agents" (ECAs) but noted that "ECAs are not people and thus not capable of being harmed" and listed various human-focused reasons for concern with the abuse, such as degrading "the business value of using ECAs." Other papers addressed topics such as user frustrations, cyberbullying, cybersex, the moral development of children, and "rudeness in email."

[79] For example, De Angeli and Carpenter (2005) stated that their paper was "a preliminary attempt" to address the lack of research on "negative outcomes" of HRI, including "moral and ethical issues." They highlighted "an urgent need to explore the requirements for the establishment and negotiation of a cyber-etiquette to regulate the interaction between humans and artificial entities" and asked whether "respect for 'machines' [will] grow along with their abilities, or will the abuse spiral upward thanks to a perception of a developing risk of inter-'species' conflict?" The paper's references are to publications about user experiences or human-computer interaction. De Angeli's (2006) paper in the second conference explicitly noted that, ordinarily, the concept of "verbal abuse… should not apply to unanimated objects, as they cannot suffer any pain," and that machines "cannot feel any pain… they are inferior, unanimated objects." These comments suggest that De Angeli and Carpenter's (2005) paper was not likely motivated by concern about the artificial entities themselves that are abused.

Explicitly following up on De Angeli and Carpenter (2005), Brahnam (2006) explored "the effect gendered embodiment has on user verbal abuse." The motivations are not stated, but given Brahnam's (2005) comment that "ECAs are not people and thus not capable of being harmed," it seems likely that Brahnam's (2006) focus was on shedding light on human gender issues. Krenn and Gstrein (2006) studied "an online dating community where users topic, Bartneck et al. (2005b) tested how willing participants were to administer electric shocks to a robot when instructed to do so, and found that "participants showed compassion for the robot but the experimenter's urges were always enough to make them continue… until the maximum voltage was reached."[80]

Christopher Bartneck's publication history demonstrates how topics that have implications for the moral consideration of artificial entities can arise out of other topics in HCI or HRI. Bartneck had previously written about "Affective Expressions of Machines" (e.g. Bartneck, 2000), human interaction with artificial entities that express emotions (e.g. Bartneck, 2003), sci-fi treatment of social robots (Bartneck, 2004), and a wide array of topics relating to HRI but not moral consideration per se. Bartneck was certainly aware of some of the prior literature on legal rights for artificial entities.[81] However, Bartneck's papers relevant to AI rights more frequently noted concern for human experiences than for the experiences of the artificial entities themselves,[82] and most of the references were to other studies from

---

are represented by avatars" and found evidence that "in peer-to-peer contexts abusive behaviour is rare." They noted that they were inspired by De Angeli and Carpenter (2005), "where verbal abuse of a chatterbot by human users is explained by an asymmetrical power distribution between the human user and the dumb computer generated conversational system," but otherwise do not clarify their motivations for the study. Horstmann et al.'s (2018) paper examining hesitation when "switching off a robot which exhibits lifelike behavior," cited De Angeli and Carpenter (2005), though relatively few of the other articles citing this paper seem to have focused explicitly on the moral consideration of artificial entities.

Another ambiguous contribution comes from Ruzich (2006), who explores how and why, when computers crash, "those who stare in horror at blank screens and error messages frequently frame their experiences as if they represent compressed experiences with the stages of grief as identified by Elisabeth Kubler-Ross: the initial denial of loss, bargaining, rising anger, depression, and acceptance of the loss."

[80] Most of the papers citing Bartneck et al. (2005b) seem to focus on human-robot interaction rather than moral consideration per se, although some do touch on this topic. One of the earliest was Misselhorn's (2009) paper on "empathy with inanimate objects." Next was Rosenthal-von der Pütten et al.'s (2013) "experimental study on emotional reactions towards a robot." The rest of the decade saw numerous others, such as Horstmann et al.'s (2018) paper examining hesitation when "switching off a robot which exhibits lifelike behavior."

At the following conference, Bartneck (2006) described the motivation and method of Bartneck et al.'s forthcoming (2007) experiment. The motivation seems similar to Bartneck et al.'s (2005b) paper, though Bartneck (2006) explicitly notes that "[i]t is unclear if [robots] might remain 'property' or may receive the status of sentient beings."

Brščić et al. (2015) cite Bartneck et al. (2005b) as having "first used the term 'robot abuse,'" which matches my own impression, at least among HRI research.

[81] Bartneck (2004) and Bartneck et al. (2007) make passing reference to Lehman-Wilzig and Bartneck et al. (2005b) note that "this discussion eventually leads to legal considerations of the status of robots in our society," citing Calverley (2005b) as a study having addressed such considerations. The same paper is cited in Bartneck et al. (2007).

[82] Bartneck (2003) began by noting that, "[m]any companies, universities and research institutes are working on the home of the future… A key component of ambient intelligence is the natural interaction between the home and the user." The study measured users' enjoyment of the interaction; the tendency to cooperate with the character was also measured, though there was no explicit discussion of the implications for the moral consideration of artificial entities. Similarly, Bartneck et al. (2004) noted that "[t]he ability to communicate emotions is essential for a natural interaction between characters and humans" and did not express any concern for the interests of artificial entities themselves.

Although they framed the experiment as testing the idea that "humans treat computers as social actors," Bartneck et al. (2005b) seem to consider this to imply a moral dimension too, noting that their study explores the borderline of when "we treat [robots] again like machines that can be switched off, sold or torn apart without a bad the HCI and HRI fields. Despite having published hundreds of times, few of Bartneck's later publications seem to address the moral consideration of artificial entities explicitly (Google Scholar, 2021f).[83] Recently, Bartneck and Keijsers (2020) conducted an experiment examining responses to videos of abuse of robots, but Bartneck noted in a podcast interview that his concern with robot abuse was primarily one of virtue ethics, about how this behavior "reflects… on us," rather than concern for the robots themselves (Radio New Zealand, 2020).[84] Bartneck et al. (2005b), Bartneck et al. (2007), and Bartneck and Hu (2008) did not accrue more than a handful of citations until around 2013 onwards (Google Scholar, 2021f), though a number of publications have cited these works and proceeded in a similar fashion, examining HRI from a perspective that has clear implications for the moral consideration of artificial entities (e.g. Beran et al., 2010; Briggs et al., 2014).

One remarkably close parallel is a paper by Slater et al. (2006), who, like Bartneck et al. (2005b), carried out partial replications of Stanley Milgram's (1974) experiment on obedience — which tested whether participants would obey instructions to administer what they believed to be dangerous electric shocks to another person — with artificial entities. Whereas Bartneck et al. (2005b) used a robot, Slater et al. (2006) used a virtual human. Whereas Bartneck et al. (2005b) prominently cited previous studies on social interaction with robots to explain and justify the motivation for the study, Slater et al. (2006) prominently cited studies on human reactions and interactions in virtual environments and with virtual entities. Whereas Bartneck himself had numerous previous publications about HRI, Slater had numerous previous publications about interactions in virtual environments (Google Scholar, 2021c).[85] Slater et al. (2006) did not cite any works by Bartneck, Friedman, or Kahn, though Bartneck and Hu (2008) and a number of

---

consciousness." They also present as context the idea that robots are increasingly ubiquitous and cite some previous research suggesting that humans treat computers as social actors.

Bartneck et al. (2005a) began by noting the proliferation of robots, then commenting that, "[w]ith an increasing number of robots, robot anxiety might become as important as computer anxiety is today." No mention is made of concern for negative treatment of the robots themselves.

[83] Bartneck et al. (2007) measured whether a robot's intelligence and agreeableness influenced "hesitation to switch it off," with the paper's introduction explicitly mentioning the idea that this might constitute murder. Bartneck and Hu (2008) expanded on the Bartneck et al. (2005b) paper with a follow-up study; this was published in a "Special Section on Misuse And Abuse Of Interactive Technologies" in the journal *Interaction Studies*, which followed up on the "agent abuse" workshops (Bartneck et al., 2008).

[84] Bartneck explained that "[t]hat was the basic structure of the experiment, where people would see either a human or a robot being abused and then we would ask them then, well, what is the ethical aspect, how do you feel about this? And it turned out that people did not distinguish a human or a robot, so the abusive behavior to either of them was equally dismissive. Which is interesting because… it doesn't matter if the robot has no emotions, it has no pride, it doesn't even have tiny understanding of abuse, it doesn't know, it doesn't care; you can rip out its arm, it wouldn't care, so it makes absolutely no sense to feel sorry for it… A robot is a representation… of humans… and if we act towards it, we act towards a representation, and if we act poorly, that reflects also poorly on us, so from a virtue ethics point of view, we're not doing so great if we do this, we'd actually be a much better human if we treat other humans and other representations of humans well and so I think we should be gentle to robots."

[85] The stated goals, however, were somewhat different, with Slater et al. (2006) noting their aim as being "to investigate how people would respond to such a dilemma within a virtual environment, the broader aim being to assess whether such powerful social-psychological studies could be usefully carried out within virtual environments." They concluded that "in spite of the fact that all participants knew for sure that neither the stranger nor the shocks were real, the participants who saw and heard her tended to respond to the situation at the subjective, behavioural and physiological levels as if it were real."

other studies relating to the moral consideration of artificial entities (e.g. Misselhorn, 2009; Hartmann et al., 2010; Rosenthal-von der Pütten et al., 2013) have since cited Slater et al. (2006).[86]

Table 10: Human-Computer Interaction and Human-Robot Interaction keyword searches

| Keyword | Items mentioning | % of items |
|---|---|---|
| "Human-Robot Interaction" | 81 | 30.0% |
| Kahn | 27 | 10.0% |
| Bartneck | 24 | 8.9% |
| Friedman | 23 | 8.6% |
| "Human-Computer Interaction" | 22 | 8.1% |
| Slater | 10 | 3.7% |

Subsequently, a number of HRI or HCI publications have continued to explore the abuse of robots (e.g. Nomura et al., 2015; Brščić et al., 2015). Others have addressed the moral consideration of artificial entities from alternative angles, influenced by news events or legal and ethics papers relating to AI rights (e.g. Spence et al., 2018; Lima et al., 2020).

## Social-relational ethics

In 2018, communications scholar David Gunkel published *Robot Rights*, the first book focused solely on this topic. The book is "deliberately designed to think the unthinkable by critically considering and making (or venturing to make) a serious philosophical case for the rights of robots" (p. xi).

Gunkel comments (2018, p. xiii) that his first "formal articulation" of the topic of robot rights was in the last chapter of his book *Thinking Otherwise: Philosophy, Communication, Technology* (2007), and that this was then developed in more depth in the section on "Moral Patiency" in his book *The Machine Question* (2012) and in numerous subsequent articles. However, Gunkel had also published the relevant chapter as an article in 2006. Although Gunkel (2012, 2018) would go on to provide thorough reviews of the existing literature, Gunkel's (2006) early discussion of "the machine question" contained little reference to previous writings explicitly about the moral consideration of artificial entities outside of science fiction. An exception is Gunkel's (2006) numerous citations of Hall's (2000) essay, especially the quote that "we have never considered ourselves to have 'moral' duties to our machines, or them to us."[87] Gunkel has cited Hall's (2000) essay in at least seven different publications, including quoting this particular sentence again (e.g. 2014; 2018, p. 55), suggesting that the essay may have been a key influence on Gunkel's interest in the topic.[88] Gunkel (2006) also cited Anderson et al. (2004), another foundational work in machine ethics.

---

[86] Slater et al. (2006) has been widely cited (Google Scholar, 2021c). Many of the most prominent citations seem to focus on various aspects of human interaction, whether in virtual environments or not.
[87] Gunkel (2006) took established philosophical lines of thinking, such as by René Descartes and Immanuel Kant, and applies them to machines. This sometimes draws in particular on discussion of the comparable "animal question." Gunkel noted increasing social capabilities and indistinguishability from humans, drawing on Turing, science fictions, and a number of writings.

In *The Machine Question* (2012), Gunkel critiqued the idea that moral patiency should just be subsumed within discussion of moral agency and critiqued the binary thinking around moral inclusion or exclusion based on "individual qualities." Gunkel (2012, p. 177) eschewed intentional decision-making about the moral consideration of other beings based on their capacities, favoring instead "an uncontrolled and incomprehensible exposure to the face of the Other." These arguments are similar to those developed in Gunkel's other publications, drawing heavily on the work of the philosopher Emmanuel Levinas to advance what he later referred to as a "social relational" ethic (e.g. 2018, p. 10). This stands in stark contrast to much of the previous literature arguing for moral consideration of artificial entities based on the capacities of the entities themselves.[89]

Gunkel's social-relational approach was developed in tandem with the philosopher Mark Coeckelbergh.[90] While Gunkel seems to have started addressing the moral consideration of robots a few years earlier, it seems that Coeckelbergh was the first to explicitly use the term "social-relational" in his 2010 paper: "Robot rights? Towards a social-relational justification of moral consideration." Coeckelbergh (2010) critiqued deontological, utilitarian, and virtue ethical approaches that "rest on ontological features of entities" and that "seem to belong to the realm of science-fiction or at least the far future." Instead, Coeckelbergh (2010) argued for granting "some degree of moral consideration to some intelligent social robots" by "replacing the requirement that we have certain knowledge about real ontological features of the entity by the requirement that we experience the features of the entity as they appear to us in the

---

[88] Of course, this was not the only influence. For example, Gunkel (2016) noted that he has been interested in the philosophy of technology since he was in high school.

Gunkel (2018) was apparently motivated by "some very real and pressing challenges concerning emerging technology and the current state of and future possibilities for moral reasoning" (p. xi). The preface highlighted the ongoing "robot invasion" — the increasing ubiquity of robots and smart devices and their increasing social interaction with humans (pp. ix-x).

[89] Many of the earlier writers on legal rights for artificial entities situated their writings in the context of increasing capacities of artificial entities. They sometimes argued explicitly that it was these capacities, such as the ability to suffer, that might lead the entities to merit moral consideration (e.g. McNally & Inayatullah, 1988), though others had focused more on updating the legal systems regulating human interaction in the light of new technologies (e.g. Karnow, 1994).

The transhumanist writers likewise seemed concerned by suffering or death of sentient artificial beings for their own sake (e.g. Bostrom, 2002; Bostrom, 2005; Hughes, 2005). Regarding information ethics, Floridi (2002) had explicitly defended the view that, "[t]he moral value of an entity is based on its ontology. What the entity is determines the degree of moral value it enjoys, if any, whether and how it deserves to be respected and hence what kind of moral claims it can have on the agent."

[90] Gunkel (2012) cited Coeckelbergh only on a single page within the section on "moral agency" (p. 87). Although Coeckelbergh's (2009; 2010) early papers on the topic had not initially referenced Gunkel, Coeckelbergh (2013) reviewed *The Machine Question* and Gunkel reciprocated by reviewing (2013) Coeckelbergh's *Growing Moral Relations: Critique of Moral Status Ascription* (2012), which he praised as "a significant paradigm shift in moral thinking… a real game changer," highlighting its "relational, phenomenological, and transcendental" approach. In contrast to his earlier writings, Gunkel mentioned Coeckelbergh's name 40 times in *Robot Rights* (2018), including in the acknowledgements (p. xiv) as a "brilliant 'sounding board' for bouncing around ideas, and it was due to these interactions in Vienna [at a conference the pair attended] that it first became clear to me that this book needed to be the next writing project."

context of the concrete human-robot relation and the wider social structures in which that relation is embedded." Coeckelbergh had addressed some similar themes in a 2009 paper, albeit without the "social-relational" label.[91] Unlike Gunkel's (2006) earliest treatment of the topic, Coeckelbergh (2009; 2010) explicitly referenced many different previous writings that had touched on moral consideration of artificial entities.[92]

In the same year as Gunkel's (2006) first treatment of the topic, Søraker (2006a) argued explicitly for a "A Relational Theory of Moral Status," where "information and information technology, at least in very special circumstances, ought to be ascribed moral status." As well being "[i]nspired by the East Asian way of viewing the world as consisting of mutually constitutive relationships," Søraker (2006a) drew heavily on animal rights writings. Søraker (2006a) also acknowledged and cited Luciano Floridi.[93] Unlike Coeckelbergh's (2009; 2010; 2012) or Gunkel's (2012; 2018) writings on the topic, however, Søraker's (2006a) paper garnered only a small handful of citations, perhaps partly because Søraker did not pursue the topic as vigorously in subsequent publications (Google Scholar, 2021i). Søraker (2006a) and Coeckelbergh (2009; 2010) did not cite or acknowledge one another in their publications. However, given that both authors were in the department of philosophy at the University of Twente during this time period and were both contributing to a small new research field from a similar but relatively novel perspective, it seems likely that at least one had influenced the other's thinking.

Although not detailing a novel ethical perspective as fully as Coeckelbergh (2010), Gunkel (2012), or Søraker (2006a) a number of other writers in the late '00s had addressed similar themes. For example, human-robot interaction researcher Brian R. Duffy wrote a short paper published in the *International Review of Information Ethics* (2006) noting that, "[w]ith the advent of the social machine, and particularly the social robot… the *perception* as to whether the machine has intentionality, consciousness and free-will will change. From a social interaction perspective, it becomes less of an issue whether the machine *actually* has these properties and more of an issue as to whether it *appears* to have them." Duffy (2006) added that one perspective that gives rise to "the issue of rights and duties… involves the notion of whether a human *perceives* the machine to have moral rights and duties, and incorporates the aesthetic of the machine."[94] Relatedly, a number of authors expressed concerns similar to those hinted at by the HRI literature, about how negative treatment of robots might have implications for how humans interact with one another, or with animals (e.g. Whitby, 2008; Levy, 2009; Goldie, 2010).[95] Since then, numerous other authors have picked up on similar themes.[96]

---

[91] Coeckelbergh (2009) argued for "an approach to ethics of personal robots that advocates a methodological turn from robots to humans, from mind to interaction, from intelligent thinking to social-emotional being, from reality to appearance, from right to good, from external criteria to good internal to practice, and from theory to experience and imagination." Coeckelbergh's earlier papers (e.g. 2007) had addressed ethical issues relating to artificial entities, but focused less on moral consideration of those entities. Coeckelbergh has expanded on or revisited the topic a few times subsequently (2012; 2014) and published many times on partly overlapping topics (e.g. Coeckelbergh 2011; Stahl & Coeckelbergh, 2016).

[92] For example, Coeckelbergh (2009) prominently cited Floridi & Sanders (2004). Coeckelbergh (2009; 2010) also cited books by Moravec and Kurzweil, some HRI research, and numerous contributions explicitly discussing the moral consideration of artificial entities (e.g. McNally & Inayatullah, 1988; Brooks, 2000; Asaro, 2006; Calverley, 2006; Torrance, 2008; Whitby, 2008; Levy, 2009).

[93] Søraker (2006b) did the same in a publication from the same year that focused on computer ethics. That article also cited a publication co-authored by Lawrence Solum, so Søraker may have been aware of the arguments about legal rights for artificial entities in Solum (1992).

[94] Duffy's (2006) contained relatively few references; several seem to focus on social interaction with robots, but none on moral consideration of them. Duffy had written about anthropomorphism and perceptions of social robots before (e.g. Duffy, 2003), but does not seem to have explicitly linked the topic to rights for robots. Gunkel (2018) cited Duffy (2006), though Gunkel and Coeckelbergh's earliest writings on the topic do not seem to have done so.

Table 11: Social-relational ethics keyword searches

| Keyword | Items mentioning | % of items |
|---|---|---|
| Gunkel | 73 | 28.5% |
| Coeckelbergh | 65 | 24.6% |
| Levy | 44 | 16.5% |
| "Social-relational" | 42 | 15.6% |
| Whitby | 29 | 10.7% |
| Duffy | 9 | 3.3% |
| Søraker | 3 | 1.1% |

## Moral and social psychology

Psychology has contributed to the study of artificial life and consciousness (e.g. Krach et al., 2008), human-computer interaction (Hewett et al., 1992), and human-robot interaction (Goodrich & Schultz, 2007), all of which have encouraged some interest in the moral consideration of artificial entities. There has also been some interest in studying artificial entities as part of wider psychological theory-building about how moral inclusion and exclusion work.

There are many different concepts and scales (batteries of tests intended to measure a particular attitude or psychological construct) relating to moral consideration that can be empirically examined across a range of different entity types. For example, Reed and Aquino's (2003) "moral regard for outgroups" scale included questions about a number of different groups of humans. Some scales have included various nonhumans but not artificial entities (e.g. Laham, 2009; Crimston et al., 2016),[97] but, at least two scales relevant to moral consideration have included artificial entities as well. Both were developed within a few years of each other and have been widely cited.

Firstly, to explore "whether minds are perceived along one or more dimensions," Gray et al. (2007) asked participants questions about "seven living human forms… three nonhuman animals… a dead woman,

[95] Indeed, Whitby (2008) and Goldie (2010) explicitly cited the proceedings of the second agent abuse workshop that had been held in Canada in 2006, with Goldie (2010) also citing a wider range of HRI research. In contrast, Levy (2009) cited extensively the research on artificial consciousness and on "legal rights of robots," but not HRI research. The final chapter in Levy's (2005) book had addressed some similar themes, with many of the same citations, alongside others, such as writing by Floridi and Sanders. Coeckelbergh (2010) cited Levy (2009) but not Whitby (2006).
[96] For discussion, see Gunkel (2018, pp. 133-58). Harris and Anthis (2021) note a number of additional publications adopting a social-relational perspective.
[97] Of course, subsequent authors can easily manipulate these scales to add in artificial entities themselves.

God, and a sociable robot (Kismet)."[98] Published in *Science*, Gray et al.'s (2007) discussion of their study is very brief, so their motivation for including a social robot in the scale is unclear. They cited Turing and Dennett in their second sentence as examples of authors who have assumed "that mind perception occurs on one dimension"; their inclusion of a robot as one of the studied entity types could be due to their interest in the topic having been sparked partly by these two thinkers, both of who prominently discussed the capabilities of AI (see "artificial life and consciousness" above).[99] Some subsequent authors have continued to use robots as an entity type when exploring issues related to moral agency and patiency (e.g. Ward et al., 2013),[100] while others have chosen not to do so (e.g. Piazza et al., 2014).[101]

Secondly, Waytz et al.'s (2010) "Individual Differences in Anthropomorphism Questionnaire" (IDAQ) asked about views on the capabilities of a number of nonhuman entities, both natural (e.g. animals, clouds) and artificial (e.g. robots, computers). Though developed by psychologists and published in psychology journals, both Waytz et al.'s (2010) paper and the earlier, more theoretical paper that it built upon (Epley et al., 2007) included a number of references from HRI and HCI, such as prior work on perceptions and anthropomorphism of robots. Both papers explicitly noted consequences of their theory and studies of anthropomorphism for HCI and "Moral Care and Concern" for nonhumans.

At a similar time, some of the psychological research around dehumanization included robots or other "automata" (e.g. Haslam, 2006; Loughnan & Haslam, 2007).[102] This literature focused more on the humans being dehumanized through comparison to or representation as automata than on robot rights, though this of course has some implications for how and why artificial entities are excluded from moral consideration. Indeed, this stream of research has often been cited alongside discussions of mind perception and anthropomorphism to explain and justify the focus of psychological research that investigates various aspects of the moral consideration of artificial entities.[103]

---

[98] Gray et al. (2007) did not directly assess moral consideration, but they noted that their two identified dimensions of "agency" and "experience" relate "to Aristotle's classical distinction between moral agents (whose actions can be morally right or wrong) and moral patients (who can have moral right or wrong done to them). Agency is linked to moral agency and hence to responsibility, whereas Experience is linked to moral patiency and hence to rights and privileges."

[99] In a subsequent paper, Gray and Wegner (2009) examined the distinction between moral agency and moral patiency more fully, briefly citing Floridi and Sanders (2004) for the idea that "moral agency [could] be ascribed to… mechanical agents, such as robots or computers." They also cited other possible sources of interest in artificial entities, such as Haslam's (2006) "integrative review" of "dehumanization" which included brief discussion of "[t]echnology in general and computers in particular." However, it is unclear whether knowledge of these papers influenced the earlier Gray et al. (2007) publication.

[100] Ward et al. (2013) found through four experiments that "observing intentional harm to an unconscious entity—a vegetative patient, a robot, or a corpse—leads to augmented attribution of mind to that entity." One of the co-authors on this paper (who was also the lead author's PhD supervisor), Daniel M. Wegner, had been a co-author of the Gray et al. (2007) paper, which is cited prominently in the introduction.

[101] Piazza et al. (2014) were certainly aware of Gray et al. (2007), since they cited that paper, other papers by its co-authors, and personal correspondence with Heather Gray. Nevertheless, they chose to test their hypothesis that "harmfulness… is an equally if not more important determinant of moral standing" than moral patiency or agency through "four studies using non-human animals as targets."

[102] This literature appears to have at least some precedent that was several decades older. For example, Haslam (2006) discussed contributions from 1983 and 1984 in support of the comment that "[t]echnology in general and computers in particular are a common theme in work on dehumanization."

[103] See, for example, Starmans and Friedman's (2016) paper exploring whether "autonomy makes entities less ownable," which cited Haslam (2006) and Loughnan and Haslam (2007) alongside Gray et al. (2007) in the introduction. Their third experiment compared vignettes describing a human, an alien, and a robot, asking

Table 12: Moral and social psychology keyword searches

| Keyword | Items mentioning | % of items |
|---|---|---|
| Psychology | 137 | 50.7% |
| Wegner [a co-author of Gray et al. (2007)] | 33 | 12.3% |
| Waytz | 28 | 10.4% |
| Haslam | 14 | 5.2% |

## Synthesis and proliferation

In more recent years, authors have continued to refine and develop ideas about the moral consideration of artificial entities and to conduct new relevant empirical research. Aggregating across the various streams of literature discussed above and the intersections between them, the number of scholarly publications on the topic seems to have been growing exponentially in the 21st century (Figure 1).

From the mid '10s onwards, it was no longer reasonable to claim that the topic as a whole had not been addressed at all; new contributions have tended to cite one or more of the relevant streams of research,[104] though of course some earlier contributions had done this as well.[105]

Even where a publication has garnered attention for addressing a seemingly new and surprising topic, there has sometimes been discussion of similar ideas among earlier contributions. For example, whether robots should be slaves was discussed decades earlier than Joanna Bryson's (2010) controversial article on the topic in science fiction (e.g. Čapek's 1921 play *R.U.R.*; Chu, 2010), at least one public-facing article (Modern Mechanix, 1957), and academic writing on legal rights for artificial entities (e.g. Lehman-Wilzig, 1981; LaChat, 1986).

The '10s also saw an increase in the prevalence of publications explicitly arguing against the moral consideration of artificial entities. There had been some earlier arguments to this effect (e.g. Drozdek, participants to judge whether someone could own the entity in question, and whether behavior relating to owning the entity was morally acceptable. See also Swiderska and Küster (2018), who "investigated if the presence of a facial wound enhanced the perception of mental capacities (experience and agency) in response to images of robotic and human-like avatars, compared to unharmed avatars." They cited Haslam (2006) briefly after discussion of the implications of Epley et al. (2007), Waytz et al. (2010), Gray et al. (2007), and Gray and Wegner (2009).

---

[104] John Danaher's (2020) exposition and defense of "ethical behaviourism" is a good example. This theory "holds that robots can have significant moral status if they are *roughly performatively equivalent* to other entities that have significant moral status." Danaher notes that "[v]ariations of this theory are        hinted at in the writings of others… but it is believed that this article is the first to explicitly name it, and provide an extended defence of it." Danaher acknowledges not only the precursors of this specific line of thinking (e.g. Sparrow, 2004; Levy, 2009), but also other authors who have "already defended the claim that we should take the moral status of robots seriously" (e.g. Gunkel, 2018).

[105] For example, Sparrow (2004) proposed a test for when machines have achieved "moral standing comparable to a human," referencing Putnam, Kurzweil, Moravec, and Floridi. Magnani (2005) argued that technological "things" are better construed as "moral mediators" than as moral agents or patients, drawing primarily on Kantian ethics, but referencing literature on animal rights, legal personhood for artificial entities, and information ethics.

1994; Birmingham, 2008), but mostly the idea had simply been ignored or marginalized, as in the early machine ethics and roboethics publications, rather than explicitly critiqued.[106]

Table 13: Recent contributions keyword searches

| Keyword | Items mentioning | % of items |
|---|---|---|
| Bryson | 52 | 19.5% |
| Darling | 43 | 16.0% |
| Danaher | 26 | 9.7% |
| Richardson | 20 | 7.5% |
| Schwitzgebel | 16 | 5.9% |

Research on AI rights and other moral consideration of artificial entities has received a number of thorough literature reviews (e.g. Gunkel, 2018; Harris & Anthis, 2021). Several papers have called for integration of the empirical research from HCI, HRI, and social psychology with moral questions relevant to AI rights (Vanman & Kappas, 2019; Harris & Anthis, 2021). Indeed, a number of empirical research projects have been inspired by or noted their relevance to ongoing ethical discussions (e.g. Spence et al., 2018; Lima et al., 2020; Küster et al., 2021). Other contributions have also explicitly sought to integrate seemingly disparate or conflicting strands of ethical and legal reasoning about the moral consideration of artificial entities (e.g. Gellers, 2020).

In the 21st century, there have also been a number of news stories relevant to AI rights, such as the 2006 paper commissioned by the UK "Horizon Scanning Centre" suggesting that robots could be granted rights in 20 to 50 years, South Korea's proposed "robot ethics charter" in 2007, a 2017 European Parliament resolution that recommended the granting of legal status to "electronic persons," and the granting of Saudi Arabian citizenship to the robot Sophia in 2017 (see Harris, 2021).

These events seem to have encouraged at least some academic discussion. Certainly, a number of authors mention them (see Table 14). Occasionally, authors explicitly cite these events as a motivation for their research or interest in the topic, such as Bennett and Daly (2020) framing their work as addressing the questions raised by the European Parliament Committee on Legal Affairs' report. In other cases, the events may be one of several influences on the authors, or just a way to help justify their research as seeming current and important. For example, shortly after the 2006 report commissioned by the Horizon Scanning Centre, a symposium was organized on the question of "Robots & Rights: Will Artificial Intelligence Change The Meaning of Human Rights?" featuring talks on the moral status of artificial entities by Nick Bostrom and Steve Torrance (James & Scott, 2008). The opening sentence of the introduction to the symposium refers to the Horizon Scanning Centre report (James & Scott, 2008), though it does not explicitly claim that this was the key spark for the symposium to be organized.[107]

---

[106] For further discussion, see "Dismissal of the Importance of Moral Consideration of Artificial Entities" in Harris and Anthis (2021) and note that few of the items given a score lower than three out of five for "Argues for moral consideration?" in Table 7 were published before 2010. Gunkel's (2018) section on "Robot Rights or the Unthinkable" includes a number of references from prior to 2010, but the contributions summarized in the chapter "S1 !S2: Although Robots Can Have Rights, Robots Should Not Have Rights" are almost all from 2010 or later.

Table 14: News events keyword searches

| Keyword | Items mentioning | % of items |
| --- | --- | --- |
| "European Parliament" | 30 | 11.1% |
| Sophia | 30 | 11.1% |
| "Robot ethics charter" | 6 | 2.2% |
| "Horizon Scanning" | 5 | 1.9% |

# Discussion

## Why has interest in this topic grown substantially in recent years?

- Certain contributors may have inspired others to publish on the topic.

The "Results" section above identifies a handful of initial authors who seem to have played a key role in sparking discussion relevant to AI rights in each new stream of research, such as Floridi for information ethics, Bostrom for transhumanism, effective altruism, and longtermism, and Gunkel and Coeckelbergh for social-relational ethics. Perhaps, then, some of the subsequent contributors who cited these authors were encouraged to address the topic because those writings sparked their interest in AI rights, or the publication of those items reassured them that it was possible (and sufficiently academically respectable) to publish about it.

This seems especially plausible given that the beginnings of exponential growth some time between the late '90s and mid-'00s (Figure 1) coincides reasonably well with the first treatments of the topic by several streams of research (Figure 2). This hypothesis could be tested further through interviews with later contributors who cited those pioneering works. Of course, even if correct, this hypothetical answer to our question would then beg another question: why did those pioneering authors themselves begin to address the moral consideration of artificial entities? Again, interviews (this time with the pioneering authors) may be helpful for further exploration.

- The gradual, accumulating ubiquity of AI and robotic technology may have encouraged increased academic interest.

---

[107] The next paragraph quotes Bill Gates, providing an alternative possible current affairs spark for interest in the topic, so it is unclear what the primary cause was. Gunkel (2018, pp. 193-4) claims that, "[l]ike the April 2007 Dana Centre event and preceding press conference at the Science Media Center (see chapter 1), this BioCenter (http://www.bioethics.ac.uk/) symposium was also developed in direct response to the Ipsos MORI document that was commissioned and published by the UK Office of Science and Innovation's Horizon Scanning Centre." No citation for this claim is provided.

A common theme in the introductions of and justifications for relevant publications is that the number, technological sophistication, and social integration of robots, AIs, computers, and other artificial entities is increasing (e.g. Lehman-Wilzig, 1981; Willick, 1983; Hall, 2000; Bartneck et al., 2005b). Some of these contributors and others (e.g. Freitas, 1985; McNally & Inayatullah, 1988; Bostrom, 2014) have been motivated by predictions about further developments in these trends. We might therefore hypothesize that academic interest in the topic has been stimulated by ongoing developments in the underlying technology.

Indeed, bursts of technical publications on AI in the 1950s and '60s, artificial life in the '90s, and synthetic biology in the '00s seem to have sparked ethical discussions, where some of the contributors seem to have been largely unaware of previous, adjacent ethical discussions.[108]

Additionally, the "Results" section above details how several new streams of relevant research from the 1980s onwards seem to have arisen independently of one another, such as Floridi's information ethics and the early transhumanist writers not citing each other or citing the previous research on legal rights for artificial entities. Even *within* the categories of research there was sometimes little interaction, such as the absence of cross-citation amongst the earliest contributors to discussion on each of legal rights for artificial entities, HCI and HRI (where relevant to AI rights), and social-relational ethics.[109] If these different publications addressing similar topics did indeed arise independently of one another, it suggests that there were one or more underlying factors encouraging academic interest in the topic. The development and spread of relevant technologies is a plausible candidate for being such an underlying factor.[110]

However, the timing of the beginnings of exponential growth in publications on the moral consideration of artificial entities — seemingly from around the beginning of the 21st century (Figure 1) — does not match up very well to the timing and shape of technological progress. For example, there seems to have only been linear growth in industrial robot installations and AI job postings in the '10s (Zhang et al., 2021),[111] whereas exponential growth in computing power began decades earlier, in the 20th century (Roser & Ritchie, 2013). This suggests that while this factor may well have contributed to the growth of research on AI rights and other moral consideration of artificial entities, it cannot single-handedly explain it.

- Relevant news events may have encouraged increased academic interest.

As noted in the "Synthesis and proliferation" subsection above, there have been a number of news events in the 21st century relevant to AI rights, and these have sometimes been mentioned by academic contributors to discussion on this topic. However, only a relatively small proportion of recent publications explicitly mention these events (Table 14). Additionally, the first relevant news event mentioned by multiple different publications was in 2006, whereas the exponential growth in publications seems to have begun prior to that (Figure 1). A particular news story also seems intuitively more likely to encourage a spike in publications than the start of an exponential growth trend.

---

[108] See the section on "Artificial life and consciousness," especially footnote 15.

[109] Lehman-Wilzig (1981), Willick (1983), Freitas (1985), and Solum (1992) did not cite one another, Slater et al. (2006) did not cite previous relevant HCI and HRI works by Bartneck, Friedman, or Kahn, and Gunkel, Coeckelbergh, Duffy, and Søraker developed somewhat similar ideas without any citation of each other in their earliest writings (though Søraker and Coeckelbergh may have had communication). See the relevant subsections above.

[110] It is not, however, the only one; see the point below on "The growth in research on this topic reflects wider trends in academic research."

[111] Of the many figures in Zhang et al.'s (2021) report, there is one that seems somewhat close to the trend in publications identified in Harris and Anthis (2021): "U.S. government total contract spending on AI, FY 2001-20."

- The growth in research on this topic reflects wider trends in academic research.

If the growth in academic publications in general — i.e. across any and all topics — has a similar timing and shape to the growth in interest in AI rights and other moral consideration of artificial entities, then we need not seek explanations for growth that are unique to this specific topic. There is some evidence that this is indeed the case; Fire and Guestrin's (2019) analysis of the Microsoft Academic Graph dataset identified exponential growth in the number of published academic papers throughout the 20th and early 21st century, and Ware and Mabe (2015) identified exponential growth in the numbers of researchers, journals, and journal articles, although their methodology for assessing the number of articles is unclear.[112]

At a more granular level, however, the prevalence of certain topics can presumably deviate from wider trends in publishing. For example, Zhang et al. (2021) report "the number of peer-reviewed AI publications, 2000-19"; the growth appears to have been exponential in the '10s, but not the '00s. There was a similar pattern in the "number of paper titles mentioning ethics keywords at AI conferences, 2000-19."

So it was not inevitable that the number of relevant publications would increase exponentially as soon as some of the earliest contributors had touched on the topic of the moral consideration of artificial entities.[113] But science fiction, artificial life and consciousness, environmental ethics, and animal ethics all had some indirect implications for the moral consideration of artificial entities, even if they were not always stated explicitly. So it seems unsurprising that, in the context of exponential growth of academic publications, at least some scholars would begin to explore these implications more thoroughly and formally. Indeed, even though several of the new streams of relevant research from the 1980s onwards seem to have arisen largely independently of each other, they often owed something to one or more of these earlier, adjacent topics.[114]

---

[112] In contrast, Google Scholar searches limited by year to each of 1990, 1991, 1992 and so on until 2021 suggest a publication pattern that looks quite unlike the growth in research on AI rights (see the spreadsheet "Google Scholar searches for "e", "robot", and ""artificial intelligence"""). The search for "e" was used as a proxy for all items in Google Scholar, although this does seem to focus on items that have "e" as a standalone word in the title or authors' names; it may be a fairly random selection of publications but not the full set of publications on Google Scholar in that year. It is also not clear whether the "About X results" comments provided by Google Scholar do indeed represent all items on the database for each year.

[113] Consider by analogy that any number of unusual and niche topics might make it through the peer review process, but that their success in doing so does not guarantee the emergence of a research field around that topic.

[114] Floridi and the early writers on legal rights for artificial entities both seem to have drawn heavily on environmental ethics and, to a lesser extent, animal ethics, while the early transhumanist writers drew on ideas and research about artificial life and consciousness. Machine ethics then drew on transhumanism, and social-relational ethics in turn drew on machine ethics and several other previous streams of discussion. Many drew on science fiction. Some of the earliest relevant contributions from HCI and HRI may have arisen more independently. Relevant contributions from moral and social psychology seem to have been influenced by research on artificial life and consciousness and by HCI and HRI. See the relevant subsections of the "Results" section for further detail.

# Which levers can be pulled on to further increase interest in this topic?

- Adopt publication strategies similar to those of the most successful previous contributors, focusing either on integration into other topics or on persistently revisiting AI rights.

There seem to be two separate models for how the most notable and widely cited contributors to AI rights research have achieved influence.

Some, like Nick Bostrom, Mel Slater (and co-authors), and Lawrence Solum have published relatively few items specifically on this topic, but where they have done so, they have integrated the research into debates or topics of interest to a broader audience. They've mostly picked up citations for those other reasons and topics, rather than their discussion of the moral consideration of artificial entities. They've also tended to have strong academic credentials or publication track record relevant to those other topics, which may be a necessary condition for success in pursuing this model of achieving influence.[115]

Others, like David Gunkel and Luciano Floridi, published directly on this topic numerous times, continuing to build upon and revisit it. Many of their individual contributions attracted limited attention in the first few years after publication,[116] but through persistent revisiting of the topic (and the passage of time) these authors have nonetheless accumulated impressive numbers of citations across their various publications relevant to AI rights. These authors continue to pursue other academic interests, however, and a substantial fraction of the interest in these authors (Floridi more so than Gunkel) seems to focus on how their work touches on other topics and questions, rather than its direct implications for the moral consideration of artificial entities.

Of course, these two models of paths to influence are simplifications. Some influential contributors, like Christoper Bartneck and Mark Coeckelbergh, fall in between these two extremes. There may be other publication strategies that could be even more successful, and it is possible that someone could adopt one of these strategies and still not achieve much influence.[117] Nevertheless, new contributors could take inspiration from these two pathways to achieving academic influence — which seem to have been quite successful in at least some cases — when seeking to maximize the impact of their own research.

- Engage with adjacent academic fields and debates.

As noted above, a number of contributors have accrued citations from papers that addressed but did not focus solely on the moral consideration of artificial entities. Early contributions that addressed the moral consideration of artificial entities more directly without reference to other debates often languished in relative obscurity, at least for many years (e.g. Lehman-Wilzig, 1981; Willick, 1983; Freitas, 1985; McNally & Inayatullah, 1988). This suggests that engaging with adjacent academic fields and debates may be helpful for contributors to be able to increase the impact of their research relevant to AI rights. Relatedly, there is reason to believe that Fields' first exposure to academic discussion relevant to AI rights may have been at an AI conference,[118] perhaps encouraging them to write their 1987 article.

---

[115] Of course, it may not be a necessary condition. For example, their success in research relevant to AI rights and in other topics may both just be attributable to underlying factors such as high intelligence or persuasive writing style. See footnote 37 for a discussion of factors potentially contributing to Solum's success.

[116] See the spreadsheet "[Google Scholar citations for key authors](#)."

[117] For example, Harris and Anthis (2021) included four publications co-authored by Dennis Küster and Aleksandra Swiderska with a combined total of only 10 Google Scholar citations at the time of checking in mid-2020, as well as five publications by Robin Mackenzie that had a combined total of 23 citations. Some reasonably high-quality contributions to the field that have adopted neither strategy seem to have attracted little attention (e.g. Søraker, 2006a).

Although it seems coherent to distinguish between moral patiency and moral agency (e.g. Floridi, 1999; Gray et al., 2007; Gunkel, 2012), many successful publications have discussed both areas. For instance, much of the relevant literature in transhumanism, effective altruism, and longtermism has focused on threats posed to humans by intelligent artificial agents but has included some brief discussion of artificial entities as moral patients. Many writings address legal rights for artificial entities in tandem with discussion of those entities' legal responsibilities to humans or each other. Before Gunkel (2018) wrote *Robot Rights*, he wrote (2012) *The Machine Question* with roughly equal weight to questions of moral agency and moral patiency. Even Floridi, who has often referred to information ethics as a "patient-oriented" ethics, has been cited numerous times by contributors interested in AI rights for his 2004 article co-authored with Jeff Sanders "On the Morality of Artificial Agents"; 32 of the items in Harris and Anthis' (2021) systematic searches (12.1%) have cited that article. Indeed, for some ethical frameworks, there is little meaningful distinction between agency and patiency.[119] Similarly, some arguments both for (e.g. Levy, 2009) and against (e.g. Bryson, 2010) the moral consideration of artificial entities seem to be motivated by concern for indirect effects on human society. So contributors may be able to tie AI rights issues back to human concerns, discuss both the moral patiency and moral agency of artificial entities, or discuss both legal rights and legal responsibilities; doing so may increase the reach of their publications.

Artificial consciousness, environmental ethics, and animal ethics all had potentially important ramifications for the moral consideration of artificial entities. These implications were remarked upon at the time, including by some of the key thinkers who developed these ideas, but the discussion was often brief. Later, machine ethics and roboethics had great potential for including discussion relevant to AI rights, but some of the early contributors seem to have decided to mostly set aside such discussion. It seems plausible that if some academics had been willing to address these implications more thoroughly, AI rights research might have picked up pace much earlier than it did. There may be field-building potential from monitoring the emergence and development of new, adjacent academic fields and reaching out to their contributors to encourage discussion of the moral consideration of artificial entities.

As well as providing opportunities to advertise publications relevant to AI rights, engagement with adjacent fields and debates provides opportunities for inspiration and feedback. Floridi (2013) and Gunkel (2018) acknowledge discussion at conferences that had no explicit focus on AI rights as having been influential in shaping the development of their books.[120] Additionally, several authors first presented

---

[118] See footnote 30.

[119] Floridi (2002) summarizes Kant's ethical views as following this pattern. See also Wareham (2013) and Laukyte (2017).

[120] Gunkel (2018, pp. xiii-xiv) credits the "form and configuration" of *Robot Rights* to the *Robophilosophy/TRANSOR 2016: What Social Robots Can and Should Do?* conference at the University of Aarhus, Denmark. Despite the conference's focus on what *robots* can and should do, Gunkel and Coeckelbergh were not the only contributors to consider how *humans* should treat robots: other presentations included Dennis Küster and

initial drafts of their earliest relevant papers at such conferences (e.g. Putnam, 1960; Lehman-Wilzig, 1981; Floridi, 1999).

- Create specialized resources for research on AI rights and other moral consideration of artificial entities, such as journals, conferences, and research institutions.

While the above points attest to the usefulness of engagement with adjacent fields and debates (e.g. by attending conferences, citing relevant publications), in order to grow further, it seems likely that AI rights research also needs access to its own specialized "organizational resources" (Frickel & Gross, 2005) such as research institutions, university departments, journals, and conferences (Muehlhauser, 2017; Animal Ethics, 2021). With a few exceptions (e.g. *The Machine Question: AI, Ethics and Moral Responsibility* symposium at the AISB / IACAP 2012 World Congress; Gunkel et al., 2012), the history of AI rights research reveals a striking lack of such specialized resources, events, and institutions. Indeed, it is only recently that whole books dedicated solely to the topic have emerged (Gunkel, 2018; Gellers, 2020; Gordon, 2020).[121]

The creation of such specialized resources could also help to guard against the possibility that, as they intentionally engage with adjacent academic fields and debates, researchers drift away from their exploration of the moral consideration of artificial entities.

- Explore legal rights for artificial entities.

Detailed discussion of the legal rights of artificial entities was arguably the first area of academic enquiry to focus in much depth on the moral consideration of artificial entities. Articles that touch on the moral consideration of artificial entities from a legal perspective seem to more frequently accrue a substantial number of citations (e.g. Lehman-Wilzig, 1981; McNally & Inayatullah, 1988; Solum, 1992; Karnow, 1994; Allen & Widdison, 1996; Chopra & White, 2004; Calverley, 2008).[122] Additionally, in recent years, there have been a number of news stories related to legal rights of artificial entities (Harris, 2021). This

Aleksandra Świderska's (2016) "Moral Patients: What Drives the Perceptions of Moral Actions Towards Humans and Robots?" and Maciej Musiał's (2016) "Magical Thinking and Empathy Towards Robots." This is a fairly similar spread of topics to the first conference of the series (Seibt et al., 2014), which had included contributions from Gunkel and Coeckelbergh and discussions ranging from robots' agency to social interaction to emotional capacities. Although the third conference of the series (Coeckelbergh et al., 2018) was more bereft of discussion of this topic, the fourth conference (Nørskov et al., 2021) contained a workshop entitled "Should Robots Have Standing? The Moral and Legal Status of Social Robots," with six contributing presentations (two by Gunkel).

Floridi (2013, p. xviii) notes that, "[t]he CEPE (Computer Ethics Philosophical Enquiries) meetings, organized by the International Society for Ethics and Information Technology, and the CAP (Computing and Philosophy) meetings, organized by the International Association for Computing and Philosophy, provided stimulating and fruitful venues to test some of the ideas presented in this and in the previous volume."

[121] Gunkel (2012) is a borderline earlier example, with the moral patiency of artificial entities being the focus of about half of the book, and their moral agency being the focus of the other half. Floridi (2013) includes discussion of moral patiency when explaining information ethics, but also addresses many other topics.

[122] From the items included in Harris and Anthis' (2021) systematic searches whose "Primary framework or moral schema used" was categorized as "Legal precedent," the median number of citations tracked by Google Scholar (at the time of checking, in mid-2020), was 9, which compares to a median of 4 from the full sample of 294 items. Six out of 33 "legal precedent" (18%) articles had 50 citations or more, compared to 34 out of 294 (12%) in the full sample.

could be due to differences in the referencing norms between different academic fields, but otherwise weakly suggests that exploration of legal topics is more likely to attract interest and have immediate relevance to public policy than more abstract philosophical or psychological topics.

# Limitations

This report has relied extensively on inferences about authors' intellectual influences based on explicit mentions and citations in their published works. These inferences may be incorrect, since there are a number of factors that may affect how an author portrays their influences.

For example, in order to increase the chances that their manuscript is accepted for publication by a journal or cited by other researchers, an author may make guesses about what others would consider to be most appealing and compelling, then discuss some ideas more or less extensively than they would like to. Scholars are somewhat incentivized to present their works as novel contributions, and so not to cite works with a substantial amount of overlap. Authors might also accidentally omit mention of previous publications or ideas that have influenced their own thinking.

There are a few instances where a likely connection between authors has not been mentioned, although we cannot know in any individual case why not. One example is the works of Mark Coeckelbergh and Johnny Hartz Søraker, who were both advancing novel "relational" perspectives on the moral consideration of artificial entities while in the department of philosophy at the University of Twente, but who do not cite or acknowledge each other's work. Another is how Nick Bostrom gained attention for the ideas that suggest our world is likely a simulation, but a similar point had been made earlier by fellow transhumanist Hans Moravec.[123]

These examples suggest that the absence of mentions of particular publications does not prove that the author was not influenced by those publications. But there are also some reasons why the opposite may be true at times; that an author might mention publications that had barely influenced their own thinking.

For example, they may be incentivized to cite foundational works in their field or works on adjacent, partly overlapping topics, in order to reassure publishers that there will be interest in their research. Alternatively, someone might come up with an idea relatively independently, but then conduct an initial literature review in order to contextualize their ideas; citing the publications that they identify would falsely convey the impression that their thinking had been influenced by those publications.

Since the relevance of identified publications was sometimes filtered by the title alone, it is likely that I have missed publications that contained relevant discussion but did not advertise this clearly in the title. Additionally, citations of included publications were often identified using the "Cited by…" tool on Google Scholar, but this tool seems to be imperfect, sometimes omitting items that I know to have cited the publication being reviewed.

---

[123] See footnote 43.

This report initially used Harris and Anthis' (2021) literature review as its basis, which relied on systematic searches using keywords in English language. This has likely led to a vast underrepresentation of relevant content published in other languages. There is likely at least some work written in German, Italian, and other European languages. For example, Gunkel (2018) discussed some German-language publications that I did not see referenced in any other works (e.g. Schweighofer, 2001).

This language restriction has likely also led to a substantial neglect of relevant writings by Asian scholars. For Western scholars exploring the moral consideration of artificial entities, Asian religions and philosophies have variously been the focus of their research (e.g. Robertson, 2014), an influence on their own ethical perspectives (e.g. McNally & Inayatullah, 1988), a chance discovery, or an afterthought, if they are mentioned at all.[124] However, very few items have been identified in this report that were written by Asian scholars themselves, and there may well be many more relevant publications.

This report has not sought to explore in depth the longer-term intellectual origins for academic discussion of the moral consideration of artificial entities, such as the precedents provided by various moral philosophies developed during the Enlightenment.

As I have discussed at length elsewhere (Harris, 2019), assessing causation from historical evidence is difficult; "we should not place too much weight on hypothesized historical cause and effect relationships in general," or on "the strategic knowledge gained from any individual historical case study." The commentary in the discussion section should therefore be treated as one interpretation of the identified evidence, rather than as established fact.

The keyword searches are limited to the items included from Harris and Anthis' (2021) systematic searches. Those searches did not include all research papers with relevance to the topic. For example, the thematic discussion in this report includes a number of publications that could arguably have merited inclusion in that review, if they had been identified by the systematic searches.

The items identified in each keyword search were not manually checked to ensure that they did indeed refer to the keyword in the manner that was assumed. For example, the search for "environment" may have picked up mentions of that word that have nothing to do with environmental ethics (e.g. how a robot interacts with its "environment") or just because they were published in — or cited another item that was published in — a journal with "environment" in its title.

Similarly, where multiple authors who might have been cited in the included publications share a surname (as is the case for at least the surnames Singer, Friedman, and Anderson), then the keyword searches might overrepresent the number of citations of that author. In contrast, if an author has a name that is sometimes misspelled by others (e.g. Putnam, Freitas, Lehman-Wilzig), then the searches might underrepresent the number of citations of them.

---

[124] For example, Floridi (2013, p. xiv) notes that, "[a]pparently, there are also some spiritual overtones and connections to Confucianism, Buddhism, Taoism, and Shintoism" in his book. "They were unplanned and they are not based on any intended study of the corresponding sources. I was made aware of such connections by other philosophers, while working on the articles that led to this book." Gunkel (2012) makes no mention of Asian philosophy in his (2012) book The Machine Question. However, his (2018a) *Robot Rights* includes some discussion when citing other contributors and their ideas, such as Robertson (2014) and McNally and Inayatullah (1988). A subsequent article (Gunkel, 2020) responds to a recent article published in *Science and Engineering Ethics* (Zhu et al., 2020) by exploring the implications of Confucianism for "AI/robot ethics" in more depth, especially the overlap between Confucianism and the social-relational perspective that Gunkel has developed.

# Potential items for further study

What is the history of AI rights research that is written in languages other than English? This report predominantly only included publications written in English, so relevant research in other languages may have been accidentally excluded.

Given the difficulty in assessing causation through historical evidence and in making inferences about authors' intellectual influences based solely on explicit mentions in their published works, it would be helpful to supplement this report with interviews of researchers and other stakeholders.

Previous studies and theoretical papers have identified certain features as potentially important for the emergence of "scientific/intellectual movements" (e.g. Frickel & Gross, 2005; Animal Ethics, 2021). A literature review of such contributions could be used to generate a list of potentially important features. The history of AI rights research could then be assessed against this list: which features appear to be present and which missing?

Histories of other research fields could be useful for better understanding which levers can be pulled on to further increase interest in AI rights. Such studies could focus on the research fields that have the most overlap in content and context (e.g. AI alignment, AI ethics, animal ethics) or that have achieved success most rapidly (e.g. computer science, cognitive science, synthetic biology).

There are numerous alternative historical research projects that could help to achieve the underlying goal of this report — to better understand how to encourage an expansion of humanity's moral circle to encompass artificial sentient beings. For example, rather than focusing on academic research fields, historical studies could focus on technological developments that have already created or substantially altered sentient life, such as cloning, factory farming, and genetic editing.

# References

Aarhus University. (2021). Robophilosophy 2016 / TRANSOR 2016. https://conferences.au.dk/robo-philosophy/previous-conferences/rp2016/. Accessed 22 September 2021

Abelson, R. (1966). Persons, P-Predicates, and Robots. *American Philosophical Quarterly*, *3*(4), 306–311.

agentabuse.org. (2005). About the Abuse Interact 2005 Workshop & About Rome's Talking Sculptures. http://www.agentabuse.org/about.htm. Accessed 3 December 2021

Allen, T., & Widdison, R. (1996). Can Computers Make Contracts. *Harvard Journal of Law & Technology*, *9*, 25.

Anderson, M., & Anderson, S. L. (2011). *Machine Ethics*. Cambridge University Press.

Anderson, M., Anderson, S. L., & Armen, C. (2004). Towards Machine Ethics (p. 7). Presented at the AAAI-04 Workshop on Agent Organizations: Theory and Practice, San Jose, CA.

Animal Ethics. (2021). *Establishing a research field in natural sciences*. Oakland, CA: Animal Ethics. https://www.animal-ethics.org/establishing-new-field-natural-sciences/. Accessed 31 December 2021

Anthis, J. R., & Paez, E. (2021). Moral circle expansion: A promising strategy to impact the far future. *Futures*, *130*, 102756. https://doi.org/10.1016/j.futures.2021.102756

Armstrong, S., Sandberg, A., & Bostrom, N. (2012). Thinking Inside the Box: Controlling and Using an Oracle AI. *Minds and Machines*, *22*(4), 299–324. https://doi.org/10.1007/s11023-012-9282-2

Asaro, P. M. (2001). Hans Moravec, Robot. Mere Machine to Transcendent Mind, New York, NY: Oxford University Press, Inc., 1999, ix + 227 pp., $25.00 (cloth), ISBN 0-19-511630-5. *Minds and Machines*, *11*(1), 143–147. https://doi.org/10.1023/A:1011202314316

Asaro, P. M. (2006). What Should We Want From a Robot Ethic? *The International Review of Information Ethics*, *6*, 9–16. https://doi.org/10.29173/irie134

Barfield, W. (2005). Issues of Law for Software Agents within Virtual Environments. *Presence: Teleoperators and Virtual Environments*, *14*(6), 741–748. https://doi.org/10.1162/105474605775196607

Bartneck, C. (2000). *Affective expressions of machines*. Stan Ackermans Institute, Eindhoven. Retrieved from https://ir.canterbury.ac.nz/bitstream/handle/10092/13665/bartneckMasterThesis2000.pdf?sequence=2

Bartneck, C. (2003). Interacting with an embodied emotional character. In *Proceedings of the 2003 international conference on Designing pleasurable products and interfaces - DPPI '03* (p. 55). Presented at the the 2003 international conference, Pittsburgh, PA, USA: ACM Press. https://doi.org/10.1145/782896.782911

Bartneck, C. (2004). From Fiction to Science – A cultural reflection of social robots (p. 4). Presented at the CHI2004 Workshop on Shaping Human-Robot Interaction, Vienna.

Bartneck, C. (2006). Killing a Robot. In A. De Angeli, S. Brahnam, P. Wallis, & A. Dix (Eds.), *Misuse and Abuse of Interactive Technologies* (pp. 5–8). Presented at the CHI 2006 Conference on Human Factors in Computing Systems, Montréal Québec Canada: ACM. https://doi.org/10.1145/1125451.1125753

Bartneck, C., Brahnam, S., Angeli, A. D., & Pelachaud, C. (2008). Editorial. *Interaction Studies*, *9*(3), 397–401. https://doi.org/10.1075/is.9.3.01edi

Bartneck, C., & Hu, J. (2008). Exploring the abuse of robots. *Interaction Studies. Social Behaviour and Communication in Biological and Artificial Systems*, *9*(3), 415–433. https://doi.org/10.1075/is.9.3.04bar

Bartneck, C., J, R., & A, B. (2004). In your face, robot! The influence of a character's embodiment on how users perceive its emotional expressions. In *Proceedings of the Design and Emotion Conference*. Ankara, Turkey. https://doi.org/10.6084/m9.figshare.5160769

Bartneck, C., & Keijsers, M. (2020). The morality of abusing a robot. *Paladyn, Journal of Behavioral Robotics*, *11*(1), 271–283. https://doi.org/10.1515/pjbr-2020-0017

Bartneck, C., Nomura, T., Kanda, T., Suzuki, T., & Kato, K. (2005a). Cultural Differences in Attitudes Towards Robots. In *Proceedings of the AISB Symposium on Robot Companions: Hard*

*Problems And Open Challenges In Human-Robot Interaction* (pp. 1–4). Hatfield, UK. https://ir.canterbury.ac.nz/bitstream/handle/10092/16849/bartneckAISB2005.pdf?sequence=2

Bartneck, C., Rosalia, C., Menges, R., & Deckers, I. (2005b). Robot Abuse – A Limitation of the Media Equation. In A. De Angeli, S. Brahnam, & P. Wallis (Eds.), *Proceedings of Abuse: The darker side of Human-Computer Interaction*. http://www.agentabuse.org/Abuse_Workshop_WS5.pdf

Bartneck, C., van der Hoek, M., Mubin, O., & Al Mahmud, A. (2007). "Daisy, daisy, give me your answer do!" switching off a robot. In *2007 2nd ACM/IEEE International Conference on Human-Robot Interaction (HRI)* (pp. 217–222). Presented at the 2007 2nd ACM/IEEE International Conference on Human-Robot Interaction (HRI).

Basl, J. (2014). Machines as Moral Patients We Shouldn't Care About (Yet): The Interests and Welfare of Current Machines. *Philosophy & Technology*, *27*(1), 79–96. https://doi.org/10.1007/s13347-013-0122-y

Basl, J., & Sandler, R. (2013). The good of non-sentient entities: Organisms, artifacts, and synthetic biology. *Studies in History and Philosophy of Science Part C: Studies in History and Philosophy of Biological and Biomedical Sciences*, *44*(4, Part B), 697–705. https://doi.org/10.1016/j.shpsc.2013.05.017

Beers, D. L. (2006). *For the Prevention of Cruelty: The History and Legacy of Animal Rights Activism in the United States*. Ohio University Press.

Bepress. (2021). SelectedWorks - Curtis E.A. Karnow. https://works.bepress.com/curtis_karnow/. Accessed 18 November 2021

Bennett, B., & Daly, A. (2020). Recognising rights for robots: Can we? Will we? Should we? *Law, Innovation and Technology*, *12*(1), 60–80. https://doi.org/10.1080/17579961.2020.1727063

Beran, T. N., Ramirez-Serrano, A., Kuzyk, R., Nugent, S., & Fior, M. (2011). Would Children Help a Robot in Need? *International Journal of Social Robotics*, *3*(1), 83–93. https://doi.org/10.1007/s12369-010-0074-7

Boden, M. A. (1984). Artificial intelligence and social forecasting*. *The Journal of Mathematical Sociology*, *9*(4), 341–356. https://doi.org/10.1080/0022250X.1984.9989954

Bostrom, N. (1998). How long before superintelligence? *International Journal of Future Studies*, *2*. https://www.nickbostrom.com/superintelligence.html. Accessed 23 November 2021

Bostrom, N. (2001). Ethical Principles in the Creation of Artificial Minds. https://www.nickbostrom.com/ethics/aiethics.html. Accessed 8 May 2022

Bostrom, N. (2002). Existential risks: analyzing human extinction scenarios and related hazards. *Journal of Evolution and Technology*, *9*. https://ora.ox.ac.uk/objects/uuid:827452c3-fcba-41b8-86b0-407293e6617c. Accessed 23 November 2021

Bostrom, N. (2003). Are We Living in a Computer Simulation? *The Philosophical Quarterly*, *53*(211), 243–255. https://doi.org/10.1111/1467-9213.00309

Bostrom, N. (2005). A History of Transhumanist Thought. *Journal of Evolution and Technology*, *14*(1), 1–25.

Bostrom, N. (2008). The Simulation Argument FAQ. https://www.simulation-argument.com/faq.html. Accessed 29 December 2021

Bostrom, N., & Yudkowsky, E. (2014). The Ethics of Artificial Intelligence. In K. Frankish & W. M. Ramsey (Eds.), *The Cambridge Handbook of Artificial Intelligence* (pp. 316–334). Cambridge, UK: Cambridge University Press.

Brahnam, S. (2005). Strategies for handling customer abuse of ECAs. In A. De Angeli, S. Brahnam, & P. Wallis (Eds.), *Proceedings of Abuse: The darker side of Human-Computer Interaction*. http://www.agentabuse.org/Abuse_Workshop_WS5.pdf

Brahnam, S. (2006). Gendered Bods and Bot Abuse. In A. De Angeli, S. Brahnam, P. Wallis, & A. Dix (Eds.), *Misuse and Abuse of Interactive Technologies* (pp. 13–16). Presented at the CHI 2006 Conference on Human Factors in Computing Systems, Montréal Québec Canada: ACM. https://doi.org/10.1145/1125451.1125753

Brennan, A., & Lo, Y.-S. (2021). Environmental Ethics. In E. N. Zalta (Ed.), *The Stanford Encyclopedia of Philosophy* (Winter 2021.). Metaphysics Research Lab, Stanford University. https://plato.stanford.edu/archives/win2021/entries/ethics-environmental/. Accessed 5 October 2021

Brey, P. (2008). Do we have moral duties towards information objects? *Ethics and Information Technology*, *10*(2), 109–114. https://doi.org/10.1007/s10676-008-9170-x

Brooks, R. (2000, June 19). Will Robots Rise Up And Demand Their Rights? *Time*. http://content.time.com/time/subscriber/article/0,33009,997274,00.html. Accessed 29 November 2021

Brščić, D., Kidokoro, H., Suehiro, Y., & Kanda, T. (2015). Escaping from Children's Abuse of Social Robots. In *Proceedings of the Tenth Annual ACM/IEEE International Conference on Human-Robot Interaction* (pp. 59–66). Presented at the HRI '15: ACM/IEEE International Conference on Human-Robot Interaction, Portland Oregon USA: ACM. https://doi.org/10.1145/2696454.2696468

Bryson, J. J. (2010). Robots should be slaves. In Y. Wilks (Ed.), *Natural Language Processing* (Vol. 8, pp. 63–74). Amsterdam: John Benjamins Publishing Company. https://doi.org/10.1075/nlp.8.11bry

Calverley, D. J. (2005a). Additional Thoughts Concerning the Legal Status of a Non-biological Machine. In *Papers from the 2005 AAAI Fall Symposium* (pp. 30–37). Menlo Park, CA: The AAAI Press.

Calverley, D. J. (2005b). Toward a Method for Determining the Legal Status of a Conscious Machine. In *Proceedings of the Symposium on Next Generation Approaches to Machine Consciousness: Imagination, Development, Intersubjectivity and Embodiment* (pp. 75–84). Presented at the AISB'05: Social Intelligence and Interaction in Animals, Robots and Agents, Hatfield, UK: The Society for the Study of Artificial Intelligence and the Simulation of Behaviour. https://www.sacral.c.u-tokyo.ac.jp/pdf/Ikegami_MachineConsciousness_2005.pdf#page=86

Calverley, D. J. (2006). Android science and animal rights, does an analogy exist? *Connection Science*, *18*(4), 403–417. https://doi.org/10.1080/09540090600879711

Calverley, D. J. (2008). Imagining a non-biological machine as a legal person. *AI & SOCIETY*, *22*(4), 523–537. https://doi.org/10.1007/s00146-007-0092-7

Cameron, D. E., Bashor, C. J., & Collins, J. J. (2014). A brief history of synthetic biology. *Nature Reviews Microbiology*, *12*(5), 381–390. https://doi.org/10.1038/nrmicro3239

Capurro, R. (2006). Towards an ontological foundation of information ethics. *Ethics and Information Technology*, *8*(4), 175–186. https://doi.org/10.1007/s10676-006-9108-0

Capurro, R., & Pingel, C. (2002). Ethical issues of online communication research. *Ethics and Information Technology*, *4*(3), 189–194. https://doi.org/10.1023/A:1021372527024

Cherry, C. (1989). The possibility of computers becoming persons. A response to Dolby. *Social Epistemology*, *3*(4), 337–348. https://doi.org/10.1080/02691728908578546

Chopra, S., & White, L. (2004). Artificial Agents - Personhood in Law and Philosophy. In R. L. De Mántaras & L. Saitta (Eds.), *ECAI'04: Proceedings of the 16th European Conference on Artificial Intelligence* (pp. 635–639). Valencia, Spain. http://astrofrelat.fcaglp.unlp.edu.ar/filosofia_cientifica/media/papers/Chopra-White-Artificial_Agents-Personhood_in_Law_and_Philosophy.pdf

Chu, S.-Y. (2010). Robot Rights. In *Do Metaphors Dream of Literal Sleep?: A Science-Fictional Theory of Representation* (pp. 214–244). Cambridge, MA: Harvard University Press.

Clifford, R. D. (1996). Intellectual Property in the Era of the Creative Computer Program: Will the True Creator Please Stand Up. *Tulane Law Review*, *71*, 1675.

Coeckelbergh, M., Loh, J., & Funk, M. (2018). *Envisioning Robots in Society – Power, Politics, and Public Space: Proceedings of Robophilosophy 2018 / TRANSOR 2018*. IOS Press.

Coeckelbergh, Mark. (2007). Violent computer games, empathy, and cosmopolitanism. *Ethics and Information Technology*, *9*(3), 219–231. https://doi.org/10.1007/s10676-007-9145-3

Coeckelbergh, Mark. (2009). Personal Robots, Appearance, and Human Good: A Methodological Reflection on Roboethics. *International Journal of Social Robotics*, *1*(3), 217–221. https://doi.org/10.1007/s12369-009-0026-2

Coeckelbergh, Mark. (2010). Robot rights? Towards a social-relational justification of moral consideration. *Ethics and Information Technology*, *12*(3), 209–221. https://doi.org/10.1007/s10676-010-9235-5

Coeckelbergh, Mark. (2011). Humans, Animals, and Robots: A Phenomenological Approach to Human-Robot Relations. *International Journal of Social Robotics*, *3*(2), 197–204. https://doi.org/10.1007/s12369-010-0075-6

Coeckelbergh, Mark. (2012). *Growing Moral Relations: Critique of Moral Status Ascription*. Palgrave Macmillan.

Coeckelbergh, Mark. (2013). David J. Gunkel: The machine question: critical perspectives on AI, robots, and ethics. *Ethics and Information Technology*, *15*(3), 235–238. https://doi.org/10.1007/s10676-012-9305-y

Coeckelbergh, Mark. (2014). The Moral Standing of Machines: Towards a Relational and Non-Cartesian Moral Hermeneutics. *Philosophy & Technology*, *27*(1), 61–77. https://doi.org/10.1007/s13347-013-0133-8

Crimston, C. R., Bain, P. G., Hornsey, M. J., & Bastian, B. (2016). Moral expansiveness: Examining variability in the extension of the moral world. *Journal of Personality and Social Psychology*, *111*(4), 636–653. https://doi.org/10.1037/pspp0000086

Crippa, A. (2017). Ted talks analyses. https://rstudio-pubs-static.s3.amazonaws.com/321337_38458c80a3fb4edf8755e8bce876e822.html. Accessed 10 November 2021

Danaher, J. (2020). Welcoming Robots into the Moral Circle: A Defence of Ethical Behaviourism. *Science and Engineering Ethics*, *26*(4), 2023–2049. https://doi.org/10.1007/s11948-019-00119-x

Danto, A. C. (1960). On Consciousness in Machines. In S. Hook (Ed.), *Dimensions of Mind* (pp. 180–187). New York, NY: New York University Press.

Dator, J. (1990). It's only a paper moon. *Futures*, *22*(10), 1084–1102. https://doi.org/10.1016/0016-3287(90)90009-7

De Angeli, A. (2006). On Verbal Abuse Towards Chatterbots. In A. De Angeli, S. Brahnam, P. Wallis, & A. Dix (Eds.), *Misuse and Abuse of Interactive Technologies* (pp. 21–24). Presented at the CHI 2006 Conference on Human Factors in Computing Systems, Montréal Québec Canada: ACM. https://doi.org/10.1145/1125451.1125753

De Angeli, A., & Carpenter, R. (2005). Stupid computer! Abuse and social identities. In A. De Angeli, S. Brahnam, & P. Wallis (Eds.), *Proceedings of Abuse: The darker side of Human-Computer Interaction*. http://www.agentabuse.org/Abuse_Workshop_WS5.pdf

De Montfort University. (2021). ETHICOMP. https://www.dmu.ac.uk/research/centres-institutes/ccsr/ethicomp.aspx. Accessed 6 October 2021

Dennett, D. C. (1978). Current Issues in the Philosophy of Mind. *American Philosophical Quarterly*, *15*(4), 249–261.

Dennett, Daniel C. (1971). Intentional Systems. *The Journal of Philosophy*, *68*(4), 87–106. https://doi.org/10.2307/2025382

Dennett, Daniel C. (1994). The practical requirements for making a conscious robot | Philosophical Transactions of the Royal Society of London. Series A: Physical and Engineering Sciences. *Philosophical Transactions of the Royal Society of London. Series A: Physical and Engineering Sciences*, *349*(1689), 133–146. https://doi.org/10.1098/rsta.1994.0118

Diderot, D. (2012). D'Alembert's Dream. (I. Johnston, Trans.). http://www.blc.arizona.edu/courses/schaffer/249/Before%20Darwin%20-%20New/Diderot/Diderot,%20D'Alembert's%20Dream.htm. Accessed 25 June 2022

Dolby, R. G. A. (1989). The possibility of computers becoming persons. *Social Epistemology*, *3*(4), 321–336. https://doi.org/10.1080/02691728908578545

Douglas, T., & Savulescu, J. (2010). Synthetic biology and the ethics of knowledge. *Journal of medical ethics*, *36*(11), 687–693. https://doi.org/10.1136/jme.2010.038232

Doyle, T. (2010). A Critique of Information Ethics. *Knowledge, Technology & Policy*, *23*(1), 163–175. https://doi.org/10.1007/s12130-010-9104-x

Drozdek, A. (1994). To 'the possibility of computers becoming persons' (1989). *Social Epistemology*, *8*(2), 177–197. https://doi.org/10.1080/02691729408578742

Duffy, B. R. (2003). Anthropomorphism and the social robot. *Robotics and Autonomous Systems*, *42*(3), 177–190. https://doi.org/10.1016/S0921-8890(02)00374-3

Duffy, B. R. (2006). Fundamental Issues in Social Robotics. *The International Review of Information Ethics*, *6*, 31–36. https://doi.org/10.29173/irie137

Elton, M. (2000). Should Vegetarians Play Video Games? *Philosophical Papers*, *29*(1), 21–42. https://doi.org/10.1080/05568640009506605

Epley, N., Waytz, A., & Cacioppo, J. T. (2007). On seeing human: A threefactor theory of anthropomorphism. *Psychological Review*, *114*(4), 864–886. https://doi.org/10.1037/0033-295X.114.4.864

Farmer, J. D., & Belin, A. d'A. (1990). *Artificial life: The coming evolution* (No. LA-UR-90-378; CONF-891131-). Los Alamos National Lab. (LANL), Los Alamos, NM (United States). https://www.osti.gov/biblio/7043104. Accessed 29 November 2021

Fiedler, F. A., & Reynolds, G. H. (1993). Legal Problems of Nanotechnology: An Overview. *Southern California Interdisciplinary Law Journal*, *3*, 593.

Fields, C. (1987). Human-computer interaction: A critical synthesis. *Social Epistemology*, *1*(1), 5–25. https://doi.org/10.1080/02691728708578410

Fire, M., & Guestrin, C. (2019). Over-optimization of academic publishing metrics: observing Goodhart's Law in action. *GigaScience*, *8*(6), giz053. https://doi.org/10.1093/gigascience/giz053

Floridi, L. (1996a). Brave.Net.World: the Internet as a disinformation superhighway? *The Electronic Library*, *14*(6), 509–514. https://doi.org/10.1108/eb045517

Floridi, L. (1996b). Internet: Which Future for Organized Knowledge, Frankenstein or Pygmalion? *The Information Society*, *12*(1), 5–16. https://doi.org/10.1080/019722496129675

Floridi, L. (1998a). Does Information Have a Moral Worth in Itself? In *Computer Ethics: Philosophical Enquiry (CEPE'98) in Association with the ACM SIG on Computers and Society*. London School of Economics and Political Science, London. https://doi.org/10.2139/ssrn.144548

Floridi, L. (1998b). Information Ethics: On the Philosophical Foundation of Computer Ethics. In J. van den Hoven, S. Rogerson, T. W. Bynum, & D. Gotterbarn (Eds.), *Proceedings of the Fourth International Conference on Ethical Issues of Information Technology*. Rotterdam, The Netherlands.

Floridi, L. (1999). Information ethics: On the philosophical foundation of computer ethics. *Ethics and Information Technology*, *1*(1), 33–52. https://doi.org/10.1023/A:1010018611096

Floridi, L. (2002). On the intrinsic value of information objects and the infosphere. *Ethics and Information Technology*, *4*(4), 287–304. https://doi.org/10.1023/A:1021342422699

Floridi, L. (2006). Information ethics, its nature and scope. *ACM SIGCAS Computers and Society*, *36*(3), 21–36. https://doi.org/10.1145/1195716.1195719

Floridi, L. (2010a). *Information: A Very Short Introduction*. OUP Oxford.

Floridi, L. (2010b). *The Cambridge Handbook of Information and Computer Ethics*. Cambridge University Press.

Floridi, L. (2011). *The fourth technological revolution*. TEDxMaastricht. https://www.youtube.com/watch?v=c-kJsyU8tgI&ab_channel=TEDxTalks. Accessed 10 November 2021

Floridi, L. (2013). *The Ethics of Information*. OUP Oxford.

Floridi, L. (2017a). Robots, Jobs, Taxes, and Responsibilities. *Philosophy & Technology*, *30*(1), 1–4. https://doi.org/10.1007/s13347-017-0257-3

Floridi, L. (2017b, February 22). Roman law offers a better guide to robot rights than sci-fi. *Financial Times*. https://www.ft.com/content/99d60326-f85d-11e6-bd4e-68d53499ed71. Accessed 10 November 2021

Floridi, L., & Sanders, J. W. (2001). Artificial evil and the foundation of computer ethics. *Ethics and Information Technology*, *3*(1), 55–66. https://doi.org/10.1023/A:1011440125207

Floridi, L., & Sanders, J. W. (2002). Mapping the foundationalist debate in computer ethics. *Ethics and Information Technology*, *4*(1), 1–9. https://doi.org/10.1023/A:1015209807065

Floridi, L., & Sanders, J. W. (2004). On the Morality of Artificial Agents. *Minds and Machines*, *14*(3), 349–379. https://doi.org/10.1023/B:MIND.0000035461.63578.9d

Floridi, L., & Taddeo, M. (2018). Romans would have denied robots legal personhood. *Nature*, *557*(7705), 309–309. https://doi.org/10.1038/d41586-018-05154-5

Freier, N. G. (2008). Children attribute moral standing to a personified agent. In *Proceeding of the twenty-sixth annual CHI conference on Human factors in computing systems - CHI '08* (pp. 343–352). Presented at the Proceeding of the twenty-sixth annual CHI conference, Florence, Italy: ACM Press. https://doi.org/10.1145/1357054.1357113

Freitas, R. A. (1985). Legal Rights of Robots. *Student Lawyer*, *13*, 54–56.

Frickel, S., & Gross, N. (2005). A General Theory of Scientific/Intellectual Movements. *American Sociological Review*, *70*(2), 204–232. https://doi.org/10.1177/000312240507000202

Friedman, B. (1995). It's the computer's fault: reasoning about computers as moral agents. In *Conference Companion on Human Factors in Computing Systems* (pp. 226–227). New York, NY, USA: Association for Computing Machinery. https://doi.org/10.1145/223355.223537

Friedman, B., Kahn, P. H., & Hagman, J. (2003). Hardware companions? what online AIBO discussion forums reveal about the human-robotic relationship. In *Proceedings of the SIGCHI Conference on Human Factors in Computing Systems* (pp. 273–280). New York, NY, USA: Association for Computing Machinery. https://doi.org/10.1145/642611.642660

Froehlich, T. (2004). A brief history of information ethics. *BiD: textos universitaris de biblioteconomia i documentació*, *13*. http://bid.ub.edu/13froel2.htm. Accessed 29 December 2021

Gamez, D. (2008). Progress in machine consciousness. *Consciousness and Cognition*, *17*(3), 887–910. https://doi.org/10.1016/j.concog.2007.04.005

Gandon, F. L. (2003). Combining reactive and deliberative agents for complete ecosystems in infospheres. In *IEEE/WIC International Conference on Intelligent Agent Technology, 2003. IAT 2003.* (pp. 297–303). Presented at the IEEE/WIC International Conference on Intelligent Agent Technology, 2003. IAT 2003. https://doi.org/10.1109/IAT.2003.1241082

Gellers, J. C. (2020). *Rights for Robots: Artificial Intelligence, Animal and Environmental Law* (1st ed.). London, UK: Routledge. https://doi.org/10.4324/9780429288159

Goldie, P. (2010). The Moral Risks of Risky Technologies. In S. Roeser (Ed.), *Emotions and Risky Technologies* (pp. 127–138). Dordrecht: Springer Netherlands. https://doi.org/10.1007/978-90-481-8647-1_8

Goodrich, M. A., & Schultz, A. C. (2008). Human–Robot Interaction: A Survey. *Foundations and Trends in Human–Computer Interaction*, *1*(3), 203–275. https://doi.org/10.1561/1100000005

Google Scholar. (2021a). ("Mindcrime" OR "mind crime" OR "mind-crime") AND Bostrom. https://scholar.google.com/scholar?start=0&q=(%22Mindcrime%22+OR+%22mind+crime%22+OR+%22mind-crime%22)+AND+Bostrom&hl=en&as_sdt=0,5

Google Scholar. (2021b). Luciano Floridi. https://scholar.google.co.uk/citations?user=jZdTOaoAAAAJ&hl=en. Accessed 10 November 2021

Google Scholar. (2021c). Mel Slater.
https://scholar.google.com/citations?user=5gGSgcUAAAAJ&hl=en. Accessed 14 December 2021

Google Scholar. (2021d). suffering subroutines.
https://scholar.google.com/scholar?hl=en&as_sdt=0%2C5&q=%22suffering+subroutines%22&btnG=

Google Scholar. (2021e). Batya Friedman.
https://scholar.google.com/citations?user=dkjR4cAAAAAJ&hl=en. Accessed 15 December 2021

Google Scholar. (2021f). Christoph Bartneck.
https://scholar.google.com/citations?user=NrcTgeUAAAAJ&hl=en. Accessed 7 December 2021

Google Scholar. (2021g). Daniel C. Dennett.
https://scholar.google.com/citations?user=3FWe5OQAAAAJ&hl=en. Accessed 19 November 2021

Google Scholar. (2021h). Frankenstein unbound: Towards a legal definition of artificial intelligence.
https://scholar.google.com/citations?view_op=view_citation&hl=en&user=Yln8YccAAAAJ&citation_for_view=Yln8YccAAAAJ:9yKSN-GCB0IC. Accessed 12 November 2021

Google Scholar. (2021i). Johnny Hartz Søraker.
https://scholar.google.com/citations?user=ZiW2NWoAAAAJ&hl=en. Accessed 20 December 2021

Google Scholar. (2021j). Lawrence Solum.
https://scholar.google.com/citations?user=vXYJjpEAAAAJ&hl=en. Accessed 16 November 2021

Google Scholar. (2021k). Robert A. Freitas Jr.
https://scholar.google.com/citations?user=DpoSX2QAAAAJ&hl=en. Accessed 12 November 2021

Google Scholar. (2021l). sohail inayatullah.
https://scholar.google.com.au/citations?user=gB0Ea_wAAAAJ&hl=en. Accessed 16 November 2021

Gordon, J.-S. (Ed.). (2020). *Smart Technologies and Fundamental Rights*. Leiden, The Netherlands: Brill. https://brill.com/view/title/55392. Accessed 3 January 2022

Gray, H. M., Gray, K., & Wegner, D. M. (2007). Dimensions of Mind Perception. *Science*, *315*(5812), 619–619. https://doi.org/10.1126/science.1134475

Gray, K., & Wegner, D. M. (2009). Moral typecasting: Divergent perceptions of moral agents and moral patients. *Journal of Personality and Social Psychology*, *96*(3), 505–520. https://doi.org/10.1037/a0013748

Guither, H. D. (1998). *Animal Rights: History and Scope of a Radical Social Movement*. SIU Press.

Gunkel, D., Bryson, J., & Torrance, S. (Eds.). (2012). *The Machine Question: AI, Ethics and Moral Responsibility*. Birmingham, UK: AISB. https://citeseerx.ist.psu.edu/viewdoc/download?doi=10.1.1.446.9723&rep=rep1&type=pdf

Gunkel, D. J. (2006). The Machine Question: Ethics, Alterity, and Technology. *Explorations in Media Ecology*, *5*(4), 259–278. https://doi.org/10.1386/eme.5.4.259_1

Gunkel, D. J. (2012). *The Machine Question: Critical Perspectives on AI, Robots, and Ethics*. MIT Press.

Gunkel, D. J. (2013). Mark Coeckelbergh: Growing moral relations: critique of moral status ascription. *Ethics and Information Technology*, *15*(3), 239–241. https://doi.org/10.1007/s10676-012-9308-8

Gunkel, D. J. (2014). A Vindication of the Rights of Machines. *Philosophy & Technology*, *27*(1), 113–132. https://doi.org/10.1007/s13347-013-0121-z

Gunkel, D. J. (2015). The Rights of Machines: Caring for Robotic Care-Givers. In S. P. van Rysewyk & M. Pontier (Eds.), *Machine Medical Ethics* (pp. 151–166). Cham: Springer International Publishing. https://doi.org/10.1007/978-3-319-08108-3_10

Gunkel, D. J. (2016). David J. Gunkel - Information. http://gunkelweb.com/info.html. Accessed 22 September 2021

Gunkel, D. J. (2018). *Robot Rights*. Cambridge, MA: MIT Press.

Gunkel, D. J. (2020). Shifting Perspectives. *Science and Engineering Ethics*, *26*(5), 2527–2532. https://doi.org/10.1007/s11948-020-00247-9

Hajdin, M. (1987). *Agents, Patients, and Moral Discourse*. McGill University. Retrieved from https://central.bac-lac.gc.ca/.item?id=TC-QMM-75751&op=pdf&app=Library&oclc_number=897472150

Hale, B. (2009). Technology, the Environment and the Moral Considerability of Artefacts. In J. K. B. Olsen, E. Selinger, & S. Riis (Eds.), *New Waves in Philosophy of Technology* (pp. 216–240). London: Palgrave Macmillan UK. https://doi.org/10.1057/9780230227279_11

Hall, J. S. (2000). Ethics for Machines. https://www.kurzweilai.net/ethics-for-machines. Accessed 22 September 2021

Harris, J. (2019, May 17). What can the farmed animal movement learn from history? *Sentience Institute*. http://www.sentienceinstitute.org/blog/what-can-the-farmed-animal-movement-learn-from-history. Accessed 31 December 2021

Harris, J. (2021). The Importance of Artificial Sentience. *Sentience Institute*. http://www.sentienceinstitute.org/blog/the-importance-of-artificial-sentience. Accessed 30 December 2021

Harris, J., & Anthis, J. R. (2021). The Moral Consideration of Artificial Entities: A Literature Review. *Science and Engineering Ethics*, *27*(4), 53. https://doi.org/10.1007/s11948-021-00331-8

Harrison, P. (1992). Descartes on Animals. *The Philosophical Quarterly*, *42*(167), 219–227. https://doi.org/10.2307/2220217

Hartmann, T., Toz, E., & Brandon, M. (2010). Just a Game? Unjustified Virtual Violence Produces Guilt in Empathetic Players. *Media Psychology*, *13*(4), 339–363. https://doi.org/10.1080/15213269.2010.524912

Harvard University. (2021). Hilary Putnam Bibliography. https://philosophy.fas.harvard.edu/people/hilary-putnam. Accessed 18 November 2021

Haslam, N. (2006). Dehumanization: An Integrative Review. *Personality and Social Psychology Review*, *10*(3), 252–264. https://doi.org/10.1207/s15327957pspr1003_4

Herrick, B. (2002). Evolution Paradigms and Constitutional Rights: The Imminent Danger of Artificial Intelligence. *Student Scholarship*, 50.

Hewett, T. T., Baecker, R., Card, S., Carey, T., Gasen, J., Mantei, M., et al. (1992). *ACM SIGCHI Curricula for Human-Computer Interaction*. New York, NY: Association for Computing Machinery.

Himma, K. E. (2004). There's something about Mary: The moral value of things qua information objects. *Ethics and Information Technology*, *6*(3), 145–159. https://doi.org/10.1007/s10676-004-3804-4

Holland, O., & Goodman, R. (2003). Robots With Internal Models A Route to Machine Consciousness? *Journal of Consciousness Studies*, *10*(4–5), 77–109.

Hook, S. (1960). A Pragmatic Note. In S. Hook (Ed.), *Dimensions of Mind* (pp. 202–207). New York, NY: New York University Press.

Horstmann, A. C., Bock, N., Linhuber, E., Szczuka, J. M., Straßmann, C., & Krämer, N. C. (2018). Do a robot's social skills and its objection discourage interactants from switching the robot off? *PLOS ONE*, *13*(7), e0201581. https://doi.org/10.1371/journal.pone.0201581

Hu, S. D. (1987). What Software Engineers and Managers Need to Know. In S. D. Hu (Ed.), *Expert Systems for Software Engineers and Managers* (pp. 38–64). Boston, MA: Springer US. https://doi.org/10.1007/978-1-4613-1065-5_3

Hughes, J. J. (2005). *Report on the 2005 Interests and Beliefs Survey of the Members of the World Transhumanist Association* (p. 16). World Transhumanist Association.

Inayatullah, S. (2001a). The Rights of Robot: Inclusion, Courts and Unexpected Futures. *Journal of Futures Studies*, *6*(2), 93–102.

Inayatullah, S. (2001b). The Rights of Your Robots: Exclusion and Inclusion in History and Future. https://www.kurzweilai.net/the-rights-of-your-robots-exclusion-and-inclusion-in-history-and-future. Accessed 16 November 2021

James, M., & Scott, K. (2008). *Robots & Rights: Will Artificial Intelligence Change The Meaning Of Human Rights?* London, UK: BioCentre. https://www.bioethics.ac.uk/cmsfiles/files/resources/biocentre_symposium_report__robots_and_rights_150108.pdf

Jenkins, P. (2006). *Historical Simulations - Motivational, Ethical and Legal Issues* (SSRN Scholarly Paper No. ID 929327). Rochester, NY: Social Science Research Network. https://papers.ssrn.com/abstract=929327. Accessed 16 November 2021

Kahn, P. H., Friedman, B., Perez-Granados, D. R., & Freier, N. G. (2004). Robotic pets in the lives of preschool children. In *CHI '04 Extended Abstracts on Human Factors in Computing Systems* (pp. 1449–1452). New York, NY, USA: Association for Computing Machinery. https://doi.org/10.1145/985921.986087

Kahn, P. H., Ishiguro, H., Friedman, B., & Kanda, T. (2006). What is a Human? - Toward Psychological Benchmarks in the Field of Human-Robot Interaction. In *ROMAN 2006 - The 15th IEEE International Symposium on Robot and Human Interactive Communication* (pp. 364–371). Presented at the ROMAN 2006 - The 15th IEEE International Symposium on Robot and Human Interactive Communication. https://doi.org/10.1109/ROMAN.2006.314461

Kahn, P. H., Kanda, T., Ishiguro, H., Freier, N. G., Severson, R. L., Gill, B. T., et al. (2012). "Robovie, you'll have to go into the closet now": Children's social and moral relationships with a humanoid robot. *Developmental Psychology*, *48*(2), 303–314. https://doi.org/10.1037/a0027033

Kak, S. (2021). The Limits to Machine Consciousness. *Journal of Artificial Intelligence and Consciousness*, 1–14. https://doi.org/10.1142/S2705078521500193

Karnow, C. E. A. (1994). The Encrypted Self: Fleshing out the Rights of Electronic Personalities. *John Marshall Journal of Computer and Information Law*, *13*, 1.

Karnow, C. E. A. (1996). Liability for Distributed Artificial Intelligences. *Berkeley Technology Law Journal*, *11*(1), 147–204.

Kester, C. M. (1993). Is There a Person in That Body: An Argument for the Priority of Persons and the Need for a New Legal Paradigm. *Georgetown Law Journal*, *82*, 1643.

Kim, J. (2005). Making Right(s) Decision: Artificial Life and Rights Reconsidered. Presented at the DiGRA Conference. https://citeseerx.ist.psu.edu/viewdoc/download?doi=10.1.1.96.8255&rep=rep1&type=pdf

Kim, J., & Petrina, S. (2006). Artificial life rights: Facing moral dilemmas through The Sims. *Educational Insights*, *10*(2), 12.

Krach, S., Hegel, F., Wrede, B., Sagerer, G., Binkofski, F., & Kircher, T. (2008). Can Machines Think? Interaction and Perspective Taking with Robots Investigated via fMRI. *PLOS ONE*, *3*(7), e2597. https://doi.org/10.1371/journal.pone.0002597

Krebs, S. (2006). On the Anticipation of Ethical Conflicts between Humans and Robots in Japanese Mangas. *The International Review of Information Ethics*, *6*, 63–68. https://doi.org/10.29173/irie141

Krenn, B., & Gstrein, E. (2006). The Human Behind: Strategies Against Agent Abuse. In A. De Angeli, S. Brahnam, P. Wallis, & A. Dix (Eds.), *Misuse and Abuse of Interactive Technologies* (pp. 33–36). Presented at the CHI 2006 Conference on Human Factors in Computing Systems, Montréal Québec Canada: ACM. https://doi.org/10.1145/1125451.1125753

Krogh, C. (1996). The rights of agents. In M. Wooldridge, J. P. Müller, & M. Tambe (Eds.), *Intelligent Agents II Agent Theories, Architectures, and Languages* (pp. 1–16). Berlin, Heidelberg: Springer. https://doi.org/10.1007/3540608052_55

Kurzweil, R. (1999). *The Age of Spiritual Machines: When Computers Exceed Human Intelligence*. New York, NY: Viking.

Küster, D., & Świderska, A. (2016). Moral Patients: What Drives the Perceptions of Moral Actions Towards Humans and Robots? In J. Seibt, M. Nørskov, & S. S. Andersen (Eds.), *What Social Robots Can and Should Do: Proceedings of Robophilosophy 2016 / TRANSOR 2016* (pp. 340–343). Amsterdam, Netherlands: IOS Press.

Küster, D., Swiderska, A., & Gunkel, D. (2021). I saw it on YouTube! How online videos shape perceptions of mind, morality, and fears about robots. *New Media & Society*, *23*(11), 3312–3331. https://doi.org/10.1177/1461444820954199

LaChat, M. R. (1986). Artificial Intelligence and Ethics: An Exercise in the Moral Imagination. *AI Magazine*, *7*(2), 70–70. https://doi.org/10.1609/aimag.v7i2.540

Laham, S. M. (2009). Expanding the moral circle: Inclusion and exclusion mindsets and the circle of moral regard. *Journal of Experimental Social Psychology*, *45*(1), 250–253. https://doi.org/10.1016/j.jesp.2008.08.012

Laukyte, M. (2017). Artificial agents among us: Should we recognize them as agents proper? *Ethics and Information Technology*, *19*(1), 1–17. https://doi.org/10.1007/s10676-016-9411-3

lawyers.com. (2022). Marshal S. Willick, Esq. https://www.lawyers.com/las-vegas/nevada/marshal-s-willick-esq-1067567-a/. Accessed 12 January 2022

Lehman-Wilzig, S. N. (1981). Frankenstein unbound: Towards a legal definition of artificial intelligence. *Futures*, *13*(6), 442–457. https://doi.org/10.1016/0016-3287(81)90100-2

Leopold, A. (1949). *A Sand County Almanac, and Sketches Here and There*. New York, NY: Oxford University Press.

Levy, D. (2005). *Robots Unlimited: Life in a Virtual Age*. CRC Press.

Levy, D. (2009). The Ethical Treatment of Artificially Conscious Robots. *International Journal of Social Robotics*, *1*(3), 209–216. https://doi.org/10.1007/s12369-009-0022-6

Lichocki, P., Kahn, P. H., & Billard, A. (2011). A Survey of the Robotics Ethical Landscape. *IEEE Robotics & Automation Magazine*, *18*(1), 39–50.

Lima, G., Kim, C., Ryu, S., Jeon, C., & Cha, M. (2020). Collecting the Public Perception of AI and Robot Rights. *Proceedings of the ACM on Human-Computer Interaction*, *4*(CSCW2), 135:1–135:24. https://doi.org/10.1145/3415206

Loughnan, S., & Haslam, N. (2007). Animals and Androids: Implicit Associations Between Social Categories and Nonhumans. *Psychological Science*, *18*(2), 116–121. https://doi.org/10.1111/j.1467-9280.2007.01858.x

Lycan, W. G. (1985). Abortion and the Civil Rights of Machines. In N. T. Potter & M. Timmons (Eds.), *Morality and Universality: Essays on Ethical Universalizability* (pp. 139–156). Dordrecht: Springer Netherlands. https://doi.org/10.1007/978-94-009-5285-0_7

MacAskill, W. (2019). The Definition of Effective Altruism. In *Effective Altruism* (pp. 10–28). Oxford University Press. https://doi.org/10.1093/oso/9780198841364.003.0001

MacAskill, W. (2022). *What We Owe the Future*. [Unpublished manuscript]. https://www.williammacaskill.com/what-we-owe-the-future

MacDorman, K. F., & Cowley, S. J. (2006). Long-term relationships as a benchmark for robot personhood. In *ROMAN 2006 - The 15th IEEE International Symposium on Robot and Human Interactive Communication* (pp. 378–383). Presented at the ROMAN 2006 - The 15th IEEE International Symposium on Robot and Human Interactive Communication. https://doi.org/10.1109/ROMAN.2006.314463

Magnani, L. (2005). Technological Artifacts as Moral Carriers and Mediators. In *Machine ethics, papers from AAAI fall symposium technical report FS-05-06* (pp. 62–69). https://www.aaai.org/Papers/Symposia/Fall/2005/FS-05-06/FS05-06-009.pdf

Magnuson, M. A. (2014). What is Transhumanism? *What is Transhumanism?* https://whatistranshumanism.org/. Accessed 23 November 2021

Matthews, D. (2014). This guy thinks killing video game characters is immoral. *Vox*. https://www.vox.com/2014/4/23/5643418/this-guy-thinks-killing-video-game-characters-is-immoral. Accessed 11 March 2022

McCarthy, J., Minsky, M. L., Rochester, N., & Shannon, C. E. (2006). A Proposal for the Dartmouth Summer Research Project on Artificial Intelligence, August 31, 1955. *AI Magazine*, *27*(4), 12–12. https://doi.org/10.1609/aimag.v27i4.1904

McCorduck, P. (2004). *Machines Who Think: A Personal Inquiry into the History and Prospects of Artificial Intelligence*. Boca Raton, FL: CRC Press.

McNally, P., & Inayatullah, S. (1988). The rights of robots: Technology, culture and law in the 21st century. *Futures*, *20*(2), 119–136. https://doi.org/10.1016/0016-3287(88)90019-5

Melson, G. F., Kahn, P. H., Beck, A., Friedman, B., Roberts, T., Garrett, E., & Gill, B. T. (2009). Children's behavior toward and understanding of robotic and living dogs. *Journal of Applied Developmental Psychology*, *30*(2), 92–102. https://doi.org/10.1016/j.appdev.2008.10.011

Melson, G. F., Kahn, P. H., Jr., Beck, A., & Friedman, B. (2009). Robotic Pets in Human Lives: Implications for the Human–Animal Bond and for Human Relationships with Personified Technologies. *Journal of Social Issues*, *65*(3), 545–567. https://doi.org/10.1111/j.1540-4560.2009.01613.x

Metzinger, T. (2013). Two Principles for Robot Ethics. In J.-P. Günther & E. Hilgendorf (Eds.), *Robotik und Gesetzgebung* (pp. 263-302.). Baden-Baden, Germany: Nomos. https://doi.org/10.5771/9783845242200-263

Miah, A. (2009). A Critical History of Posthumanism. In B. Gordijn & R. Chadwick (Eds.), *Medical Enhancement and Posthumanity* (pp. 71–94). Dordrecht: Springer Netherlands. https://doi.org/10.1007/978-1-4020-8852-0_6

Milgram, S. (1974). *Obedience to Authority: An Experimental View*. New York, NY: Harper & Row. https://repository.library.georgetown.edu/handle/10822/766828. Accessed 14 December 2021

Minsky, M. (1994). Will Robots Inherit the Earth? *Scientific American*, *271*(4), 108–113.

Minsky, M. L. (1991). Conscious Machines. In *National Research Council of Canada, 75th Anniversary Symposium on Science in Society*. http://www.aurellem.org/6.868/resources/conscious-machines.html

Misselhorn, C. (2009). Empathy with Inanimate Objects and the Uncanny Valley. *Minds and Machines*, *19*(3), 345. https://doi.org/10.1007/s11023-009-9158-2

Mittelstadt, B. D., & Floridi, L. (2016). The Ethics of Big Data: Current and Foreseeable Issues in Biomedical Contexts. In B. D. Mittelstadt & L. Floridi (Eds.), *The Ethics of Biomedical Big Data* (pp. 445–480). Cham: Springer International Publishing. https://doi.org/10.1007/978-3-319-33525-4_19

Modern Mechanix. (1957). You'll Own "Slaves" by 1965. https://www.impactlab.com/2008/04/14/you'll-own-"slaves"-by-1965/. Accessed 29 December 2021

Moravec, H. (1988). *Mind Children: The Future of Robot and Human Intelligence*. Harvard University Press.

Moravec, H. (1998). When will computer hardware match the human brain? *Journal of Evolution and Technology*, *1*, 12.

Moravec, H. P. (2000). *Robot: Mere Machine to Transcendent Mind*. Oxford University Press.

Muehlhauser, L. (2017). Some Case Studies in Early Field Growth. *Open Philanthropy*. https://www.openphilanthropy.org/research/history-of-philanthropy/some-case-studies-early-field-growth. Accessed 31 December 2021

Musial, M. (2016). Magical Thinking and Empathy Towards Robots. In J. Seibt, M. Nørskov, & S. S. Andersen (Eds.), *What Social Robots Can and Should Do: Proceedings of Robophilosophy 2016 / TRANSOR 2016* (pp. 347–356). Amsterdam, Netherlands: IOS Press.

Nilsson, N. J. (2009). *The Quest for Artificial Intelligence: A History of Ideas and Achievements*. Cambridge: Cambridge University Press. https://doi.org/10.1017/CBO9780511819346

Nomura, T., Kanda, T., & Suzuki, T. (2006). Experimental investigation into influence of negative attitudes toward robots on human–robot interaction. *AI & SOCIETY*, *20*(2), 138–150. https://doi.org/10.1007/s00146-005-0012-7

Nomura, T., Uratani, T., Kanda, T., Matsumoto, K., Kidokoro, H., Suehiro, Y., & Yamada, S. (2015). Why Do Children Abuse Robots? In *Proceedings of the Tenth Annual ACM/IEEE International Conference on Human-Robot Interaction Extended Abstracts* (pp. 63–64). Presented at the HRI '15: ACM/IEEE International Conference on Human-Robot Interaction, Portland Oregon USA: ACM. https://doi.org/10.1145/2701973.2701977

Nørskov, M., Seibt, J., & Quick, O. S. (2021). *Culturally Sustainable Social Robotics: Proceedings of Robophilosophy 2020*. IOS Press.

Oppy, G., & Dowe, D. (2021). The Turing Test. In E. N. Zalta (Ed.), *The Stanford Encyclopedia of Philosophy* (Winter 2021.). Metaphysics Research Lab, Stanford University. https://plato.stanford.edu/archives/win2021/entriesuring-test/. Accessed 18 November 2021

Pauketat, J. V. (2021). *The Terminology of Artificial Sentience*. PsyArXiv. https://doi.org/10.31234/osf.io/sujwf

Petersen, S. (2007). The ethics of robot servitude. *Journal of Experimental & Theoretical Artificial Intelligence*, *19*(1), 43–54. https://doi.org/10.1080/09528130601116139

Petrina, S., Volk, K., & Kim, S. (2004). Technology and Rights. *International Journal of Technology and Design Education*, *14*(3), 181–204. https://doi.org/10.1007/s10798-004-0809-6

PETRL. (2015). People for the Ethical Treatment of Reinforcement Learners. http://www.petrl.org/. Accessed 25 November 2021

Platt, C. (1995). Superhumanism. *Wired*. https://www.wired.com/1995/10/moravec/. Accessed 29 December 2021

Putnam, H. (1960). Minds and Machines. In S. Hook (Ed.), *Dimensions of Mind* (pp. 148–180). New York, NY: New York University Press.

Putnam, H. (1964). Robots: Machines or Artificially Created Life? *The Journal of Philosophy*, *61*(21), 668–691. https://doi.org/10.2307/2023045

Radio New Zealand. (2020). The Morality of Abusing A Robot. *Nights*. https://www.rnz.co.nz/national/programmes/nights/audio/2018757787/the-morality-of-abusing-a-robot. Accessed 7 December 2021

Reed, A. II., & Aquino, K. F. (2003). Moral identity and the expanding circle of moral regard toward out-groups. *Journal of Personality and Social Psychology*, *84*(6), 1270–1286. https://doi.org/10.1037/0022-3514.84.6.1270

Regan, T. (2004). *The Case for Animal Rights*. University of California Press.

Reggia, J. (2013). The rise of machine consciousness: Studying consciousness with computational models. *Neural Networks*, *44*, 112–131. https://doi.org/10.1016/j.neunet.2013.03.011

Robertson, J. (2014). HUMAN RIGHTS VS. ROBOT RIGHTS: Forecasts from Japan. *Critical Asian Studies*, *46*(4), 571–598. https://doi.org/10.1080/14672715.2014.960707

Rogerson, N. B. F., S. (2001). A moral approach to electronic patient records. *Medical Informatics and the Internet in Medicine*, *26*(3), 219–234. https://doi.org/10.1080/14639230110076412

Rorvik, D. M. (1979). *As Man Becomes Machine: The Evolution of the Cyborg*. London, UK: Sphere Books Limited.

Rosenthal-von der Pütten, A. M., Krämer, N. C., Hoffmann, L., Sobieraj, S., & Eimler, S. C. (2013). An Experimental Study on Emotional Reactions Towards a Robot. *International Journal of Social Robotics*, *5*(1), 17–34. https://doi.org/10.1007/s12369-012-0173-8

Roser, M., & Ritchie, H. (2013). Technological Progress. *Our World in Data*. https://ourworldindata.org/technological-progress. Accessed 31 December 2021

Ruzich, C. M. (2006). With Deepest Sympathy: Understanding Computer Crashes, Grief, and Loss. In A. De Angeli, S. Brahnam, P. Wallis, & A. Dix (Eds.), *Misuse and Abuse of Interactive Technologies* (pp. 37–40). Presented at the CHI 2006 Conference on Human Factors in Computing Systems, Montréal Québec Canada: ACM. https://doi.org/10.1145/1125451.1125753

Ryder, R.D. (1992). Painism: Ethics, Animal Rights and Environmentalism. *Global Bioethics*, *5*(4), 27–35. https://doi.org/10.1080/11287462.1992.10800621

Ryder, Richard D. (1975). *Victims of Science: The Use of Animals in Research*. London, UK: Davis Poynter. Accessed 22 December 2021

Sapontzis, S. F. (1981). A Critique of Personhood. *Ethics*, *91*(4), 607–618.

Savulescu, J., & Bostrom, N. (2009). *Human Enhancement*. OUP Oxford.

Scheutz, M., & Crowell, C. R. (2007). The Burden of Embodied Autonomy: Some Reflections on the Social and Ethical Implications of Autonomous Robots. In *Proceedings of Workshop on Roboethics at ICRA 2007*. Rome, Italy. https://citeseerx.ist.psu.edu/viewdoc/download?doi=10.1.1.115.6076&rep=rep1&type=pdf

Schmidt, M., Ganguli-Mitra, A., Torgersen, H., Kelle, A., Deplazes, A., & Biller-Andorno, N. (2009). A priority paper for the societal and ethical aspects of synthetic biology. *Systems and Synthetic Biology*, *3*(1), 3. https://doi.org/10.1007/s11693-009-9034-7

Schweighofer, E. (2001). Vorüberlegungen zu künstlichen Personen: autonome Roboter und intelligente Softwareagenten. *Jusletter IT*, (IRIS). https://jusletter-it.weblaw.ch/issues/2001/IRIS/TB-3_I4_Schweighofer.html__ONCE&login=false. Accessed 31 December 2021

Scriven, M. (1960). The Compleat Robot: A Prolegomena to Androidology. In S. Hook (Ed.), *Dimensions of Mind* (pp. 118–147). New York, NY: New York University Press.

Searle, J. (2009). Chinese room argument. *Scholarpedia*, *4*(8), 3100. https://doi.org/10.4249/scholarpedia.3100

Searle, J. R. (1980). Minds, brains, and programs. *Behavioral and Brain Sciences*, *3*(3), 417–424. https://doi.org/10.1017/S0140525X00005756

Seibt, J., Nørskov, M., & Hakli, R. (2014). *Sociable Robots and the Future of Social Relations: Proceedings of Robo-Philosophy 2014*. IOS Press.

Severson, R. J. (1997). *The Principles of Information Ethics*. M.E. Sharpe.

Shulman, C., & Bostrom, N. (2021). Sharing the World with Digital Minds. In *Rethinking Moral Status* (pp. 306–326). Oxford University Press. https://doi.org/10.1093/oso/9780192894076.003.0018

Simon, M. A. (1969). Could There Be a Conscious Automaton? *American Philosophical Quarterly*, *6*(1), 71–78.

Singer, P. (1995). *Animal Liberation*. Random House.

Singer, P., & Sagan, A. (2009, December 14). When robots have feelings. *The Guardian*. https://www.theguardian.com/commentisfree/2009/dec/14/rage-against-machines-robots. Accessed 22 December 2021

Siponen, M. (2000). Is Polyinstantation Morally Blameworthy? *AMCIS 2000 Proceedings*. https://aisel.aisnet.org/amcis2000/227

Slater, M., Antley, A., Davison, A., Swapp, D., Guger, C., Barker, C., et al. (2006). A Virtual Reprise of the Stanley Milgram Obedience Experiments. *PLOS ONE*, *1*(1), e39. https://doi.org/10.1371/journal.pone.0000039

Solum, L. B. (1992). Legal Personhood for Artificial Intelligences. *North Carolina Law Review*, *70*, 1231–1287.

Solum, L. B. (2001). To Our Children's Children's Children: The Problems of Intergenerational Ethics. *Loyola of Los Angeles Law Review*, *35*, 163.

Solum, L. B. (2014). Artificial Meaning. *Washington Law Review*, *89*, 69.

Solum, L. B. (2019). *Artificially Intelligent Law* (SSRN Scholarly Paper No. ID 3337696). Rochester, NY: Social Science Research Network. https://doi.org/10.2139/ssrn.3337696

Søraker, J. H. (2006a). The Moral Status of Information and Information Technologies: A Relational Theory of Moral Status. In S. Hongladarom & C. Ess (Eds.), *Information Technology Ethics: Cultural Perspectives* (pp. 1–19). Hershey, PA: Idea Group Reference.

Søraker, J. H. (2006b). The role of pragmatic arguments in computer ethics. *Ethics and Information Technology*, *8*(3), 121–130. https://doi.org/10.1007/s10676-006-9119-x

Sotala, K., & Gloor, L. (2017). Superintelligence As a Cause or Cure For Risks of Astronomical Suffering. *Informatica*, *41*(4). https://www.informatica.si/index.php/informatica/article/view/1877. Accessed 25 November 2021

Sparrow, R. (2004). The Turing Triage Test. *Ethics and Information Technology*, *6*(4), 203–213. https://doi.org/10.1007/s10676-004-6491-2

Spence, P. R., Edwards, A., & Edwards, C. (2018). Attitudes, Prior Interaction, and Petitioner Credibility Predict Support for Considering the Rights of Robots. In *Companion of the 2018 ACM/IEEE International Conference on Human-Robot Interaction* (pp. 243–244). New York, NY, USA: Association for Computing Machinery. https://doi.org/10.1145/3173386.3177071

Spennemann, D. H. R. (2007). On the Cultural Heritage of Robots. *International Journal of Heritage Studies*, *13*(1), 4–21. https://doi.org/10.1080/13527250601010828

Stahl, B. C., & Coeckelbergh, M. (2016). Ethics of healthcare robotics: Towards responsible research and innovation. *Robotics and Autonomous Systems*, *86*, 152–161. https://doi.org/10.1016/j.robot.2016.08.018

Starmans, C., & Friedman, O. (2016). If I am free, you can't own me: Autonomy makes entities less ownable. *Cognition*, *148*, 145–153. https://doi.org/10.1016/j.cognition.2015.11.001

Steinert, S. (2013). The Five Robots—A Taxonomy for Roboethics. *International Journal of Social Robotics*, *6*, 249–260. https://doi.org/10.1007/s12369-013-0221-z

Stone, C. D. (1972). Should Trees Have Standing--Toward Legal Rights for Natural Objects. *Southern California Law Review*, *45*, 450–501.

Sudia, F. W. (2001). A Jurisprudence of Artilects: Blueprint for a Synthetic Citizen. *Journal of Futures Studies*, *6*(2), 65–80.

Sullins, J. P. (2005). Ethics and Artificial life: From Modeling to Moral Agents. *Ethics and Information Technology*, *7*(3), 139–148. https://doi.org/10.1007/s10676-006-0003-5

Sullins, J. P. (2009). Artificial Moral Agency in Technoethics. In R. Luppicini & R. Adell (Eds.), *Handbook of Research on Technoethics* (pp. 205–221). Information Science Reference. https://doi.org/10.4018/978-1-60566-022-6.ch014

Swiderska, A., & Küster, D. (2018). Avatars in Pain: Visible Harm Enhances Mind Perception in Humans and Robots. *Perception*, *47*(12), 1139–1152. https://doi.org/10.1177/0301006618809919

Tavani, H. T. (2002). The uniqueness debate in computer ethics: What exactly is at issue, and why does it matter? *Ethics and Information Technology*, *4*(1), 37–54. https://doi.org/10.1023/A:1015283808882

Taylor, P. W. (1981). The Ethics of Respect for Nature. *Environmental Ethics, 3*(3), 197-218.

Taylor, P. W. (2011). *Respect for Nature: A Theory of Environmental Ethics*. Princeton, NJ: Princeton University Press. https://press.princeton.edu/books/paperback/9780691150246/respect-for-nature. Accessed 22 December 2021

The American Society for the Prevention of Cruelty to Robots. (1999a). Frequently Asked Questions. http://www.aspcr.com/newcss_faq.html. Accessed 17 November 2021

The American Society for the Prevention of Cruelty to Robots. (1999b). What is a Robot? http://www.aspcr.com/newcss_robots.html. Accessed 17 November 2021

Thompson, D. (1965). Can a machine be conscious?1. *The British Journal for the Philosophy of Science*, *16*(61), 33–43. https://doi.org/10.1093/bjps/XVI.61.33

Thornhill, J. (2017, March 3). Philosopher Daniel Dennett on AI, robots and religion. *Financial Times*. https://www.ft.com/content/96187a7a-fce5-11e6-96f8-3700c5664d30. Accessed 19 November 2021

Tomasik, B. (2011). Risks of Astronomical Future Suffering. *Center on Long-Term Risk*. https://longtermrisk.org/files/risks-of-astronomical-future-suffering.pdf

Tomasik, B. (2014). Do Artificial Reinforcement-Learning Agents Matter Morally? *arXiv:1410.8233 [cs]*. http://arxiv.org/abs/1410.8233. Accessed 29 November 2021

Tomasik, B. (2015). A Dialogue on Suffering Subroutines. *Center on Long-Term Risk*. https://longtermrisk.org/a-dialogue-on-suffering-subroutines/. Accessed 25 November 2021

Tonkens, R. (2012). Out of character: on the creation of virtuous machines. *Ethics and Information Technology*, *14*(2), 137–149. https://doi.org/10.1007/s10676-012-9290-1

Torrance, S. (1984). Editor's Introduction: Philosophy and AI: Some Issues. In S. Torrance (Ed.), *The Mind and the Machine: Philosophical Aspects of Artificial Intelligence* (pp. 11–28). Chichester, UK: Ellis Horwood.

Torrance, S. (2000). Towards an Ethics for Epersons. *AISB Quarterly*, 38–41.

Torrance, S. (2005). A Robust View of Machine Ethics. In *Papers from the 2005 AAAI Fall Symposium* (pp. 88–93). Menlo Park, CA: The AAAI Press.

Torrance, S. (2008). Ethics and Consciousness in Artificial Agents. *AI and Society*, *22*(4), 495–521. https://doi.org/10.1007/s00146-007-0091-8

Torrance, S., Tamburrini, G., & Datteri, E. (2006). The Ethical Status of Artificial Agents – With and Without Consciousness. In *Ethics of Human Interaction with Robotic, Bionic and AI Systems: Concepts and Policies*. Naples, Italy: Italian Institute for Philosophical Studies.

Treiblmaier, H., Madlberger, M., Knotzer, N., & Pollach, I. (2004). Evaluating personalization and customization from an ethical point of view: an empirical study. In *37th Annual Hawaii International Conference on System Sciences, 2004. Proceedings of the* (p. 10 pp.-). Presented at the 37th Annual Hawaii International Conference on System Sciences, 2004. Proceedings of the. https://doi.org/10.1109/HICSS.2004.1265434

Uzgalis, B. (2002). Information Informs the Field: A Conversation with Luciano Floridi. *APA Newsletters: Newsletter on Philosophy and Computers*, *2*(1), 72–77.

Vanman, E. J., & Kappas, A. (2019). "Danger, Will Robinson!" The challenges of social robots for intergroup relations. *Social and Personality Psychology Compass*, *13*(8), e12489. https://doi.org/10.1111/spc3.12489

Versenyi, L. (1974). Can Robots be Moral? *Ethics*, *84*(3), 248–259. https://doi.org/10.1086/291922

Veruggio, G. (2006). The EURON Roboethics Roadmap. In *2006 6th IEEE-RAS International Conference on Humanoid Robots* (pp. 612–617). Presented at the 2006 6th IEEE-RAS International Conference on Humanoid Robots. https://doi.org/10.1109/ICHR.2006.321337

Vigderson, T. (1994). Hamlet II: The Sequel: The Rights of Authors vs. Computer-Generated Read-Alike Works. *Loyola of Los Angeles Law Review*, *28*, 401.

Vize, B. (2011). *Do Androids Dream of Electric Shocks?* Victoria University of Wellington. Retrieved from http://researcharchive.vuw.ac.nz/xmlui/bitstream/handle/10063/1686/thesis.pdf?sequence=2

Volkman, R. (2010). Why Information Ethics Must Begin with Virtue Ethics. *Metaphilosophy*, *41*(3), 380–401. https://doi.org/10.1111/j.1467-9973.2010.01638.x

Walker, M. (2006). A Moral Paradox in the Creation of Artificial Intelligence: Mary Poppins 3000s of the World Unite. In *Human implications of human-robot Interaction: Papers from the AAAI workshop*. https://www.aaai.org/Papers/Workshops/2006/WS-06-09/WS06-09-005.pdf

Wallach, W., & Allen, C. (2008). *Moral Machines: Teaching Robots Right from Wrong*. Oxford University Press.

Ward, A. F., Olsen, A. S., & Wegner, D. M. (2013). The Harm-Made Mind: Observing Victimization Augments Attribution of Minds to Vegetative Patients, Robots, and the Dead. *Psychological Science*, *24*(8), 1437–1445. https://doi.org/10.1177/0956797612472343

Ware, M., & Mabe, M. (2015). *The STM Report: An overview of scientific and scholarly journal publishing* (p. 181). The Hague, The Netherlands: International Association of Scientific, Technical and Medical Publishers. https://digitalcommons.unl.edu/cgi/viewcontent.cgi?article=1008&context=scholcom

Wareham, C. (1AD). On the Moral Equality of Artificial Agents. In R. Luppicini (Ed.), *Moral, ethical, and social dilemmas in the age of technology: Theories and practice* (pp. 70–78). IGI Global. https://www.igi-global.com/gateway/chapter/www.igi-global.com/gateway/chapter/73611. Accessed 3 January 2022

Warwick, K. (2010). Implications and consequences of robots with biological brains. *Ethics and Information Technology*, *12*(3), 223–234. https://doi.org/10.1007/s10676-010-9218-6

Watanabe, S. (1960). Comments on Key Issues. In S. Hook (Ed.), *Dimensions of Mind* (pp. 143–147). New York, NY: New York University Press.

Waytz, A., Cacioppo, J., & Epley, N. (2010). Who Sees Human?: The Stability and Importance of Individual Differences in Anthropomorphism. *Perspectives on Psychological Science*, *5*(3), 219–232. https://doi.org/10.1177/1745691610369336

Wheeler, M. (2008). God's Machines: Descartes on the Mechanization of Mind. In P. Husbands, O. Holland, & M. Wheeler (Eds.), *The Mechanical Mind in History* (pp. 307–330). The MIT Press. https://doi.org/10.7551/mitpress/9780262083775.003.0013

Whitby, B. (1996). The Potential Moral Duties and Rights of Intelligent Artifacts. In *Reflections on Artificial Intelligence* (pp. 93–105). Intellect Books.

Whitby, B. (2008). Sometimes it's hard to be a robot: A call for action on the ethics of abusing artificial agents. *Interacting with Computers*, *20*(3), 326–333. Presented at the Interacting with Computers. https://doi.org/10.1016/j.intcom.2008.02.002

White, B. (1993). Sacrificial Rights: The Conflict between Free Exercise of Religion and Animal Rights. *St. John's Journal of Legal Commentary*, *9*, 835.

White, L. (1967). The Historical Roots of Our Ecologic Crisis. *Science*, *155*(3767), 1203–1207.

White, L. (1973). Continuing the Conversation. In I. G. Barbour (Ed.), *Western Man and Environmental Ethics: Attitudes Toward Nature and Technology* (pp. 55–64). Addison-Wesley Publishing Company.

Wiener, N. (1960). The Brain and the Machine (summary). In S. Hook (Ed.), *Dimensions of Mind* (pp. 113–117). New York, NY: New York University Press.

Wikipedia. (2021). List of fictional robots and androids. In *Wikipedia*. https://en.wikipedia.org/w/index.php?title=List_of_fictional_robots_and_androids&oldid=1052639591. Accessed 17 November 2021

Wilks, Y. (1975). Putnam and Clarke and Mind and Body. *The British Journal for the Philosophy of Science*, *26*(3), 213–225. https://doi.org/10.1093/bjps/26.3.213

Wilks, Y. (1985). Responsible Computers? In A. Joshi (Ed.), *Proceedings of the Ninth International Joint Conference on Artificial Intelligence* (pp. 1279–1280). Los Altos, CA: Kaufmann.

Wilks, Y. (1998). Liability and Consent. In A. Narayanan & M. Bennun (Eds.), *Law, Computer Science, and Artificial Intelligence*. Intellect Books.

Willick, M. (1985). Constitutional Law and Artificial Intelligence: The Potential Legal Recognition of Computers as "Persons". In A. Joshi (Ed.), *Proceedings of the Ninth International Joint Conference on Artificial Intelligence* (pp. 1271–1273). Los Altos, CA: Kaufmann.

Willick, M. S. (1983). Artificial Intelligence: Some Legal Approaches and Implications. *AI Magazine*, *4*(2), 5–5. https://doi.org/10.1609/aimag.v4i2.392

World Scientific. (2021). Journal of Artificial Intelligence and Consciousness. https://www.worldscientific.com/page/jaic/aims-scope

Yampolskiy, R., & Fox, J. (2013). Safety Engineering for Artificial General Intelligence. *Topoi*, *32*(2), 217–226. https://doi.org/10.1007/s11245-012-9128-9

York, P. F. (2005). Respect for the World: Universal Ethics and the Morality of Terraforming. https://www.tesionline.it/tesi/respect-for-the-world-universal-ethics-and-the-morality-of-terraforming/15012. Accessed 11 November 2021

Young, P. R. (1991). *Persons and artificial intelligence*. The Catholic University of America. Retrieved from https://www.proquest.com/openview/e780faa1e13d815919c3aa2bbb353892/1?pq-origsite=gscholar&cbl=18750&diss=y

Yudkowsky, E. (1996). Staring Into The Singularity. http://www.fairpoint.net/~jpierce/staring_into_the_singularity.htm. Accessed 23 November 2021

Yudkowsky, E. (2008). Artificial Intelligence as a Positive and Negative Factor in Global Risk. In N. Bostrom & M. M. Cirkovic (Eds.), *Global Catastrophic Risks* (pp. 308–345). Oxford, UK: OUP Oxford.

Yudkowsky, E. (2015). Mindcrime. *Arbital*. https://arbital.greaterwrong.com/p/mindcrime?l=6v. Accessed 29 November 2021

Zancanaro, M., & Leonardi, C. (2005). A trouble shared is a troubled halved: Disruptive and self-help patterns of usage for co-located interfaces. In A. De Angeli, S. Brahnam, & P. Wallis (Eds.), *Proceedings of Abuse: The darker side of Human-Computer Interaction*. http://www.agentabuse.org/Abuse_Workshop_WS5.pdf

Zhang, D., Mishra, S., Brynjolfsson, E., Etchemendy, J., Ganguli, D., Grosz, B., et al. (2021). *The AI Index 2021 Annual Report,*. Stanford, CA: AI Index Steering Committee, Human-Centered AI Institute, Stanford University. https://aiindex.stanford.edu/wp-content/uploads/2021/11/2021-AI-Index-Report_Master.pdf. Accessed 31 December 2021

Zhu, Q., Williams, T., Jackson, B., & Wen, R. (2020). Blame-Laden Moral Rebukes and the Morally Competent Robot: A Confucian Ethical Perspective. *Science and Engineering Ethics*, *26*(5), 2511–2526. https://doi.org/10.1007/s11948-020-00246-w

Ziesche, S., & Yampolskiy, R. (2019). Towards AI Welfare Science and Policies. *Big Data and Cognitive Computing*, *3*(1), 2. https://doi.org/10.3390/bdcc3010002

# Acknowledgements

Many thanks to Abby Sarfas, Ali Ladak, Elise Bohan, Jacy Reese Anthis, Thomas Moynihan, and Joshua Gellers for providing feedback on earlier drafts of this article.

# Funding

No specific financial support was received for this article.

# Data availability

The full results of the keyword searches and the list of included items are provided in a separate spreadsheet.